\documentclass[12pt]{puthesis}
\newcommand{\proquestmode}{}

\title{Networks in a World Unknown:\\Public WhatsApp Groups in the Venezuelan Refugee Crisis}

\submitted{June 2020}  
\copyrightyear{2020}  
\author{Adam Chang}
\adviser{Miklos Z. Racz}  
\department{Operations Research and Financial Engineering}

\usepackage{graphicx}
\usepackage{subcaption}

\usepackage{verbatim}
\usepackage{listings}
\usepackage{wrapfig}

\usepackage{amsmath}

\usepackage{multirow}
\usepackage{longtable}

\usepackage{booktabs}

\usepackage{url}

\usepackage{breakurl}
\usepackage[breaklinks]{hyperref}

\usepackage{listings}
\ifdefined\printmode

\usepackage{url}

\else

\ifdefined\proquestmode

\usepackage{hyperref}
\hypersetup{bookmarksnumbered}

\makeatletter
\hypersetup{pdftitle=\@title,pdfauthor=\@author}
\makeatother

\else

\usepackage{hyperref}
\hypersetup{colorlinks,bookmarksnumbered}

\makeatletter
\hypersetup{pdftitle={Networks in a World Unknown: Public WhatsApp Groups in the Venezuelan Refugee Crisis},pdfauthor=\@author}
\makeatother

\fi 
\fi 

\ifodd 0
\renewcommand{\maketitlepage}{}
\renewcommand*{\makecopyrightpage}{}
\renewcommand*{\makeabstract}{}

\else

\abstract{
By early March 2020, five million Venezuelans had fled their home country after its complete economic and institutional collapse, and over 1.6 million have migrated to Colombia. Migrants struggle to start their lives over in Colombia, having arrived with few economic resources, and often no legal documentation, in cities with little to offer them. Venezuelan migrants, however, rely heavily on mobile phones and social media networks as lifelines for information, opportunities, and resources---making WhatsApp both a critical tool for migrants' settlement and integration, as well as an invaluable source of data through which we can better understand migrant experiences.

This thesis explores the dynamics of public WhatsApp groups used by Venezuelan migrants to Colombia, and what they can tell us about how migrants use and share information. We center our research on information spread and trust, especially as they intersect with concentration and geographic heterogeneity within groups. We analyze messages and memberships broadly, then explore interaction within groups, fake news and economic scams, and effects of the coronavirus pandemic. Our results have a range of policy implications, from reflections on Colombia's decision to shut its borders amidst the coronavirus pandemic, to understandings of how aid organizations can effectively share information over social media channels.

}

\acknowledgements{
My most important thank you is \textit{gracias}, to the Venezuelans and Colombians who helped me begin to understand this crisis much better than I ever could have otherwise. Applying mathematics and computer science to social issues is fraught with phenomenal opportunities but also phenomenal dangers, lest we forget contexts and cultures. I'm hopeful for the former, but only because of how openly you shared your stories with me. You never asked for anything in return, but I owe you everything, and I'm so, so sorry that we---the rest of the world---haven't done more.

I had the tremendous pleasure of meeting and learning from heroes who have dedicated their lives to resolving this crisis. Ana and Sebastián (International Rescue Committee); Esperanza, Jen, and the Riohacha staff (MercyCorps); María, Jorge, Manuel, and so many Maicao staff (Save the Children); and Joshua (the freelance journalist behind \textit{Muros Invisibles}): I'm in infinite admiration of what you do, and infinitely thankful that you took the time to show me your world, all from cold emails.

Miki: thank you for so thoughtfully advising, and so earnestly cheering on, this rather unconventional project. From telco behemoths that no longer share their data but instead sell it for \$100,000s, to WhatsApp trying really hard to make this research impossible, there have been quite a few roadblocks, but you've always been incredibly understanding. From the very beginning, I've felt an immense freedom to flesh out ideas, and that is the best thing I could've asked for in an adviser.

Thank you to my family for their unconditional support. And thank you to my families on campus: Princeton Running Club, Aquinas friends, Brown Co-op, and others from whom I've learned and with whom I've laughed and loved. You've given me so much to look forward to each and every day I've spent at Princeton, my home.

Finally, I am grateful to the Fred Fox Fund, the School of Engineering and Applied Science, and the Program in Latin American Studies for their generous funding of field work and various technological aspects of this project.

}

\preface{
In the title, ``a world unknown'' has three important meanings, all of which motivated and guided this project.

First and foremost, it refers to the harrowing journey undertaken by Venezuelans fleeing a country under crisis, where people now eat garbage and die of treatable diseases because there is no medicine. The migration itself is perilous, but even after migrants arrive in places like Colombia, there are few resources for them and fewer jobs. To be sure, Venezuela and Colombia share much in common, but this is still a story of migrants starting over in a place they know little about, without people or institutions they can turn to. Can WhatsApp groups make the difference, helping migrants learn about crossing the border, or helping them find employment in Colombia? Should aid organizations disseminate information over these networks?

Second, the world of Venezuelan migrants is largely unknown outside Latin America, but perplexing even to the experts who study it. The transience and vulnerability of migrants---many cross irregularly, many live in informal settlements or on the street---make traditional data collection impossible, so an unconventional data source like WhatsApp groups and messages can offer otherwise undiscoverable insight.

Finally, so much of this is a world unknown to me. I'm immensely grateful for what I've learned at Princeton: mathematical concepts like probability distributions and regularization, and of equal importance, ideas about society and human development---topics like Locke and Rawls, the Rwandan genocide and Colombian civil conflict, and power and neocolonialism. But I started with very little knowledge about migration, WhatsApp, and the Venezuelan and Colombian contexts. More than anything else, this project is an exploration---of an unimaginable humanitarian crisis, of what we can learn from digital data, of how technological means can improve the lives of the least well-off, and of how a Princeton education grounded in the technical might intersect with communities and cultures halfway around the world.

}

\dedication{
To migrants---\\
and to those who make their journeys possible.\\
There are few greater injustices than the lottery of where we are born.
}

\fi  

\begin{document}
\makefrontmatter


\chapter{Introduction\label{ch:intro}}

Across their country, Venezuelans face widespread malnutrition, extreme medication shortages, exponential inflation, and a host of related concerns, including political repression and increased crime. By early March 2020, five million Venezuelans had fled their home country after its complete ``economic, social, and institutional collapse,'' with the figure expected to top eight million by the end of 2020 \cite{mercycorps-20190813}.

More than a third of Venezuelan migrants,\footnote{The 1951 Geneva Convention on refugees, the main international instrument of refugee law, defines refugees as those ``forced to flee [their] countries because of persecution, war or violence'' \cite{unhcr-whatis}. Per the United Nations, refugees have a ``well-founded fear of persecution for reasons of race, religion, nationality, political opinion or membership in a particular social group.'' This definition doesn't apply to the majority of Venezuelans, even given the country's devastating circumstances, so we instead use the term migrant.} over 1.6 million, have permanently migrated to Colombia, with most others residing in other Latin American countries. These are only the official counts; many do not cross legally. Thousands more cross back-and-forth between Venezuela and Colombia daily, in order to earn money and to purchase food, medicine, and other essentials no longer available in Venezuela.

\section{Venezuelans in Colombia}

Venezuelan migrants are often highly educated, having left professional and academic lives behind \cite{mercycorps-20190813}, but they're forced to resort to low-skill, low-paying jobs in Colombia, to the consternation of Colombians who face increased competition for already scarce work. For this and other reasons, including perceived increases in crime, the reception towards Venezuelan migrants in Colombia has been mixed, with many migrants facing xenophobia, especially in border regions where per capita concentrations surpass 20\%.

At the same time, hundreds of thousands of Venezuelans have been regularized---granted permanent residency and employment permits---under the administrations of Colombian presidents Juan Manuel Santos and Iván Duque. In August 2019, Duque announced that Colombia would grant citizenship to more than 24,000 undocumented children born to Venezuelan refugees, proclaiming, ``Today Colombia gives this message to the world: to those who want to use xenophobia for political goals, we take the path of fraternity'' \cite{nytimes-20190805}. Still, most migrants, especially newcomers, are not regularized, forcing them into informal work that's often exploitative.

On top of social and legal challenges to settlement, acceptance, and integration, migrants face a slew of economic difficulties. Migration journeys often involve robberies and violence, especially if migrants enter Colombia through \textit{trochas}, irregular border crossings controlled by criminal syndicates. Consequently, migrants come with few material possessions, often lacking legal documents or even the means to pay for them, and must eke out survival in border cities overwhelmed by migrants and their complicated needs. Many migrants sleep outside in public, and either beg or work informal hawking jobs on the street \cite{mercycorps-20190813}. Women, in particular, are often forced to resort to selling their bodies and/or parts thereof, with both prostitution and the sale of hair to wigmakers common practices along the border \cite{pri-20190204}.

\section{Digital Aspects of the Migrant Crisis}

Cell phones and social media networks serve as lifelines of information and resources for migrants during their arduous journeys. As Oscar Pérez, the president of the Unión Venezolana en Perú, a nonprofit that assists the settlement of migrants, says, ``The Venezuelan...has finished with that old adage that the best friend of man is the dog. For a Venezuelan, his best friend is the cellphone'' \cite{azpbs-20190430}. Mariangie Tarzona, a Venezuelan migrant who arrived in Lima in February 2017, describes her experience of resettlement by recounting, ``I had Jesus in social networks.'' Questions she asked over Facebook groups included: ``How much did you spend on your route?'', ``What was the best bus that you took?'', ``Where did you go to buy the bus ticket?'', ``How much did it cost you?'', ``What was the service like on it?'', ``How long did it take you to leave Venezuela?'', ``How is the border?'', and ``How are you doing now?'' \cite{azpbs-20190430}.

Joshua Collins, a journalist who has extensively covered the Venezuelan migrant crisis in Colombia, has reported on a “network of shelters, kitchens and healthcare checkpoints” for migrants along the resettlement routes from the Colombia-Venezuela border (many migrants make the 600km journey from Cúcuta to Bogotá on foot over eight days, passing through high altitude and subzero temperatures, because they cannot afford the \$30 bus fare) \cite{jcollins-20190628}. Per Collins, before and during this journey, migrants communicate information about distances, conditions, shelter availability, and other factors in various Facebook and WhatsApp groups.

Facebook and WhatsApp groups of strangers, however, do not come without their own complications. Because of how decentralized and democratic they are, users often encounter scams and misinformation, and in general don't trust information or users from such groups. In Chapter \ref{ch:reflections}, we discuss reflections on field interviews with Venezuelan migrants in Colombia, many of whom use but place little trust in public Facebook and WhatsApp groups.

\section{Research Motivation and Overview}

As we discussed in the preface, WhatsApp groups offer information, assistance, and resources that can help migrants in their settlement and integration. More than this, however, groups can also serve as an unconventional data source through which we research migrant experiences.

In Chapter \ref{ch:reflections}, we begin with reflections on two weeks of field work, in which we spoke to migrants about how they obtain information and resources, both offline and on social media networks, and heard their perspectives on---and experiences with---aspects of the crisis like xenophobia and regularization. We also spent time with three leading aid organizations, in an attempt to learn about their responses to the crisis, and to understand how they might begin to distribute information about their programs and offerings through public WhatsApp groups.

Chapter \ref{ch:relatedwork} discusses related work on social media networks, and the limited research done so far on WhatsApp groups. In Chapter \ref{ch:datamethod}, we share our methodology for collecting data from public WhatsApp groups, focusing on various technical challenges involved in joining and scraping WhatsApp groups en masse (including the all-too-real possibility of being banned from WhatsApp), as well as limitations of our methodology and the data we collect.

Next, Chapters \ref{ch:members} and \ref{ch:messages} dive into the memberships and messages within our collection of WhatsApp groups. We construct measures for the concentration, inequality, and geographic diversity of our groups, which are important characteristics that may affect how migrants connect and share information. We examine patterns in connections between users from different Latin American countries, as well as the network structure of both groups and users. In the chapter on messages, we analyze messages of various content types, and also construct a measure for group activity robust to our being removed from certain groups.

Chapter \ref{ch:replycascades} studies replies to messages, an important marker of attention and interaction given WhatsApp's minimal feature set. We discuss the limitations of using data about replies, and propose an alternate measure---(structural) virality---to better compare interaction across content types and groups. We show how these features are correlated with the group characteristics we constructed earlier, and provide possible explanations grounded in the context of the migrant crisis.

Chapter \ref{ch:misinformation} investigates misinformation---fake news and economic scams---within WhatsApp groups, beginning with how we identify and label misinformation. We attempt to understand how user and group characteristics are linked to the prevalence of misinformation, and later apply various machine learning classifiers to the problem of automatically detecting scams. Finally, in Chapter \ref{ch:coronavirus}, we briefly put our research in the context of the coronavirus pandemic and its consequences in Colombia, which include the closure of borders and a nationwide lockdown.

These are broad topics---and this is a broad thesis---but the issues related to Venezuelan migrants in Colombia are wide-ranging, and call for investigation along multiple intersecting perspectives. Moreover, studying the Venezuelan migrant crisis through WhatsApp groups is a completely new research area,\footnote{In general, research on WhatsApp groups is scant. As of April 2020, there are a total of 12 English-language papers, with most focused on political groups in Brazil.} compelling this kind of wide-ranging exploration.

\chapter{Reflections on Field Interviews\label{ch:reflections}}

I spent two weeks in Colombia in early January 2020, between the cities of Bogotá, Riohacha, Maicao, and another city on the border. While some migrants return home to Venezuela over the Christmas/New Year period, there were still ample Venezuelans to speak with across these cities, and the migrant flow was high during the last half of my trip, as Venezuelans who went home for the holidays returned to Colombia with more family members.

My interviews were in Spanish, except for those with Joshua Collins (an American journalist) and Jen Daum (an American who is Director of Programs at the NGO MercyCorps Colombia). In Bogotá, I spoke with Venezuelan street vendors and artisans, Colombian shopkeepers, the manager at a popular remittance service to Venezuela, Joshua, Jen, and various Venezuelan delivery workers for a GrubHub-like food delivery service. In Riohacha, I joined MercyCorps Colombia for two characterization activities, in which they surveyed potential recipients for an unconditional cash transfer scheme; I spoke with MercyCorps staff members, clients participating in the characterization, and migrants left out of the program. In Maicao, I spent a day with the charity Save the Children, which maintains vast operations in the city with a full-time staff of 120+. We visited the border and a sexual health clinic run by Save, and I spent time talking with Venezuelan migrants served by Save, with Save staff members, and with Save's wonderful director María Inés Fernández. Finally, elsewhere on the border, I met with the staff of the NGO International Rescue Committee, and interviewed visitors to their sexual health clinic. I also spoke with street vendors in the city and at the border crossing, as well as various (formal and informal) mobile phone vendors.

Notes from these interviews are included in Appendix \ref{ch:appendicies:field}. Below are reflections along general themes.

\section{WhatsApp Use}

Everyone I spoke to in Colombia knew what WhatsApp was, and this was true even amongst the Venezuelan migrant population. Use of WhatsApp, however, was limited by the need to own a smartphone (those who do have a smartphone all use WhatsApp...indeed, the app is one of the primary reasons for people to purchase a smartphone). Estimates of what percentage of Venezuelan migrants have smartphones varied wildly, with semi-official sources  giving answers from ``very few'' to ``nearly everyone.'' Around half of the integrated migrants I spoke with (those who had been in Colombia for at least several months) had smartphones.

With basic smartphones priced in the \$30-40 range in Colombia, and many times more in Venezuela (motivating many Venezuelans to purchase phones in Colombia when they pickup remittances), cost was the only reason I encountered for individuals to not have a smartphone. Most Venezuelan migrants seem to have smartphones intermittently---nearly everyone had one in Venezuela, but migrants either were robbed while crossing through the \textit{trochas} or sold their phones to pay for food or buses during their migration. High rates of crime, especially in La Guajira, further deterred people from owning smartphones.

The other factor that limited smartphone and WhatsApp use was the need for a data plan. Claro, the largest operator in Colombia (50\% market share), offers service at around \$12 monthly; the smaller providers (both of which have h20\% market share) offer service at around \$8. Neither offering is terribly cheap, especially for poorer migrants who may only make \$5 daily.

For those with smartphones, WhatsApp, as one migrant I spoke to stated, is ``primordial.'' Staying in contact with family in Venezuela always tended to be migrants’ primary reason for using WhatsApp, though such communication often also took place on international calls (which, while more expensive than WhatsApp, don’t require family members to also have smartphones and WhatsApp...but they do require electricity in family members' dwellings, which is never certain in Venezuela).

Those without smartphones are not out of the loop entirely. Most maintain Facebook accounts that they use primarily to communicate with family, and are able to access these accounts at cybercafes, or on borrowed phones. Many individuals, without smartphones of their own or in their immediately family, share smartphones with neighbors; Venezuelan migrant families often live with other migrant families in the same dwelling.

In general, smartphones and WhatsApp/Facebook are very well used amongst migrants. One NGO staffer told me of instances when aid recipients, while holding smartphones, told her how they did not have enough food to eat.

\section{Use of Public WhatsApp/Facebook Groups}

When stating uses for WhatsApp, no interviewee ever outright described large migrant-centric WhatsApp groups. Yet many did cite WhatsApp as an important source of information and resources, even beyond their immediate contacts, so use of these groups is likely prevalent amongst migrants who use WhatsApp. When I asked directly about such groups, around 50\% of those with a smartphone reported being active members of public WhatsApp groups, either currently or previously. Nearly everyone knew about these groups.

Motivations for using these groups, as described by individuals for themselves or people in general,\footnote{Framing it this way may have encouraged individuals to share motivations they would be embarrassed to assign to themselves (e.g., using these groups to find romantic partners).} included finding employment (primarily), finding assistance and aid resources (primarily), reading news about Venezuela (secondarily), buying and selling items, finding housing, and finding romantic partners. Nobody I spoke to mentioned going on these groups for fun, even though several of the public WhatsApp groups we joined, which were also amongst the most active, were dedicated to memes and jokes, their names literally translating to ``Fucking Around'' or ``Venezuelan Fuckers.'' It could certainly be that members in fun/social groups are from demographics I didn’t encounter as often (e.g., younger people with more time on their lands, and likely more stable economic situations that wouldn't bring them out onto the street), but it might also have been that people only chose to report more serious uses.\footnote{If you ask me why I spend time on Facebook, I’d answer to keep in touch with friends and follow their lives, even though much of my attention on Facebook diverts to random news, memes, and so on.}

Generally, migrants in large/public groups learned about such groups from friends, or were directly added by friends.

\section{Trust Towards Public WhatsApp/Facebook Groups}

Nobody I spoke to placed significant trust in public WhatsApp groups. Yet most peopled reported that they at least knww someone (personally) who trusted these groups enough to have conducted important transactions through them, especially finding employment. Several interviewees described horrific outcomes of such endeavors, including wage theft and outright sexual exploitation. Overall, the situation seemed like a 50/50, in that there do exist legitimate opportunities in these groups (which are almost certainly low-paying, like call center work), and because of that, there exists a decent contingent of migrants who expend serious efforts using these groups to find employment and/or assistance. In general, however, migrants' perception of large/public WhatsApp groups was that they were not an honest or accountable situation, and that many migrants only participate in transactions and/or employment offers out of desperation.

General information in these groups, whether news about Venezuela or information on how to obtain regularization, is seen as more trustworthy by migrants, around the same level as hearsay on the street. Migrants do have a good understanding of the various actors that might be in play behind this kind of ``free'' information, whether it be Venezuelan opposition forces or scammers who have something to gain from false information about the regularization process.

We discuss issues of trust in Chapter \ref{ch:misinformation}, but even more basically, it might be interesting to create some kind of central reputation system for these groups (i.e., a WhatsApp bot, if it doesn’t get banned). Imagine that our bot records and publicly displays users’ transaction histories---say, after every transaction, one party activates the bot with a command, and the bot waits for the counter-party to confirm that the transaction was successful. This concept certainly requires refinement, but something as simple as this would still be better than the completely uncertain landscape in which users currently perform transactions over WhatsApp.

\section{Gender}

One of the clearest conclusions I reached was that women face extremely difficult circumstances when using public WhatsApp groups to find employment. A significant percentage of employment offers---at least 25\%---is outright advertised as sex work (most commonly, being webcam models), but greater issues abound in employment advertised as ``domestic'' work or ``assistant'' work, with most such postings only soliciting females for some reason. Even if such positions are legitimate, many women I spoke with told me that those employment offers come with romantic and/or sexual strings attached.

In spite of this, it seemed like women and men used public WhatsApp groups equally, with one source even arguing to me that women are more likely to be in these groups, since their male partners typically do odd jobs or work in construction. Those industries not being available to women, women instead resort to WhatsApp groups to find other employment or assistance.

By far, women were much more likely to have relationships with aid organizations than men, sometimes by the design/choice of NGOs. Medical clinics, for example, were usually dedicated to sexual and reproductive health, since no such care is available in Venezuela. But even in more gender-neutral programs, participants skewed heavily towards women. In Riohacha, for example, 90\% of MercyCorps’s participants the first day and 80\% of the participants the second day were women; the focus group that Save the Children invited me to speak with included ten women and one man.

The combination of these factors---women being the primary recipients of aid (aid might be shared in the household, but it was still women who showed up), and women frequenting public WhatsApp groups amidst endless unscrupulous employment offers---might make distributing information about social services in these groups an extremely valuable proposition.

\section{Migrants' Knowledge about Aid and Social Services}

Many fewer migrants took advantage of social services than I expected. Outside of the days I spent with Save the Children and MercyCorps, very few people I spoke with described any significant use of aid (of course, migrants might not be particularly keen to be seen as reliant on aid\footnote{Still, there doesn’t seem to be significant shame attached to participating in aid programs, especially given the well-known hardship of migrating from Venezuela. Never in my conversations about xenophobia, for example, were Venezuelans characterized (either by themselves or Colombians) as reliant on or taking advantage of aid. Some Colombians have complained that Venezuelan migrants are taking jobs and resources meant for them, but it’s less of a feeling of ``they’re so dependent'' than ``they’re taking what we deserve.''

In many of my interviews at MercyCorps’s programs, migrants did state that this was the first time they sought assistance, usually after I asked how they found out about MercyCorps. That could reflect them not wanting to be seen as reliant on aid, but it probably reflects more on how little they knew about aid programs.
}).

Why might this be? UNHCR, NGOs, and migrants all reported that knowledge about aid programs largely came from participation in related aid programs (programs themselves sometimes make direct referrals), so \textit{failing to make contact with aid organizations upon arrival in Colombia} would make it less likely for migrants to know where to turn to for help later on. Per staff from Save the Children and UNHCR, most incoming migrants arrive prepared for at least their first few weeks in Colombia, with enough resources (possibly through selling their smartphone) to cover food and onward transport. Consequently, fewer than 20\% of permanent migrants who cross into Cúcuta stop by the UNHCR/Red Cross aid station at the border (where there are also UNHCR shelters and a World Food Programme soup kitchen). In Maicao, the number is much, much less: the UNHCR/Red Cross station at the border is a 10’ $\times$ 20’ garden shed (compared to a fenced-in, football field-sized space in Cúcuta), which was even closed when we visited.

Other factors could also explain the low usage of aid. Even for poorer migrants, the availability of informal work likely makes turning to aid programs (for example, a soup kitchen where you wait an hour or two in line) less appealing. In two hours of selling coffee on the street, migrants earn enough to cover a meal.

More than anything else, however, the limited use of aid stems from the extremely limited availability of aid. Cash transfer schemes, for example, are mandated by the Colombian government to give no more than 252k COP (\$74) monthly to a family of four or more (families are very often a lot more than four), even if aid comes privately or from overseas. But 252k COP is a \textit{quarter of the minimum wage for one person}.

More generally, it’s commonly estimated that approximately \$50,000 in aid supports each Syrian refugee who arrives in Europe. The corresponding amount for Venezuelan migrants to Colombia, Jen Daum (MercyCorps) told me in an early conversation, is \$50.\footnote{Given these numbers, the Colombian government’s largely open-arms response to the Venezuelan migrant crisis is nothing short of exemplary.}

Participants in aid programs generally reported finding out about the assistance through acquaintances/friends, or through participation in related aid programs (for example, eating at the soup kitchen where MercyCorps hosts their characterization activity). Occasionally, migrants do report learning about aid programs from other migrants on public WhatsApp/Facebook groups, but none of the organizations I spoke to shared official information over these groups.

\section{Migrants' Knowledge about Life in Colombia}

A vast majority of the migrants I spoke to said they learned about life in Colombia through Venezuelan friends and family who had arrived months or years prior. The remainder said they came without much knowledge, simply asking questions to strangers along the way and, as they often put it, letting God show them the way.

Migrants I spoke to who had more stable employment—in restaurants and stores, or even the quite lucrative business of selling crafts—typically found their jobs through other Venezuelans they met in Colombia.

Outside of WhatsApp and Facebook groups, migrants rarely use other sources of digital information, as both migrants and NGO staff told me. WhatsApp and Facebook are what people are familiar with and used to, making it critical to distribute information over these existing and well-known platforms. Creating a new app or website, as certain NGOs responding to the migrant crisis have done, makes little sense.

\section{Xenophobia}

Xenophobia is a widespread and well-known issue, and its presence in different geographies seemed to heavily depend on the economic fortunes of Venezuelan migrants in the area (both Colombians and Venezuelans explained the link as due to crime, with worse job opportunities increasing crime and anti-Venezuelan sentiment, but I’m sure it also has to do with the presence of beggars, integration of Venezuelans into the local economy, and so on).

Unsurprisingly, the people I spoke with reported xenophobia being the worst in La Guajira (Riohacha and Maicao), an economically depressed region to begin with. Still, relations with Venezuelans were much better in Colombia than in Peru, Ecuador, and Chile (which are seen by Venezuelans as more attractive destinations with higher salaries): worse work opportunities in those countries, and greater ethnic distance (Colombia and Venezuela both being ``la tierra caliente'') were migrants’ main explanations for increased xenophobia in those countries. The difference between Colombia’s response (widespread acceptance and slight attempts at integration/regularization) to migrants, and that of Ecuador/Peru/Chile (closed borders, strict document requirements), certainly matters.

Very interestingly, explanations for xenophobia in Colombia rarely involved racial or ethnic grounds---which might be obvious, given that Colombians and Venezuelans are quite similar in appearance, both being from the browner part of South America. Both Colombians and Venezuelans typically attributed xenophobia, either from others or themselves, to criminal activity by Venezuelans or their practices like giving birth to many children. Even behind these factors, racial or ethnic factors never came out, and people instead pointed to the environment in which Venezuelans lived in—socialism, several Colombians told me, meant that Venezuelans never learned to work hard. Sometimes, people would even unknowingly blame the conditions Venezuelans grew up in (as opposed to the character of Venezuelans themselves): one taxi driver complained about how Venezuelans had so many children, and I later found out this was because no contraceptives were available in Venezuela.

\section{General Issues in the Migrant Crisis}

I was extremely surprised by the disparity in incomes between Bogotá and cities on the border. We’re talking 4-5x as much income for the same work, with street vendors able to earn \$600 monthly in the capital but only \$150 in La Guajira or Cúcuta.

Most people I spoke to near the border cited either high cost-of-living in Bogotá or unaffordable transport as the main reasons stopping them from moving to the capital. Some had more personal reasons for staying near the border, wanting to be closer to their family or their ``land'' (Venezuela). Some felt stable in their current situations, especially if they were using aid resources like the WFP soup kitchen.

Migrants who cite cost as the main obstacle seem ripe for some kind of credit-based solution, since the increase in even a single month’s income from moving to Bogotá would more than cover associated expenses. Even considering that potential migrants (everywhere in the world) often assign extremely high subjective costs to moving, it doesn’t seem like that cost would surpass the \$400/month they could gain from moving. Of course, individuals’ current stability, current community integration, and desire to be close to the border—while difficult to economically quantify—are extremely important factors and should never be ignored. Nonetheless, it still seems like too few people are choosing to migrate onwards to Bogotá. With this, it might be valuable to distribute information over WhatsApp groups about the experience, costs, and benefits of moving to the capital.

I commonly encountered Venezuelan migrants who wished to become regularized but were unable to do so because they never entered legally, more due to document requirements than anything else. All had Venezuelan ID cards, but few could afford passports in Venezuela (which, nowadays, cost from \$1,000-5,000, the bulk of that sum being bribes); gaining legal residency requires having legally entered with a passport.

The prevailing sentiment, by far, amongst everyone I spoke to, was that life in Colombia was hard but good, with its stability and economic prospects. Most were grateful to Colombia for being able to restart their lives there (and possibly even gaining legal residency), and this—more than anything else!—emphasized to me how dire things are in Venezuela. Along the border, even people earning less than \$150-200 a month were grateful for their situations, which were still much better than Venezuela; income isn't everything, but \$5 a day is an extremely difficult life.

In comparison to Colombia, countries like Peru, Ecuador, and Chile offer higher salaries. Yet many migrants have curtailed desires to migrate onward because of the difficulty of legally emigrating to those countries and finding work there. Peru, Ecuador, and Chile have much stricter document requirements, and Venezuelans face significantly greater discrimination in those countries, as we discussed in the section on xenophobia.

\chapter{Related Work\label{ch:relatedwork}}

\section{CDRs}

Originally, we sought to center our work on call detail records (CDRs), which typically contain line-by-line metadata of all phone calls and SMS messages transmitted over a network operator’s infrastructure, including origin and destination accounts (with operator account numbers and/or operator-agnostic IMEI), time and duration, nearest cell towers, and more. Like social media data, CDRs offer an innovative source of information in situations where populations may be hard to survey---for example, refugees or other vulnerable groups---or formal censuses/surveys may be inappropriate (e.g., too slow during disaster response).

CDRs would seem to offer several advantages over social media data, especially in better representing migrant populations. Active WhatsApp users skew younger, wealthier, and more educated than the typical Venezuelan migrant; phone calls and SMS messages, on the other hand, are a more traditional and much more accessible form of communication. Moreover, unlike communications data from WhatsApp, network operators provide formal access to CDRs, and thus eliminate the biases of sampling communication from public, advertised groups.

In research published to date, only one network operator in Colombia has provided access to CDRs, which were used in a 2016 paper by Bogomolov et al. examining neighborhood activity in Bogota \cite{worldbank-bogota-2016} and in a 2017 study by Florez et al. studying Bogota commuting networks using origin-destination matrices \cite{florez-bogota-2018}. At this time, the operator, Telefónica Colombia, has formally denied our request to access CDRs, as it transitions toward a business model of analyzing data in-house and selling aggregate results.\footnote{\url{https://luca-d3.com/products-services}}

Disappointing? Certainly. But WhatsApp data offers numerous and significant advantages compared to CDRs. For one thing, CDRs only include metadata, making interpretability of results difficult and precluding any analysis of what is actually being communicated. More importantly, WhatsApp is the primary medium that migrants actually use to communicate, especially for seeking information, opportunities, and resources.

\section{Social Media Network Analysis}

There is a broad literature of research that seeks to understand social interaction as it unfolds on internet sites and communications networks \cite{networks-book-2010} \cite{survey-mobile-2015}, with topics including connectivity and distance between users, the strengths of ties between users, information flow through networks, and homophily. Ediger et al., in a 2010 paper, construct undirected Twitter interaction graphs---with users as nodes---related to crisis topics (the H1N1 pandemic, and the 2009 Atlanta floods) and find the distribution of user degrees fits a power law, with media and government accounts having especially high degrees \cite{ediger-twitter-2010}. 

\section{Research on WhatsApp\label{ch:relatedwork:whatsapp}}

WhatsApp is an internet-based messaging application comparable to and more feature-rich than SMS messaging. By the early months of 2020, WhatsApp was used by over two billion daily users, who send many billions of messages every day \cite{verge-whatsapp-2020}. Beyond allowing individuals to message each other (for free and with multimedia content), WhatsApp also allows individuals to create and join group messages with up to 256 users total, a feature that has significantly fueled WhatsApp’s growth. Many groups are public and able to be accessed with links shared in various sources.\footnote{See, for example, \url{https://whatsgrouplink.com/}, or simply the Google search results for "chat.whatsapp.com" plus any query of interest.}

Relatively little attention is paid to WhatsApp compared to other social media networks. As of April 2020, Google Scholar returns only 214,000 articles about WhatsApp, compared to 6.2 million results for research on Facebook and 7.3 million on Twitter; there are even 1.1 million articles on Pinterest.

To date, formal access to WhatsApp data has remained proprietary and outside of the hands of researchers. Because of this limitation, until the previous few years, most studies of WhatsApp groups have involved qualitative methodologies, centering on surveys and interviews—for example, of students at a university who self-report WhatsApp usage (and share their personal WhatsApp data directly with researchers) \cite{seufert-whatsapp-2015}.

\subsection{Analysis of WhatsApp Public Groups}

More recently, several researchers have taken a more quantitative, systematic approach of joining public WhatsApp groups en masse. These researchers scrape various sources for links to join public WhatsApp groups, automatically join these groups on some regular (usually daily) basis, and scrape messages and data from the groups, including links to join further WhatsApp groups. Research on WhatsApp groups, then, has shifted from working \textit{with} users (through interviews and accessing their personal WhatsApp accounts) to working \textit{as} users (by joining hundreds to thousands of public WhatsApp groups).

Bursztyn and Birnbaum take such an approach to study politically-themed WhatsApp groups leading up to the 2018 Brazilian election, and analyze aspects including network metrics (constructing, for example, a graph of users as nodes and edges as co-participation in any group) and sharing of media from different sources \cite{bursztyn-whatsapp-2019}. Garimella and Tyson, in a 2018 study, research WhatsApp public groups generally, without any subject or demographic in focus (they join groups found on Google and other topic-agnostic websites) \cite{garimella-whatsapp-2018}. They analyze the distribution of messages between and within groups, the geographic distribution of users,\footnote{The geographic distribution of users in groups collected by Garimella and Tyson are quite illustrative of the biases of sampling from public WhatsApp groups advertised on Google and other highly-public sites. From their paper, “the top countries include India (25K), Pakistan (3.6K), Russia (3K), Brazil (2K) and Colombia (1K).”} and the content, language, and multimedia within messages.

A Brazilian group at the Federal University of Minas Gerais (UFMG) has started to dominate this subfield. Out of the 12 English-language studies that analyze WhatsApp public groups (as of late March 2020), six have been published by the same group, with lead investigators F. Benevenuto and J.M. Almeida, both associate professors at UFMG. They began their work around 2018, developing a system to help journalists analyze and visualize the activities of political WhatsApp groups during the historic Brazilian election that eventually put Jair Bolsonaro in power \cite{resende-system-2018}; their tool analyzed the political views and demographics of groups. Later in 2019, with Garimella, they extended this tool to India, incorporating the Perceptual Hashing (pHash) algorithm to identify re-shared images that were slightly altered \cite{melo-monitor-2019}.

Other work by UFMG researchers includes a 2019 paper that tracks replies within WhatsApp \cite{caetano-attention-2019}---which constructed directed graphs from reply cascades, characterizing and analyzing structural attributes of these graph---and a 2019 paper studying textual content in WhatsApp groups \cite{resende-textual-2019}. This latter paper attempted to identify misinformation, and separately analyzed text properties like message size, linguistic elements, and sentiment and topic analysis. The UFMG group has also studied misinformation and information spread within Brazilian WhatsApp groups \cite{melo-misinformation-2019}, as well as the content and propagation of audio messages within such groups \cite{maros-audio-2020}.

\chapter{Collecting Data\label{ch:datamethod}}

In this section, we detail our data collection methodology, limitations on WhatsApp data collection in general, limitations of our methodology, and privacy concerns. Details about technical implementations, as well as full source code, are available in Appendix \ref{ch:appendicies:implementation}.

\section{Methodology Overview}

At a high level, we collect information from WhatsApp groups by joining groups \textit{as} user. We adapt and alter the methodology of several recent researchers, as described in Chapter \ref{ch:relatedwork:whatsapp} on related work. At a high level, we:
\begin{enumerate}
  \item Search for links to join WhatsApp groups across various Facebook groups. These links are all of the form \url{chat.whatsapp.com/...}
  \item Join these WhatsApp groups, either on the WhatsApp Web interface using Python and Selenium or directly on smartphones.
  \item Continuously collect message and member information from each WhatsApp group, on the WhatsApp Web interface using Python and Selenium.
\end{enumerate}

\section{Joining Groups}

To search for WhatsApp groups to join, we first searched for Facebook groups related to Venezuelan migrants in Colombia. We included all public Facebook groups of 50,000 members or more that appeared in search results for (``Venezuela'' OR ``Venezolanos'') AND (``Colombia'' OR various large cities in Colombia\footnote{Specifically, we searched for all groups using the terms (``Venezuela'' OR ``Venezolano'') AND (``Bogota'' OR ``Medellin'' OR ``Cali'' OR ``Barranquilla'' OR ``Cartagena'' OR ``Cucuta''), which are the six largest cities in Colombia.}). We collected all WhatsApp links posted in these groups between November 1, 2019 and January 23, 2020, either posted directly or as comments/replies to other posts (most were the latter).

In total, we collected 280 unique links, and were able to join around 200 groups. A few links were broken/mistyped, but most unsuccessful links had been revoked by the owner of the WhatsApp groups---indeed, some of the links had last been posted months prior. There are several flaws inherent to this process, which center on the fact that if we don't collect and join links in real-time (as they are posted), links may be revoked by the time we attempt to join.

It’s not clear how links being revoked would bias our data, the sample not purporting to represent anything in the first place, but avoiding revoked links is certainly good for expanding the size and diversity of the dataset. Facebook posts/comments themselves may also be deleted as time passes. A better and more systematic approach to joining groups would involve continuous (or at least daily) monitoring of Facebook groups, which is made difficult by the fact that Facebook’s API doesn’t allow for automatic scraping (Facebook’s rather clunky/glitchy user interface also means that a Selenium-based approach, as we implement to collect data from WhatsApp Web, would be quite error-prone).

Joining groups from their invite links is rather simple, involving a few button clicks on WhatsApp Web, which can be automated using Selenium.

As described above, collecting and joining group links in real-time (or at least daily) would solve the issue of links being revoked. Yet joining groups every day at the same hour (more specifically, groups that likely just had invite links created) would probably raise WhatsApp’s suspicions. In this sense, any efforts to systematize our processes are also more likely to create obvious patterns, raise the suspicions of WhatsApp, and ultimately lead to adverse action against us. Randomizing join time might help—--say we collect links every day at midnight, then join them at some random time (within reasonable hours) in the next day.

\section{Interference Measures by WhatsApp}

Joining many groups, it turns out, is incredibly suspicious to WhatsApp. On multiple instances, midway through joining a list of groups, accounts became banned from WhatsApp. With high confidence, these were automatic bans, both because of their instantaneous mid-process nature and because they took place near the same chokehold each time (around the 40th to 50th group joined by that account).

The obvious solution was to limit how quickly we joined groups; we eventually found that joining no more than 30 groups in any 24 hour period was enough to stave off the auto-ban. More than this, we also found that multiple smartphones (each with their own WhatsApp account) were necessary to reduce the suspiciousness of our processes. To join 200 groups, each on at least two smartphones (as a resiliency measure), we eventually used six smartphones of different models/operating systems, each with their own phone number and WhatsApp account.

\section{Collecting Data from Groups}

Most studies in the literature collect data from groups by decrypting a WhatsApp message database that is stored locally on the smartphone\footnote{Specifically, all WhatsApp studies that explicitly mention how they collect data mention this method; the UFMG studies do not explicitly mention how they collect data.} \cite{bursztyn-whatsapp-2019} \cite{garimella-whatsapp-2018}. This published method involves the somewhat dodgy (and sometimes illegal\footnote{In the United States, the Digiital Millenium Copyright Act (DMCA) made it illegal to root Android phones. Later exemptions made it temporarily legal to root certain devices (phones but not tablets), but overall, rooting is a legal grey area in the United States as well as globally \cite{arstechnica-dmca-20121025}.}) exercise of rooting Android phones (akin to ``jailbreaking'' iPhones), which is necessary for obtaining the encryption key WhatsApp uses to secure this message database.

Our process of collecting data from groups significantly differs from this known approach. Using Selenium and WhatsApp Web, we navigate to each group in the WhatsApp Web interface, and then in each group record the group’s members and log the group’s messages. This approach is more complicated than simply decrypting the message database, since it relies on the rapidly changing and quite ``fragile''\footnote{This is a term from Kiran Garimella, author of various articles in the recent literature on WhatsApp public groups.} WhatsApp Web, but is better in certain aspects:
\begin{itemize}
  \item Our method doesn't require rooting, which is sometimes illegal but also generally dependent on a highly ad-hoc community of mobile software development engineers occasionally publishing root ``exploits.''
  \item Our method is much less likely to be rendered impossible by WhatsApp. WhatsApp could easily change how they store and secure the message database, making the known method infeasible or substantially more difficult. But access to WhatsApp Web is a given.
\end{itemize}

Admittedly, several tools to scrape messages from WhatsApp Web have already been published,\footnote{See, for example, \url{https://github.com/UoMResearchIT/whatsapp-scraper}, or \url{https://github.com/bansalsamarth/whatsapp-chats-scraper}, or \url{https://github.com/codenoid/WhatsappScraper}. More generally, these tools can be found by searching GitHub.} but our implementation is much more complete than any available method. The scripts available are almost all between 200-300 lines long, while our implementation is just short of 700 lines; quantity is not quality, but all 700 lines in our implementation are needed to fully deal with the intricacies of WhatsApp Web. Anything shorter fails to capture certain information (for example, multimedia data) or deal with extreme cases (for example, groups with thousands of daily messages).

On top of this, none of the public tools include implementation details, only (e.g.) Python scripts for scraping. Questions like how often we should scrape and with what infrastructure we should scrape remain unanswered.

We share most implementation details, and a final Python script, in Appendix \ref{ch:appendicies:implementation}. Below, we detail some more novel aspects of our implementation (hereafter referred to as ``traverseGroups,'' named after the Python script that we use to traverse groups and collect data), including frequency of data collection, infrastructure, and anti-interference measures.

\subsection{How do we uniquely identify groups?\label{ch:datamethod:identifygroups}}

A part of this process involving some sophistication was finding a way to keep track of groups and identify them uniquely.

Once we join a group, there’s no obvious and foolproof way of identifying the group uniquely. In the WhatsApp Web interface, the only information viewable about the group are title, group icon, message history, other members, and (rarely, if the group administrator created one) group description. Even in HTML, where other sites may include some kind of (hidden) unique identifier, WhatsApp doesn’t. Of course, WhatsApp’s intended use case doesn’t center on users being in many groups with identical titles.

We needed to be able to uniquely identify groups from the scant information available. Using title and/or array of other group members were our initial guesses, but those change over time and are not necessarily unique (in our case, there were multiple groups with the titles ``Venezuela,'' ``Venezolanos en Bogota,'' and ``Emprendedores'' (\textit{Entrepreneurs})). Group profile pictures (specifically, the link to the image) are a more unique alternative, since the same picture, if used as the icon for more than one group, would be uploaded to distinct links.\footnote{This is not true of all images; some commonly-used images, like emojis, are shown from base64 directly in HTML without a separate link. Strange nuances like this kept appearing on WhatsApp Web, and made the entire process of scraping WhatsApp Web quite tedious.} Yet profile pictures can also be changed, and not all groups have profile pictures (around 10-20\% of groups we joined did not).

It turned out that, buried\footnote{The profile picture link looks something like: \url{https://web.whatsapp.com/pp?e=https\%3A\%2F\%2Fpps.whatsapp.net\%2Fv\%2Ft61.24694-24\%2F71104943_726169654564928_648446692972455XXXX_n.jpg\%3Foe\%3D5E35C2D4\%26oh\%3Dc8619054cae8f2c2766c3ce819d3ea7f&t=s&u=58416572XXXX-157097XXXX\%40g.us&i=1571100754}. This identifier really is buried in there.} in the link to a group’s profile picture (if it has one), is some kind of unique group identifier, of the form ``58416572XXXX-157097XXXX''. In this string, the first half is the phone number of the group creator (which stays constant over time), and the second half appears to be some kind of unique group identifier. Even when profile pictures change, the same unique group identifier remains in the link to the new profile picture.

This solves the question of groups that have profile pictures. Some groups still don’t, so we resorted to using a cryptographic hash of the full HTML of the title (which includes both the text and any emojis in the title). So we recorded the unique identifier for each group, which we call its \textit{uid}, as either: the profile picture link's unique identifier (if the group has a profile picture), or the cryptographic hash of its title (if it doesn't have a profile picture). Combining these methods resulted in $100\%$ success in identifying groups uniquely, and $> 90\%$ success in tracking groups over time (when they add/remove profile pictures, or change titles). Still, this is not foolproof: a group's \textit{uid} may change if, for example, it removes its profile picture completely, or if it changes title without having a profile pictures.

In general, there isn’t a foolproof way to perfectly identify groups. Using as many details as possible about the groups (everything from message history to member array) would allow for greater confidence in tracking a group over time should details (like title HTML) change, but this seems overly complicated for the scope of this thesis. In Appendix \ref{ch:appendicies:implementation:groupschange}, we describe how in preprocessing data we are able to identify and link groups even when their \textit{uid}s change.

\subsection{How frequently do we collect data?}

Due to the design of WhatsApp Web, scrolling to an earlier message requires simultaneously loading all later messages, which imposes heavy resource (CPU/memory) usage. If messages 31-45 are currently loaded, scrolling to messages 15-30 would mean loading messages 15-45 (i.e., keeping messages 31-45 loaded while loading the new messages).\footnote{This differs from how WhatsApp loads groups in the sidebar, which is more efficient. In the sidebar, WhatsApp only loads 15 groups at a time.} Clearly, checking for messages every $n$ hours has cost greater than $O(n)$ each time, since the CPU/memory are strained by having to load $n$ hours of messages all at once, and cannot read/log messages as efficiently.

Using sample data, we estimated the processing time to be around $O(n^2)$. In our sample, we checked around 200 groups every three hours (for 48 hours total), and for each group-time pair, we recorded the number of new messages in that group, how long it took to read those messages, and how long it took to scroll to those messages. We ran this process on two servers, a late-model Macbook and a server from Amazon Web Services, and had approximately 2600 group-time pairs on each server.

\begin{figure}[h]
\centering
\begin{subfigure}{0.5\textwidth}
  \centering
  \includegraphics[width=\textwidth]{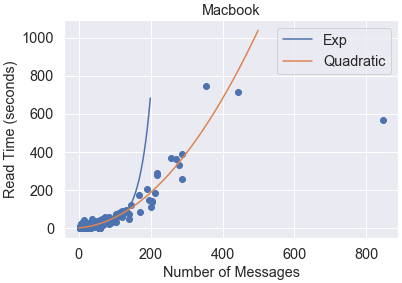}
\end{subfigure}%
\hfill
\begin{subfigure}{0.5\textwidth}
  \centering
  \includegraphics[width=\textwidth]{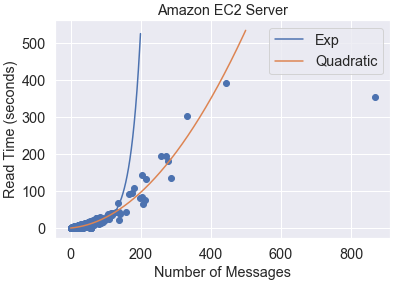}
\end{subfigure}
\caption{Read time is quadratic in number of messages; the CPU/memory are taxed by having to simultaneously load all messages at once, and consequently run slower when they actually read messages.}
\label{ch:datamethod:readtime}
\end{figure}

\pagebreak
A plot of the number of messages and read time for each group-time pair is shown in figure \ref{ch:datamethod:readtime}.

\begin{figure}[h]
\centering
\begin{subfigure}{0.5\textwidth}
  \centering
  \includegraphics[width=\textwidth]{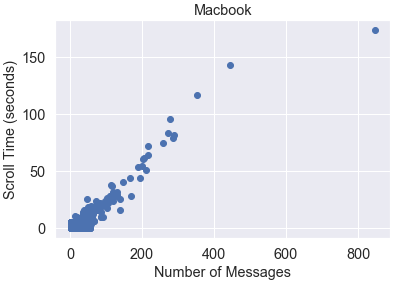}
\end{subfigure}%
\hfill
\begin{subfigure}{0.5\textwidth}
  \centering
  \includegraphics[width=\textwidth]{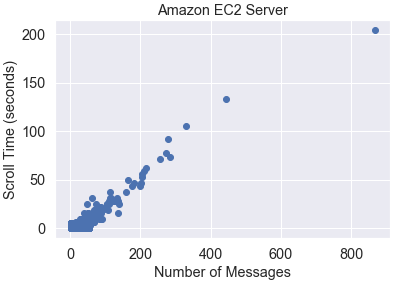}
\end{subfigure}
\caption{Scroll time is linear in number of messages; this is largely due to the delay (that we implemented) between each scroll so that earlier messages are properly loaded.}
\end{figure}

While the Amazon server was able to read messages almost twice as quickly, on both servers there was a quadratic relationship betewen number of messages and read time. The fact that computing time is $O(n^2)$ compels us to check groups in small time intervals. Some testing led us to settle on every three hours, with the workload typically being finished in an hour. Obviously, there’s no guarantee of success here—--members of every group could send 500 messages in a three hour period—--but three hours seems enough to assure success with near certainty given our data.

\subsection{What infrastructure do we use to collect data?}

The script we use to collect data from groups must run continuously in small time intervals, so we deployed it to remote servers. Running continuously naturally suggests the Unix scheduler cron. The traverseGroups Python script by itself could’ve been deployed as just a Python script, but the added complication of cron made a smarter deployment make sense. Using cron requires changing system settings that could affect other processes on the remote servers; in particular, cron kills Python and Chrome occasionally to make sure that crashed/glitched processes (which may result from glitches in WhatsApp Web) don’t affect future runs. This could be problematic when other processes are in play, as they might be on a shared server.

Enter Docker, a platform that uses OS-level virtualization to deliver software in packages called containers \cite{fink-docker-2014}. Essentially, each Docker ``container'' is designed to run one specific task; a container begins as some base image (e.g., a bare-bones Unix distribution), and onto which we can add software, and libraries/configurations needed for the software to run.

In our case, we started from a Unix distribution with Python already loaded, then installed the Selenium package, installed Chrome and chromedriver, added the traverseGroups script, and configured cron. The container functions like a dedicated virtual machine, but with much less overhead than a virtual machine.

\begin{figure}[h]
  \includegraphics[width=\textwidth]{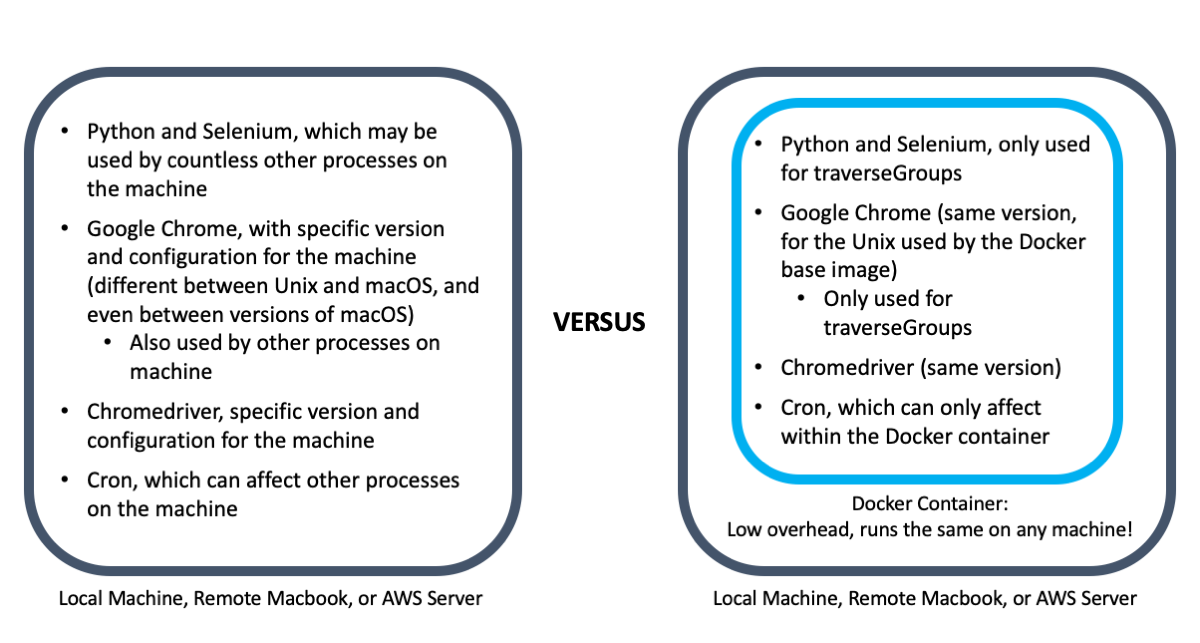}
  \caption{A diagram illustrating how Docker differs from typical deployments.}
  \label{ch:datamethod:docker_comparison}
\end{figure}

\pagebreak
Once containers are created, they can be downloaded to any machine with the Docker host, which runs on Mac, Windows, and Linux. There are a variety of benefits to Docker aside from not needing to change system settings (figure \ref{ch:datamethod:docker_comparison} illustrates the differences between Docker and typical deployments):
\begin{itemize}
  \item As long as a computer can run the Docker host, Docker containers will run exactly the same on that machine as they would on any other machine. This meant that we could instantly test code updates on the local computer, and be confident that they would work exactly the same on remote servers, without needing to upload and test code on each of them.
  \item The traverseGroups script also depends on Google Chrome and chromedriver (which allows Chrome to be controlled by Python/Selenium), which can both differ significantly across platforms. Using Docker means avoiding any problems that arise from differences in Chrome version. One version of Chrome may load certain aspects of WhatsApp Web faster, for example, and this would affect timings we have in the script.
  \item Docker makes it extremely easy for results to be replicated, and for other researchers to be able to run this code. Instead of needing to configure $\{\textrm{cron}, \textrm{Python}, \textrm{Selenium}, \textrm{Chrome}, \textrm{chromedriver}\}$, all they need to do is download the Docker host and pull our Docker container.
\end{itemize}

With this, we ultimately deployed the traverseGroups code on six separate Amazon EC2 Unix servers (one for each smartphone). We ended up settling on AWS t2.medium instances, which offer 2 ``burstable'' Intel Xeon processors and 4 GiB of memory, and each cost around \$1.11 per day to run.\footnote{\url{https://aws.amazon.com/ec2/instance-types/}} As we detail below, the traverseGroups script runs at random intervals to maintain a lower profile, so dedicating a server to each phone makes the most sense. Standardizing the timings (i.e., having the process for each account run at different set times) would allow for more efficient server use, but would appear significantly more suspicious to WhatsApp.

\subsection{How do we avoid getting banned from WhatsApp?}

Our accounts and processes are endlessly suspicious. Each account joins dozens of groups, but never messages anything or uses WhatsApp outside of these groups. Each account also logs in every few hours or so from WhatsApp Web, and looks in each group every time.

To avoid getting banned from WhatsApp, we undertake a two-fold strategy of making both accounts and processes less suspicious. WhatsApp has infinite data about each account: IP address of the phone (at every point in time), phone number, device location (if granted permission), device model and platform, and various details about how WhatsApp Web is accessed (including browser user agent, screen resolution, IP address, and more). Beyond this metadata, WhatsApp also records patterns of activities on each account, such as the groups joined (and when those groups were joined) and when WhatsApp Web is accessed.

We lower the profile of each account as much as we can, by:
\begin{itemize}
  \item Limiting how many groups each account joins. This requires more smartphones/accounts running in parallel, though there are natural limits for how low we can make this number. For one, additional smartphones are costly to operate and mtainin, but more than this, more accounts are more suspicious, especially if WhatsApp can link them in some way (it would be infeasible, for example, to maintain 30 different smartphones each with individual IP addresses). We settled on using one smartphone per every 60 or so groups (in total, six smartphones, so that the 200 groups were each joined by two separate accounts, in case of technological failure or bans).
  \item IP address masking of the smartphones. This seemed especially important during account creation and when joining groups. Yet technology news sources report that WhatsApp also flags accounts where the IP address doesn't match telephone number geography \cite{androidauthority-whatsapp-2019}, so this is more difficult than it seems; VPN addresses are likely also highly suspicious. We ended up running the phones clearnet on independent Princeton University networks---WhatsApp seems to consider Bayes' rule when flagging IP addresses,\footnote{On several occasions, WhatsApp outright banned IPs we used to create and modify accounts, but WhatsApp never banned the Princeton University IP range, on which we conducted over 90\% of our total activity.} so the high amount of legitimate WhatsApp activity from Princeton likely made our accounts less suspicious.
  \item IP address masking of the servers used to run traverseGroups and collect data. Each Amazon EC2 instance, by default, has a different public IP, motivating the use of one server per smartphone. We ran into no issues with this setup, whatever the EC2 geographies (we used servers in Amazon's us-east-2 (Ohio), us-west-1 (N. California), and us-west-2 (Oregon) regions). IP address likely matters much less when simply accessing WhatsApp, in comparison to creating accounts or joining groups.
  \item Varying device models and platforms. Six old-model Android smartphones suddenly appearing on WhatsApp is much more suspicious than six late-model iPhones, given their usage in the general population. We ended up using mostly iPhones of varying models.
  \item Varying the user agent and screen resolution of the servers that access WhatsApp Web. We spoof this data anyway, so we used a different user agent on each of the EC2s.
  \item Varying the times/intervals at which WhatsApp Web is accessed. For each account, we generated random intervals (uniform between 2.5-4.5 hours), and from those intervals, we generated cron scripts that ran the traverseGroups Python script to collect data at random times.
\end{itemize}

\section{General Limitations}

Our approach has significant drawbacks, beginning with the fact that public WhatsApp groups only represent a small and skewed sample of communications on WhatsApp, with most WhatsApp communication either private or in groups involving close acquaintances: roommates, colleagues, participants of specific social occasions, and so on \cite{church-whatsapp-2013}. Yet because of the importance WhatsApp holds in migrant experiences---and because its proprietary nature means that analysis can \textit{only} be done either with user or as user---we still consider our approach to have significant merit.

Moreover, while public groups may not be a good representation of all communications on WhatsApp, we have strong suspicions that such groups do come closer to representing migration-related communications of Venezuelans in Colombia. From a Reuters survey of social media users in nine countries (which included the US and UK, as well as Turkey, Malaysia, and Brazil), 76\% of WhatsApp users participate in groups and a vast majority of these users (around 58-65\% of \textit{all} WhatsApp users in Turkey, Malaysia, and Brazil) are “active members of groups that mostly include people they do not know” \cite{reuters-digital-2019}. We hypothesize, then, that Venezuelan migrants are indeed likely to be part of public, widely-advertised WhatsApp groups, especially since they have much greater interaction with strangers in general.

Beyond our analysis only capturing the dynamics of public WhatsApp groups, we also fall short in capturing only a certain sample of public WhatsApp groups—--those we’re able to find. The possible biases of joining groups that have been shared on Facebook or the internet (i.e., these groups are quite heavily advertised) are somewhat counteracted by the fact that migrants are also more likely to have joined these groups, compared to other public WhatsApp groups.

With these biases in mind, we proceed with the understanding that our research should focus not on how our sample represents migrant communication in general, and instead on the dynamics within our sample, and how they change. Our sample is certainly interesting in and of itself, even while it may not offer rigorous broader conclusions toward migration-related communications in general.

\section{Data Limitations\label{ch:datamethod:data-limitations}}

While our methodology grants us full access to data from WhatsApp groups, we intentionally limit what data we collect. WhatsApp offers a rich variety of content types for messages, including documents and locations, but many (e.g., documents and locations) are rarely used in our groups of interest, and add unnecessary complexity to both our data collection and analysis.

More significantly, we do not fully download multimedia content (audio recordings, images, and videos). For images and videos, we only record their cryptographic hash (and only of the thumbnail for videos); for audio recordings, we only record their duration. To reiterate, our methodology is fully capable of downloading this content (indeed, because our process relies on WhatsApp Web, it ``sees'' exactly what a user would see), but downloading content involves significant technological complexity, and would also require significiant manual analysis to reap any benefit (i.e., labeling images and audio).

Moreover, one of the principal researchers in this field has explicitly warned against downloading multimedia content,\footnote{Interestingly enough, several of his collaborators have come to center their research on multimedia content in WhatsApp groups. See \cite{resende-misinformation-2019}, \cite{maros-audio-2020}. One has to wonder.} since some groups are of an adult nature where obscene (and sometimes illegal) content is frequently shared \cite{garimella-whatsapp-2018}.

What this means is that images/videos end up useful in two ways: we know that an image was shared, and we know if the same image or video is ever shared again. A caveat is that popular images/videos are sometimes altered before being re-shared, either unknowingly (e.g., a user downloading the image in lower-resolution, or taking a screenshot, and then re-sharing) or intentionally (in an attempt to circumvent various systems, like Youtube's copyright control system \cite{kjelsrud-phash-2014} or WhatsApp's anti-spam filters).

A compromise between our methodology and downloading multimedia might involve perceptual hashes, which matches similar images \cite{weng-perceptual-2011} and audio files \cite{ozer-perceptualaudio-2005} at a high level, without needing to retain the entirety of their contents. Still, perceptual hashes do not produce any interpretability.

\section{Privacy Concerns}

The ethical considerations of working with social media data are murky, particularly in the case of WhatsApp, which is seen as a more private messaging service in comparison to Facebook and Twitter, which are more traditional social media platforms.

WhatsApp’s terms of service allow for users to access data from groups in which they are members, so our methodology is compliant with the company’s terms. WhatsApp also does not make restrictions on who can join groups. An ethical question perhaps arises from our act of joining groups (which is without pretense, though still possibly misleading if group members do not expect researchers to join). Our target groups, however, are not only public but also advertised somewhere, so it seems unlikely that we are violating expectations of privacy within these groups.

With regard to other group members (i.e., Venezuelan migrants), WhatsApp’s privacy policy “states that a user shares their messages and profile information (including phone number) with other members of the group (both for public and private groups)” \cite{garimella-whatsapp-2018}. Because our collection process is implemented as user, we collect data symmetrical to what other users are able to access. We find it likely, given WhatsApp’s intuitive user interface, that users have full knowledge of what information they and others can access, so users \textit{do} understand what data we can collect. Effectively, by joining a public WhatsApp group, users agree—--both formally and informally—--to share certain data with other members of the group, and are aware of what data is being shared (i.e., their profile information and messages).

An important question remains of delineating between these users’ agreement and consent (to join groups and share information), and their \textit{choice} (to do so), especially given our context. No users are forced to join any of our target groups, but the circumstances because of which they join our target groups—--the arduous processes of migration and resettlement—--can certainly be coercive. In other words, users may join groups intentionally but without choice as we typically understand it.\footnote{This point is heavily inspired by a brilliant essay on sex work and agency written by Lorelei Lee, an American sex worker and writer. See \cite{lee-cashconsent-2019}.} While this concern will remain paramount in our work, for now we emphasize that our research aims are wholly in line with the well-being of migrants, and that we will heavily restrict what data we share, as described below.

We will share neither identifying information about individual WhatsApp profiles nor individual messages from WhatsApp groups, because of the possible expectations of privacy within these groups--—users may expect their profile and messages to only be seen by other members of the group at that time,\footnote{Though again, we reiterate that any person with internet access and a mobile phone number would have been able to access all of this information, legally and in accordance with WhatsApp’s terms of service.} with the total number of such members capped at 256. When sharing aggregate results from WhatsApp, we will take care to ensure that no individuals can be singled out from data.

\chapter{Initial Analysis of Members\label{ch:members}}

We begin our analysis by exploring the network of members in our WhatsApp groups. Principally, we define membership in each group as having sent a message to the group during the 53 days we collected data.

In total, we recorded 7,860 users participating in 174 groups. Nearly all of these users (7,377 users, or 93.85\%) joined only one group, but 434 were in two groups, 36 were in three groups, 11 were in four groups (of which four had CO phone numbers and two had VZ numbers), and 2 were in five groups (one with BR phone number, one with VZ number).

\begin{figure}[h]
\centering
\includegraphics[width=0.45\textwidth]{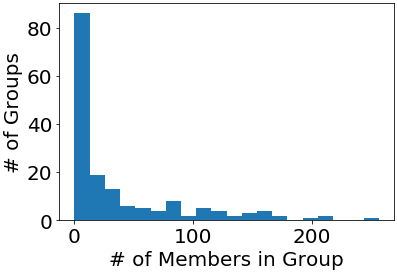}
\caption{The number of members in groups is heavily skewed; 44\% of groups involve 10 or fewer participants.}
\label{figure:members:hist_size}
\end{figure}

Figure \ref{figure:members:hist_size} shows a histogram of group sizes, which follows a power-law distribution with a long tail. 44\% of groups involve 10 or fewer active members, and 59\% involve 20 or fewer, but 9.8\% of groups involve over 150 members.\footnote{British anthropologist Robin Dunbar famously proposed 150 as a theoretical limit for a person's social network size \cite{dunbar-2002}; this number was supported by a 2011 study of communications over Twitter \cite{goncalves-2011}, and likely explains, to some extent, WhatsApp's hard limit of 256 members per group.}

\section{Comparing Users by Country}

We preface this section by conceding that telephone country codes are an imperfect approximation of location, and a much worse approximation of nationality. WhatsApp only requires a user's telephone number when signing up and for occasional verification purposes; it's certainly possible for, say, a South African traveling in Peru to sign up for WhatsApp with a French phone number.

Yet we argue that in our context, the telephone numbers of WhatsApp users telephone number are a satisfactory, even helpful, approximation of their geographies. We found in field work that most users in Colombia access WhatsApp over a mobile network, at least part of the time; Venezuelan network SIMs don't work in Colombia, so we expect users to have a phone number that matches what country they're in. Whether or not they register this number on WhatsApp is debatable, but we find it likely for two reasons. First, WhatsApp occasionally requires verification through SMS sent to the account number; second, we also found in field work that WhatsApp users occasionally reverted to SMS/phone calls when their data bundles ran out (users would want their WhatsApp contacts to still be able to contact them, compelling them to register their active phone number on WhatsApp).

With this in mind, we compute the proportion of each group with phone numbers from Colombia (CO), Venezuela (VZ), and other countries including Ecuador (EC), Chile (CL), and Peru (PE). In nearly all groups, fewer than 10\% of members were from EC, CL, or PE. In figure \ref{figure:members:pVZ_vs_pCO} below, we show that while most groups have fewer than 25\% of members from VZ, a decent number of groups have member bases that are 25-50\% VZ, and some groups are even over 90\% VZ. In contrast, there are many groups that are 75-100\% CO.

\begin{figure}[h]
  \centering
  \includegraphics[width=0.45\textwidth]{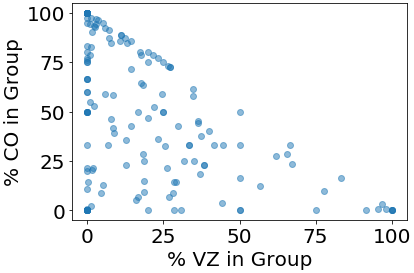}
  \caption{Most groups involve a large percentage of CO members and few VZ members.}
  \label{figure:members:pVZ_vs_pCO}
\end{figure}

\begin{figure}[h]
\centering
\begin{subfigure}{0.45\textwidth}
  \centering
  \includegraphics[width=\textwidth]{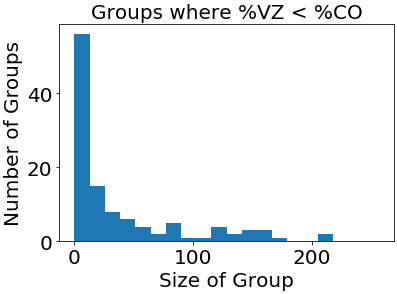}
\end{subfigure}%
\hfill
\begin{subfigure}{0.45\textwidth}
  \centering
  \includegraphics[width=\textwidth]{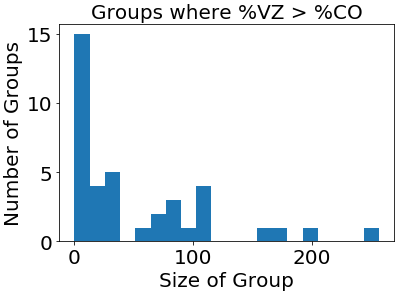}
\end{subfigure}
\caption{The distribution of sizes of CO-dominant groups is close to the overall distribution of group size, but groups with more VZs tend to be larger.}
\label{figure:members:hist_size_moreCOVZ}
\end{figure}

In figure \ref{figure:members:hist_size_moreCOVZ}, we present histograms of group sizes, for groups where there are more CO members than VZ members, and for groups with more CO members than VZ members. Of groups with more CO members, the histogram of sizes is nearly identical (with scaling) to the histogram of group sizes in general, while for groups with more VZ members, groups tend to be larger.

\subsection{Who's connected to who?}

For the rest of this chapter, we treat users as \textit{connected} if they've participated in the same group. Figure \ref{figure:members:hist_connToCOVZ_ECPE} below shows that Ecuadorian and Peruvian users are around equally well-connected to CO and VZ users, with the vast majority of users from both countries having connections to 50 users or fewer from either CO or VZ.

\begin{figure}[h]
\centering
\begin{subfigure}{0.45\textwidth}
  \centering
  \includegraphics[width=\textwidth]{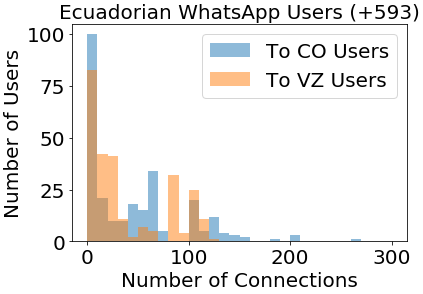}
\end{subfigure}%
\hfill
\begin{subfigure}{0.45\textwidth}
  \centering
  \includegraphics[width=\textwidth]{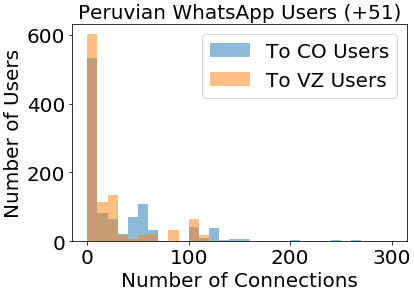}
\end{subfigure}
\caption{Users from Ecuador and Peru are equally well-connected to CO and VZ users.}
\label{figure:members:hist_connToCOVZ_ECPE}
\end{figure}

\begin{figure}[h]
\centering
\begin{subfigure}{0.45\textwidth}
  \centering
  \includegraphics[width=\textwidth]{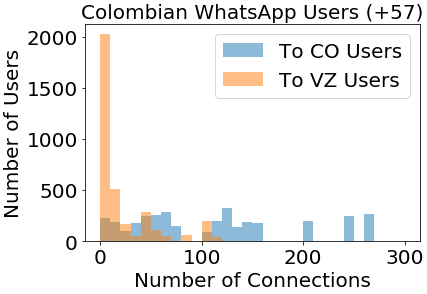}
\end{subfigure}%
\hfill
\begin{subfigure}{0.45\textwidth}
  \centering
  \includegraphics[width=\textwidth]{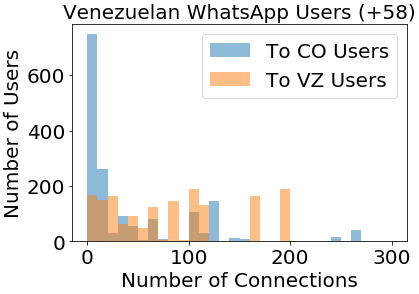}
\end{subfigure}
\caption{Users from Colombia are much better connected to other users from Colombia; Venezuelan users are much better connected to others from Venezuela..}
\label{figure:members:hist_connToCOVZ_COVZ}
\end{figure}

Next, figure \ref{figure:members:hist_connToCOVZ_COVZ} shows that CO users strongly tend to be well-connected with other CO users, and poorly connected with VZ users; the opposite relation holds for VZ users. This gives credence to our hypothesis that WhatsApp users use phone numbers from their current locations; if VZ migrants retained +58 (Venezuelan) phone numbers after migrating to Colombia, we wouldn't see nearly as strong a relation here.

\begin{figure}[h]
\centering
\includegraphics[width=0.45\textwidth]{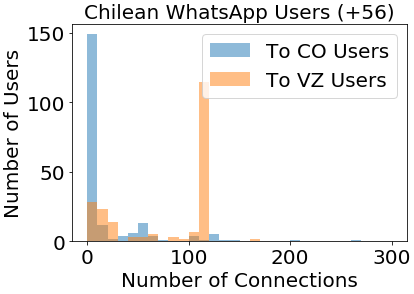}
\caption{Chilean users appear extremely well-connected to VZ users, and much more poorly connected to CO users.}
\label{figure:members:hist_connTo_CL}
\end{figure}

A strange nuance appears in the connections of CL users. In figure \ref{figure:members:hist_connTo_CL}, CL users appear extremely well-connected to VZ users, and much more poorly connected to CO users. This pattern appears somewhat suspicious to us, meriting a look at the actual data, which reveals that this disparity was driven by one 251-member group, which included 113 CLs and 111 VZs.

But more than mere coincidence, geography explains this disparity better than anything else. Colombia shares a (very porous) land border with Peru and Ecuador but not Chile; Santiago, the Chilean capital, is over 2,600 miles away from Bogotá (Quito, EC and Lima, PE are 441 and 1,167 miles respectively). So while migrants to EC and PE are likely to have recently spent time in CO, migrants to CL are much further along in their journeys---so we expect them to retain relatively weaker ties to CO users. The 251-member group described above, indeed, is titled \textit{Venezolanos por migrar} (``Venezuelans for migrating'').

\begin{figure}[h]
\centering
\begin{subfigure}{0.45\textwidth}
  \centering
  \includegraphics[width=\textwidth]{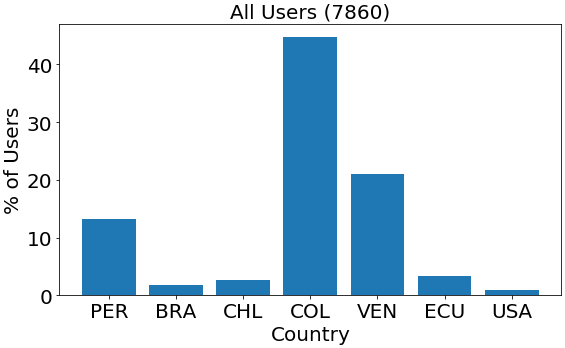}
\end{subfigure}%
\hfill
\begin{subfigure}{0.45\textwidth}
  \centering
  \includegraphics[width=\textwidth]{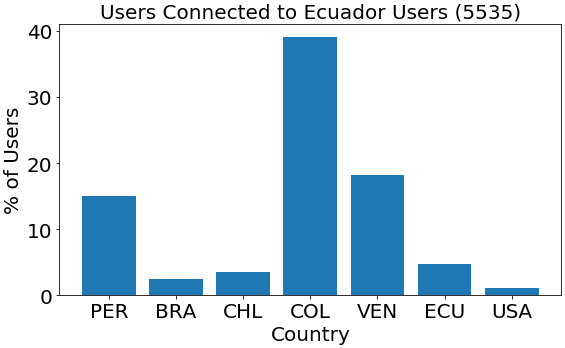}
\end{subfigure}
\begin{subfigure}{0.45\textwidth}
  \centering
  \includegraphics[width=\textwidth]{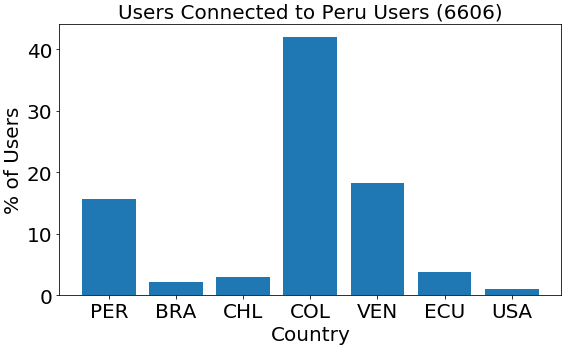}
\end{subfigure}%
\hfill
\begin{subfigure}{0.45\textwidth}
  \centering
  \includegraphics[width=\textwidth]{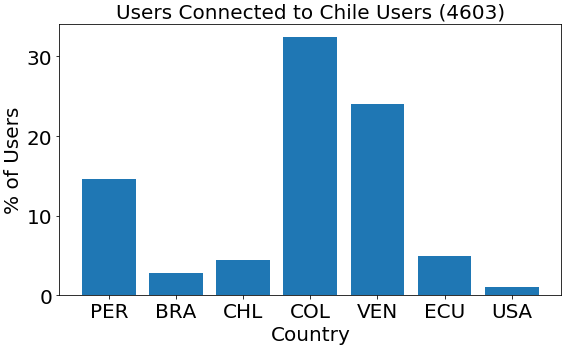}
\end{subfigure}
\caption{There are around twice as many CO users as VZ users, and this is also true amongst users connected to EC users, and users connected to PE users. But there are disproportionately more VZs amongst users connected to CL users.}
\label{figure:members:bar_withConnTo}
\end{figure}

An alternative look at this phenomenon shows that the geographical hypothesis holds, even without the effects of the CL/VZ-heavy 251-member group. The four graphs in figure \ref{figure:members:bar_withConnTo} depict country breakdowns of users in general, of users connected to EC users, of users connected to PE users, and of users connected to CL users.

The first three graphs are all quite similar, but the fourth reveals that relatively more VZ users (compared to their presence in the overall population) are connected to CL users. Aggregating connections in this way (looking at all users connected to CL users, as opposed to CO/VZ connections per individual CL user) decisively reduces the influence of the 251-member group described above. For the CL graph in figure \ref{figure:members:bar_withConnTo} to follow the pattern of the other graphs, nearly 460 fewer VZs (10\% of 4603) would have to be connected to CL users.

\section{Geographical Diversity within Groups}

Next, we investigate the geographical diversity of groups, using users' telephone country codes as an approximation of current (national) location. Given the signficance of xenophobia in the experience of Venezuelan migrants to Colombia, as well as migrants' hesitation towards trusting public WhatsApp groups, quantifying how ``cross-border'' and transnational each group is can allow us to better understand relationships and activity within groups.

Conventionally, two main indicies are used to measure diversity: (Shannon) entropy and the Simpson index \cite{jost-2006}. In our case, let there be users from countries $1, \dots, m$ in a group, so that $p_i$ of users are from country $i$. Then the entropy is calculated as $-\sum_{i=1}^m p_i \log p_i$, while the Simpson index is calculated as $\sum_{i=1}^m p_i^2$.

\begin{figure}[h]
\centering
\includegraphics[width=0.45\textwidth]{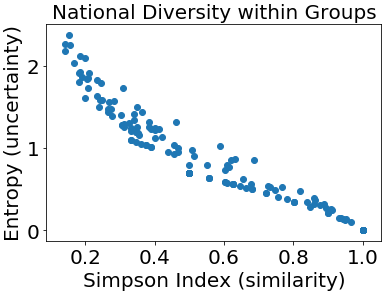}
\caption{The relationship between entropy and Simpson index is close to linear, but not quite.}
\label{figure:members:scatter_simpson_entropy}
\end{figure}

Figure \ref{figure:members:scatter_simpson_entropy} shows that these two indicies are very closely correlated across our groups, as they should be (Pearson $r = -0.962$). A slight nuance enters in that because entropy and the Simpson index measure different things, they may not always be so closely related. Intuitively, the Simpson index gives the probability of two users drawn from a group being from the same country; Shannon entropy, on the other hand, is more a measure of uncertainty, representing the average number of bits needed to convey which country a user is from.

We default to using entropy in later parts of this thesis, simply by how we frame our research interest. If an aid organization is deciding which groups to send a message to, it makes more sense to consider the uncertainty of geography in the group---approximating the uncertainty of where the message may end up---rather than the similarity or diversity of users. In any case, Shannon entropy and the Simpson index exhibit close to a linear relation across our groups, so it shouldn't matter.

\begin{figure}[h]
\centering
\includegraphics[width=0.45\textwidth]{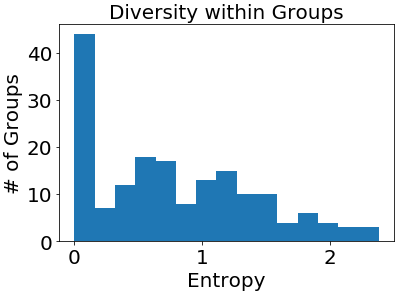}
\caption{Histogram of group entropies.}
\label{figure:members:hist_entropy}
\end{figure}

The histogram in figure \ref{figure:members:hist_entropy} illustrates how homogeneous some groups are.

\subsection{Correlates of Diversity}

\begin{figure}[h]
\centering
\begin{subfigure}{0.45\textwidth}
  \centering
  \includegraphics[width=\textwidth]{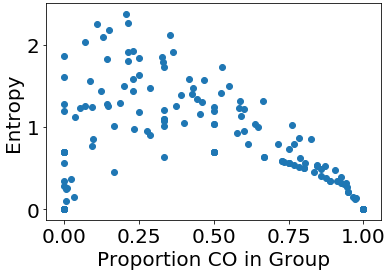}
\end{subfigure}%
\hfill
\begin{subfigure}{0.45\textwidth}
  \centering
  \includegraphics[width=\textwidth]{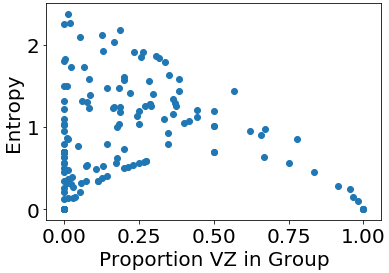}
\end{subfigure}
\caption{Scatter plots of entropies against proportions CO and proportion VZ of groups.}
\label{figure:members:scatter_pCOVZ_entropy}
\end{figure}

Figure \ref{figure:members:scatter_pCOVZ_entropy} plots entropy of each group against proportion of users from CO and proportion of VZ users, respectively. Both graphs are characteristically bounded below by the minimum entropy curve $-p \log p - (1-p) \log(1-p)$ (this is approximately $4p(1-p)$, which is Bernoulli variance scaled to 1); the minimum entropy curve is obtained if users are from only two countries with proportions $p, 1-p$ (naturally, this curve peaks at $p = 0.5$).

We see that there are many more heavily CO groups than there are heavily VZ groups, but more starkly, groups with few VZs are relatively homogeneous, while groups with few COs are relatively diverse. Entropy and proportion CO are moderately negatively correlated (Pearson $r = -0.50$, $p < 0.001$), while the entropy and proportion VZ aren't correlated ($r = 0.08$, $p = 0.31$). In individual regressions with entropy as the dependent variable, the OLS coefficient on proportion VZ is 0.20, and a much more drastic -0.88 on proportion CO, with the same $p$-values.

\begin{figure}[h]
\centering
\begin{subfigure}{0.45\textwidth}
  \centering
  \includegraphics[width=\textwidth]{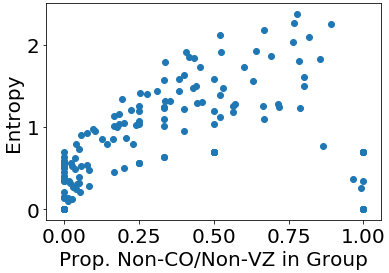}
  \caption{Scatter plot of entropies against proportion of users from neither CO nor VZ..}
  \label{figure:members:scatter_p3rd_entropy}
\end{subfigure}%
\hfill
\begin{subfigure}{0.45\textwidth}
  \centering
  \includegraphics[width=\textwidth]{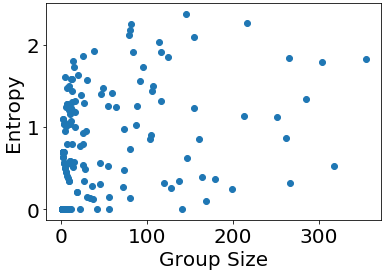}
  \caption{Scatter plot of entropies against group sizes.}
  \label{figure:members:scatter_size_entropy}
\end{subfigure}
\caption{Scatter plots with entropy.}
\end{figure}

Figure \ref{figure:members:scatter_p3rd_entropy} plots entropy against the proportion of users from neither CO nor VZ. Generally, groups with more 3rd country users are more diverse (Pearson $r = 0.51$, $p < 0.001$), but the effect is diminished by a few groups that are mostly 3rd country users yet very homogeneous; we can imagine these as Peru-centered groups, Ecuador-centered groups, etc. Finally, figure \ref{figure:members:scatter_size_entropy} plots entropy against group sizes; larger groups tend to be more diverse, but the effect is weak.

To restate that entropies and Simpson indicies are nearly interchangeable, all of the results above were equally statistically significant/insignificant when using Simpson index, and correlations/OLS coefficients were in the direction we'd expect. For example, the Pearson $r$ between entropy and proportion CO was $-0.50$ ($p < 0.001)$, while the Pearson $r$ between Simpson index and proportion CO was $0.53$ ($p < 0.001$).

\section{Network Properties}

As we saw in Chapter \ref{ch:relatedwork}, the network structure of users and groups offers important insights on how information propagates on WhatsApp.

\subsection{Network Properties of Groups}

We first construct an undirected graph with groups as nodes, connecting groups if they share a user in common. Of 174 groups, 107 are connected to at least one other group, and the largest connected component involves 86 groups (49.4\%). This is reasonable; work like that by Resende et al. (2019) found varying sizes of largest connected components (LCCs): 25 groups of 136 groups related to a Brazilian truckers' strike (18.4\%), and 206 of 333 political groups related to the 2018 election (61.9\%) \cite{resende-misinformation-2019}.

Taking an alternate look, our network of groups is actually \textit{really} well connected. Only 107 groups (of 174) are connected to any other group, so we might imagine that the remaining 67 might simply never be connected, for whatever reason---they might be, for example, dedicated business channels where only group administrators can send messages (this restriction is possible on WhatsApp). So of groups that \textit{are} connected to other groups, over 80\% are in the largest connected component!

\begin{figure}[h]
\centering
\includegraphics[width=\textwidth]{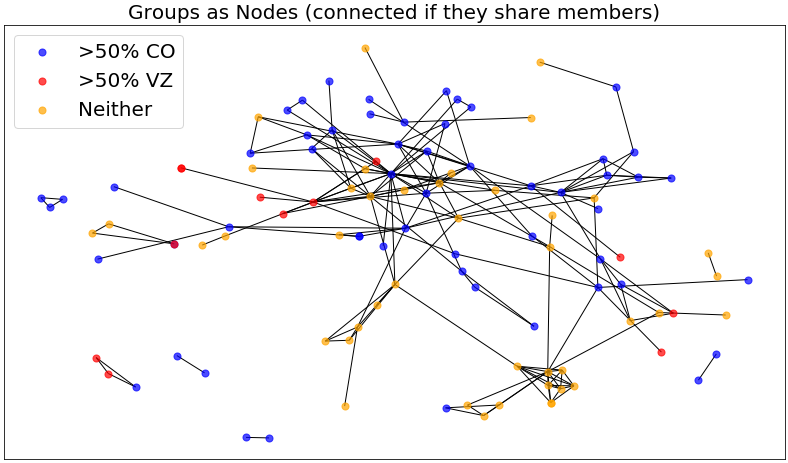}
\caption{Visualization of group network. We only include the 107 groups that are connected to any other group.}
\label{figure:members:networkx_groups}
\end{figure}

We present in figure \ref{figure:members:networkx_groups} a visualization of this graph with groups as nodes. The behemoth LCC is clear here, and, as expected, we see that there aren't any other connected components of significant size (indeed, the second LCC involves four groups). We shade groups by whether they're mostly CO users, mostly VZ users, or neither.

It appears that the main hubs connecting groups in the LCC are not VZ-dominant groups (i.e., either blue or yellow). For groups in the LCC, we calculate their average (shortest path) distance to all other groups as a measure of their centrality (a perfectly central group would have average-shortest-path 1). Of the 16 most central groups in the LCC, indeed, only one is VZ-dominant (it's easily identifiable on the graph).

Averaged across all groups, however, the centrality of CO-dominant groups, VZ-dominant groups, and groups that are neither CO-dominant nor VZ-dominant are all quite similar. The average shortest path distances are 3.487, 3.684, and 3.590 respectively; neither ANOVA nor a $t$-test between CO-dominant and VZ-dominant groups were significant.

For groups in the LCC, their centrality was moderately positively correlated to their size (Pearson $r = -0.27$ between average-shortest-path and size, $p = 0.01$), though the effect is small. From an OLS regression with just these variables, an increase in group membership by 100, on average, is linked to a reduction in average-shortest-path by 0.2.

\begin{figure}[h]
\centering
\includegraphics[width=0.45\textwidth]{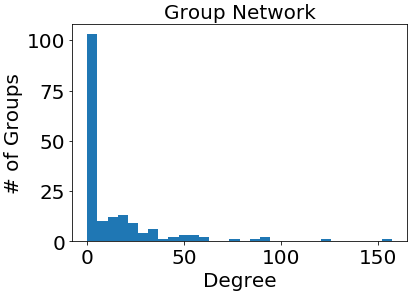}
\caption{Histogram of group degrees.}
\label{figure:members:hist_degree}
\end{figure}

Figure \ref{figure:members:hist_degree} shows a histogram of group degrees. Given that 67 groups are not connected to any other groups, this highly-skewed distribution is unsurprising; 60\% of groups have degree less than 5, but 20.7\% have degree larger than 20. The average degree of all groups is 13.1, and the two extremely well-connected groups, with degrees 123 and 157, are both general/just-for-fun groups (one about salsa, the other a general interest group for Cúcuta).

Unsurprisingly, degree is moderately positively correlated with both size and entropy of groups; larger and more diverse groups are immediately adjacent to more groups. The relationship with both is significant even when controlling for one another: an OLS regression of group degree on size and entropy yields coefficients $0.134$ on size ($p < 0.001$) and $6.45$ on entropy ($p = 0.01$). Even when dropping highly-connected groups (e.g., groups with degree $> 70$), the relationship holds---the coefficients about halve, but remain signficant.

Finally, we examine the clustering coefficient of groups, which is the probability that for any two groups connected to a group, those other two groups are also connected. At node $i$ with degree $d_i$, the clustering coefficient $C_i = \frac{\textrm{\# of triangles involving $i$}}{\textrm{\# of possible triangles involving $i$}} = \frac{\textrm{\# of triangles involving $i$}}{d_i \times (d_i - 1) / 2}$ \cite{survey-mobile-2015}; the clustering coefficient represents the presence of triadic closure around a group---the tendency for two nodes both to a third node to themselves connect \cite{networks-book-2010}.

\begin{figure}[h]
\centering
\includegraphics[width=0.45\textwidth]{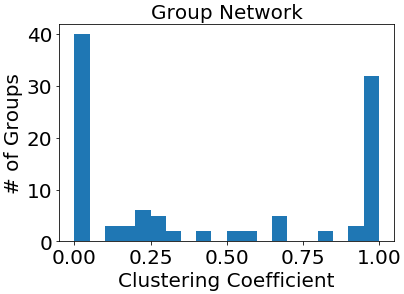}
\caption{Distribution of group clustering coefficients (only including groups with degree $\geq 2$).}
\label{figure:members:hist_clustering}
\end{figure}

In our scenario, the clustering coefficient at a group meaningfully approximates interactions around that group. Figure \ref{figure:members:hist_clustering} shows a histogram of group clustering coefficients. Of 107 groups with degree at least 2, around 40 have clustering coefficient 0 and around 30 have clustering coefficient 1, while the remainder of groups fall in between. Clustering coefficient is weakly negatively correlated to group size ($Pearson r = -0.29, p < 0.01$).

\subsubsection{Three Class Graphs}

As one perspective on the centrality and distribution of groups in this network, we classify groups into three classes based on various metrics---size, proportion CO, proportion VZ, proportion non-CO/non-VZ, and entropy. For each metric, we consider groups in the 0-30th percentiles of all groups for that metric, groups in the 30th-70th percentiles, and groups in the 70th-100th percentiles. For group size, for example, groups are categorized based on whether they have $\leq 5$ members, have 6-46 members, or have $> 46$ members.

We graph the largest connected component in the group network below, shading groups by their classification.

\begin{figure}[h]
\centering
\includegraphics[width=\textwidth]{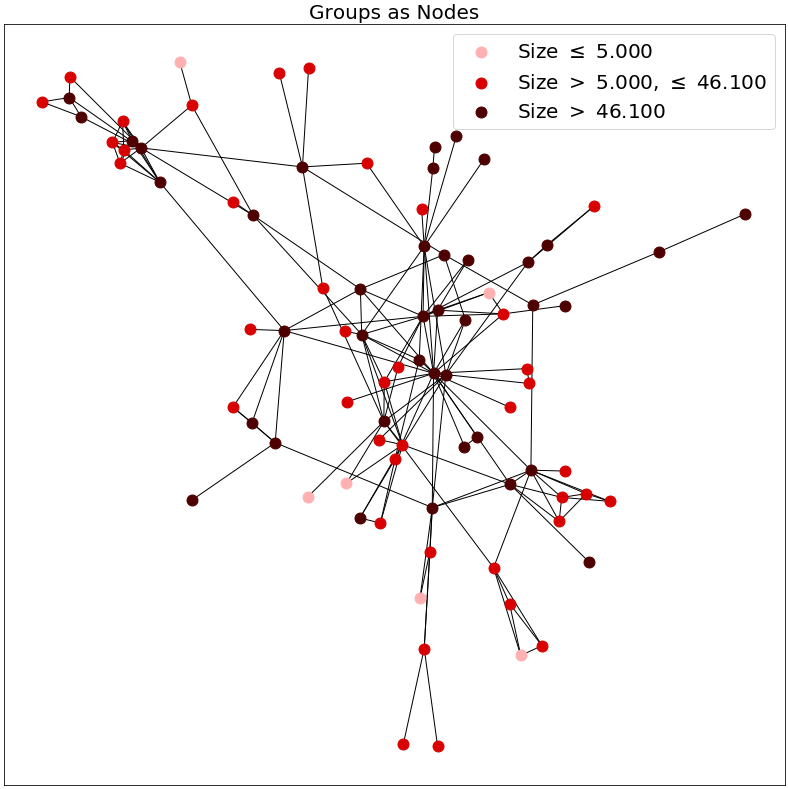}
\caption{Visualization of group network, shading groups by their size.}
\label{figure:members:networkx_3cls_Size}
\end{figure}

In figure \ref{figure:members:networkx_3cls_Size}, large groups seem a lot more central. Indeed, the smallest groups have average-shortest-path 3.93, medium-sized groups (6 to 46 members) have average-shortest-path 3.69, and the largest groups have average shortest path 3.35. Both ANOVA (across all three classes) and a $t$-test (between the smallest/largest groups) yielded $p < 0.05$.

This is an important characterization, since we might imagine a trade-off between sending messages to a group with many active participants, and a group with few active participants. An aid organization, for example, might consider disseminating information to a more active group, at the risk of being crowded out by the many active participants. While attention/interaction is a different story, one we discuss in Chapter \ref{ch:replycascades}, it's clear that messages sent to the group with many active participants require fewer steps to be disseminated more broadly.

\begin{figure}[h]
\centering
\includegraphics[width=\textwidth]{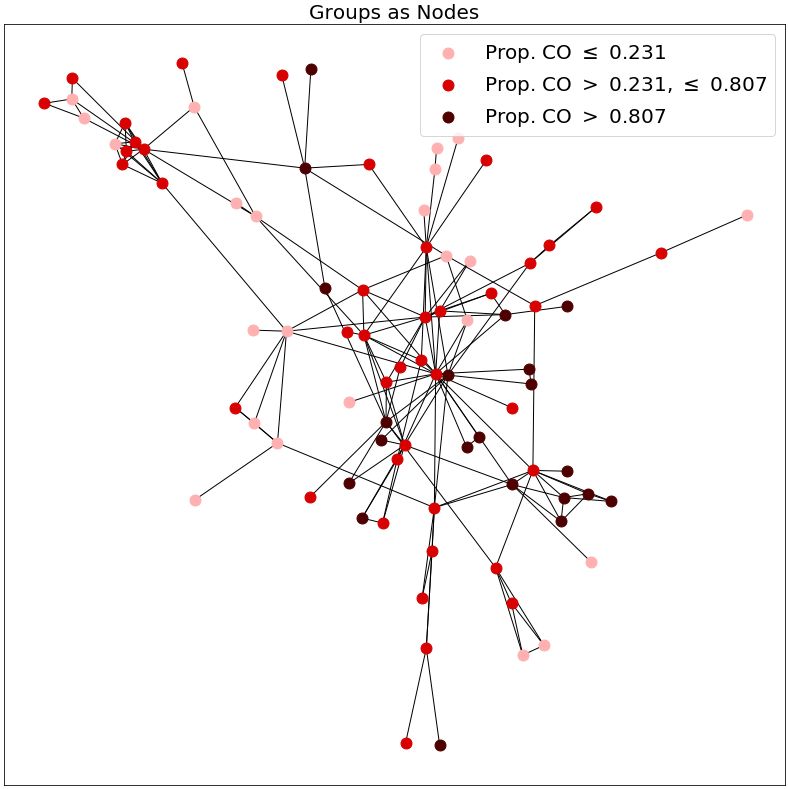}
\caption{Visualization of group network, shading groups by their proportion of CO members.}
\label{figure:members:networkx_3cls_pCO}
\end{figure}

Figure \ref{figure:members:networkx_3cls_pCO} classifies groups by what proportion of their members are CO. CO-dominant groups are slightly more central, but this relationship is weak. For brevity, we exclude the graphs where we classified groups by proportion VZ, by proportion non-CO and non-VZ, and by entropy; under none of those classifications was there a statistically significant difference in centrality.

\subsection{Network Properties of Users}

Finally, we construct an undirected graph with users as nodes, connecting users if they are both part of any group. Of 7,860 users in our graph, 5,693 users (72.4\%) are part of the largest connected component; this kind of ``giant'' connected component has been shown in nearly every social network. In \cite{resende-misinformation-2019}, 8,934 of 10,860 WhatsApp users (82.3\%) were in the LCC; \cite{bursztyn-whatsapp-2019} calculates LCC sizes as $\sim 80\%$ and $\sim 95\%$ of the networks in two samples; the text \textit{Networks, Crowds, and Markets} explains that most large, complex networks \textit{should} have exactly one ``giant component'' \cite{networks-book-2010}.

\begin{figure}[h]
\centering
\includegraphics[width=\textwidth]{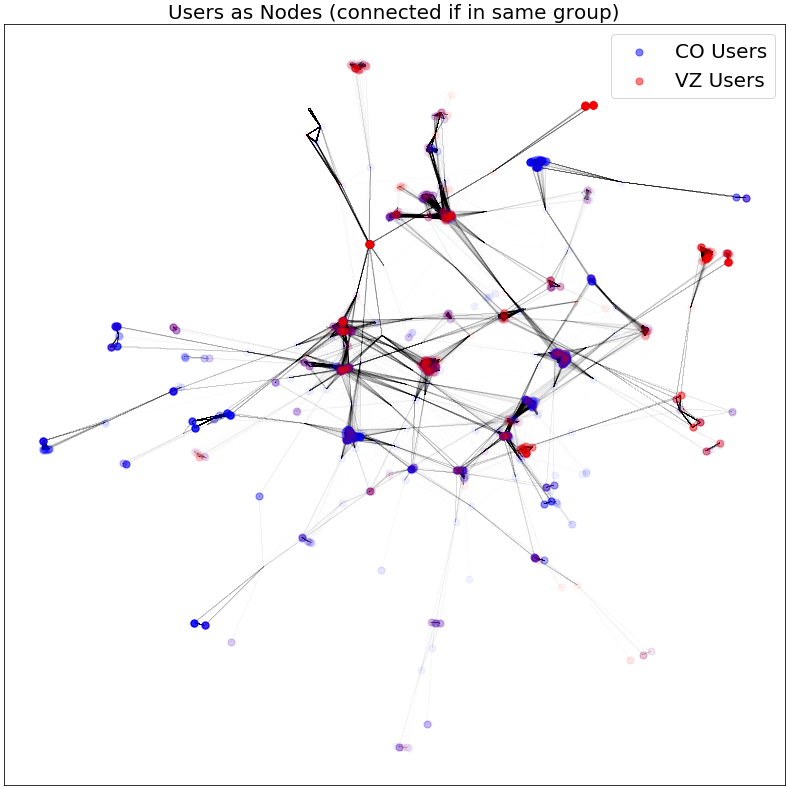}
\caption{Visualization of user network.}
\label{figure:members:networkx_co_vz}
\end{figure}

\begin{figure}[h]
\centering
\includegraphics[width=\textwidth]{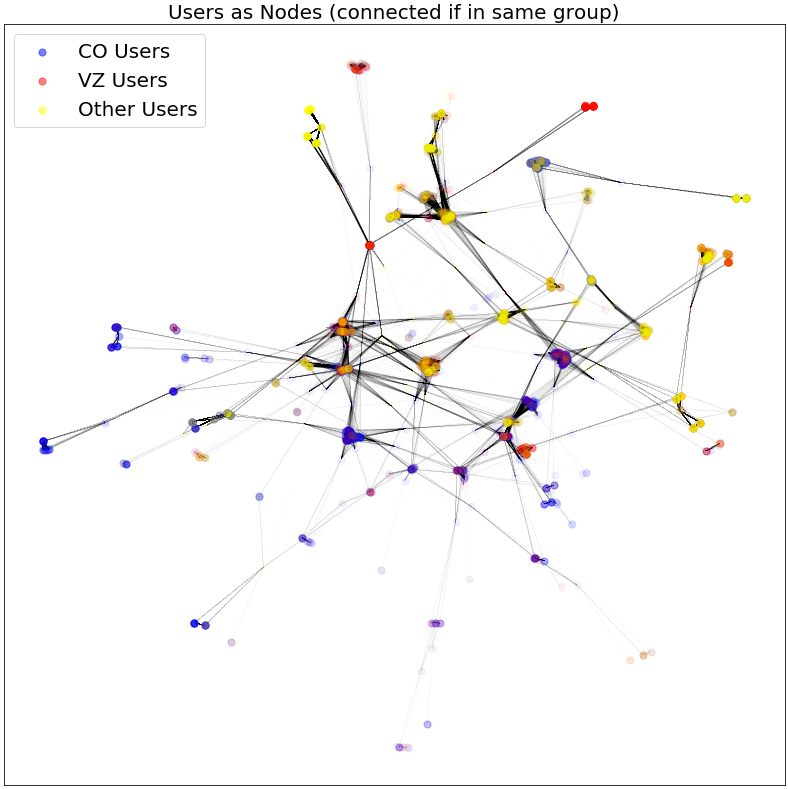}
\caption{Visualization of user network.}
\label{figure:members:networkx_co_vz_other}
\end{figure}

Figures \ref{figure:members:networkx_co_vz} and \ref{figure:members:networkx_co_vz_other} attempt to visualize this LCC, a difficult (and rather futile) task given that this graph includes 5,693 nodes (we settled on coloring nodes with very low opacity). Still, it certainly appears that the more central clusters in each graph are more diverse than clusters on the outskirts. We skip over any analysis here, since our exploration of the network of groups covered most relevant aspects.

\begin{figure}[h]
\centering
\includegraphics[width=\textwidth]{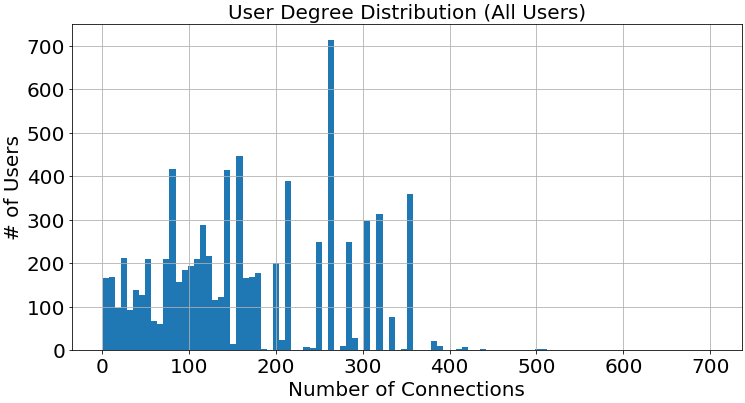}
\includegraphics[width=\textwidth]{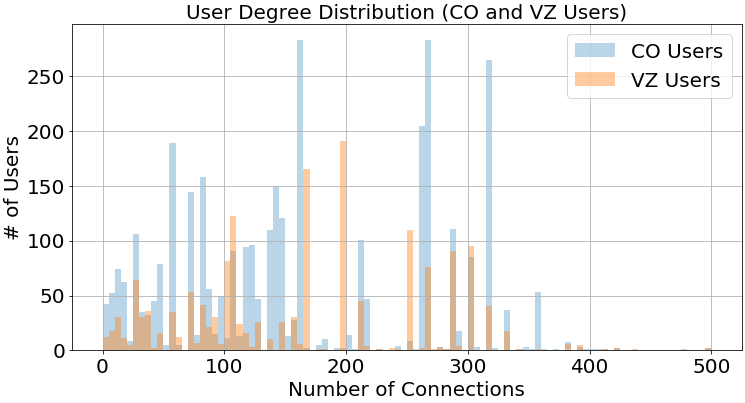}
\caption{Histogram of degrees, of all users and of VZ vs. CO users.}
\label{figure:members:degree_dist}
\end{figure}

We do graph in figure \ref{figure:members:degree_dist} the distributions of user degrees. One user (from Mexico of all places) has degree 701, while of the 20 users with highest degree, seven are from VZ and six are from CO. The mean and median user degree are close, at 167.8 and 155.0 respectively; 70.6\% of users have degree over 100, and 35.7\% of users have degree over 200. The distribution of degrees of CO users does not differ significantly from that of VZ users (their means are 162.0 and 166.0 respectively).

We note that while distributions of user degree in social networks often follow the power law, this is not the case for us. Indeed, our measure of connection between users is relatively weak---only requiring them to be in a group together---so we should not expect the exponential distributions that have been observed elsewhere.

\chapter{Initial Analysis of Messages\label{ch:messages}}

In this section, we provide an overview of messages from our dataset.

We recorded 171,634 messages between February 13th, 2020 and April 5th, 2020 (inclusive) from 174 unique groups and 7860 unique users. These messages included 101,414 messages with text (59.1\%); 38,455 messages with images (22.4\%); 8,918 audio messages (5.2\%), and 15,596 videos (9.1\%);\footnote{Because messages could have included multiple content types---images with captions, for example---and because some messages were of miscellaneous content types (documents, location pins, etc.), these proportions do not add to 100\%.} 28,886 messages (16.8\%) included emojis, which could have been sent with text or by themselves.

\section{Descriptive Statistics}

\subsection{Text Messages}

\begin{figure}[h]
\centering
\begin{subfigure}{0.45\textwidth}
  \centering
  \includegraphics[width=\textwidth]{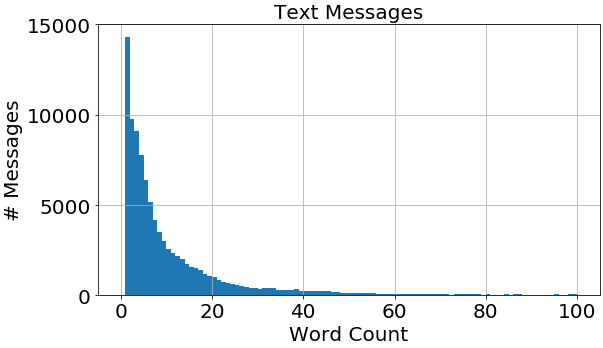}
  \caption{Histogram of word count of WhatsApp text messages.}
  \label{figure:messages:word_count}
\end{subfigure}%
\hfill
\begin{subfigure}{0.45\textwidth}
  \centering
  \includegraphics[width=\textwidth]{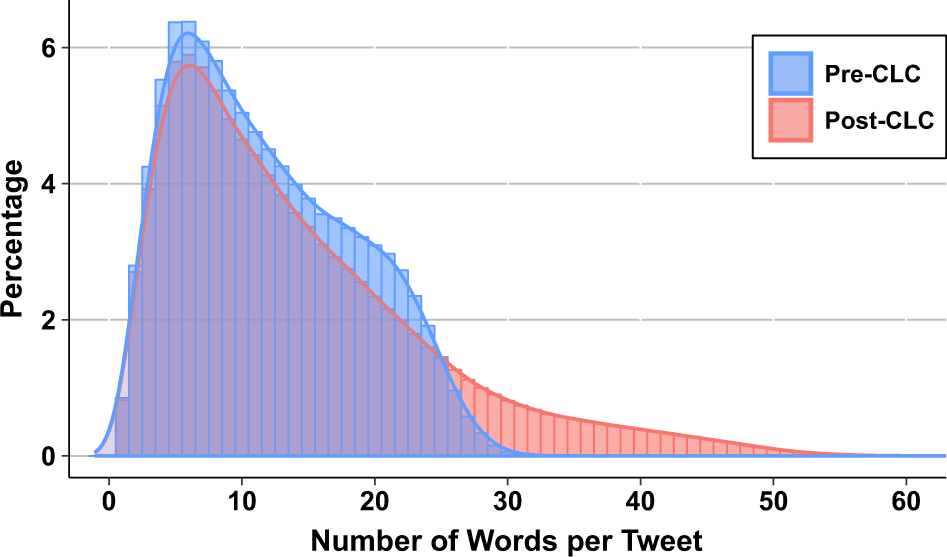}
  \caption{Density estimation of word count of tweets, from \cite{boot-tweets-2019}.}
  \label{figure:messages:twitter_wordcount}
\end{subfigure}
\caption{Comparison of distributions of word count, between our WhatsApp dataset and Twitter.}
\end{figure}

For the 101,414 messages with text, figure \ref{figure:messages:word_count} shows a histogram of word counts, which nearly perfectly follows a power law distribution. 14.1\% of text messages are exactly one word, 25\% of messages are three words or fewer, and 75\% of messages are $\leq 16$ words. The tail of this distribution, as we expect, is extremely long, with 5.1\% of messages over 100 words and 1.1\% of messages over 500 words.

Our distribution of word count contrasts nicely with figure \ref{figure:messages:twitter_wordcount}, which estimates the probability density of word counts in tweets; on Twitter, word count peaks around 10 words, and doesn't sharply fall until 30-40 words. Thus, even though our dataset consists entirely of public groups, it's clear that the lengths of messages are more similar to what we'd expect in SMS and private WhatsApp conversations, rather than a public forum like Twitter. In particular, this might mean that official actors (e.g., governments and aid organizations) who distribute information over public WhatsApp groups should pay attention to message length, since even a 20 word message would be longer than 79.8\% of other messages.

\begin{figure}[h]
\centering
\begin{subfigure}{0.45\textwidth}
  \centering
  \includegraphics[width=\textwidth]{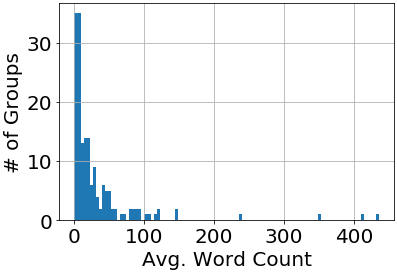}
  \caption{Histogram of average word count of various groups.}
  \label{figure:messages:group_word_count}
\end{subfigure}%
\hfill
\begin{subfigure}{0.45\textwidth}
  \centering
  \includegraphics[width=\textwidth]{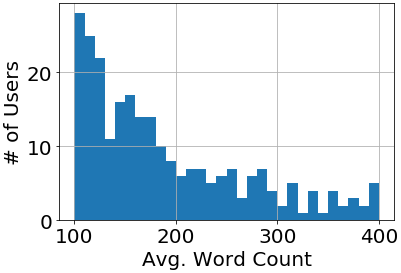}
  \caption{Histogram of average word count of users; we only include users with average word count over 100.}
  \label{figure:messages:user_word_count}
\end{subfigure}
\caption{Some groups and some users have lengthy messages.}
\end{figure}

\pagebreak
Yet the reception towards longer messages likely differs across groups. Indeed, in figure \ref{figure:messages:group_word_count}, we notice that some groups have sharply longer messages, on average, than others; the same is true of users, as shown in figure \ref{figure:messages:user_word_count}.

\begin{figure}[h]
\centering
\includegraphics[width=0.45\textwidth]{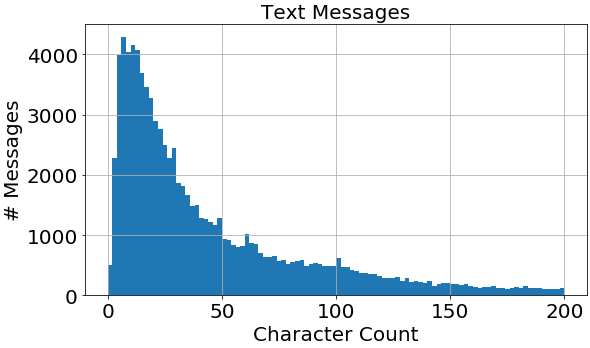}
\caption{Histogram of character count of WhatsApp text messages.}
\label{figure:messages:char_count}
\end{figure}

In figure \ref{figure:messages:char_count}, we see that the distribution of character counts of messages, like the distribution of word counts, also follows an exponential distribution, with a peak at around 10 characters. We had found that 5.1\% of messages are over 100 words. Average word length in Spanish is around 5.22 words,\footnote{\url{http://www.puchu.net/doc/Average_Word_Length}} and 5.7\% messages are over $5.22 \times 100$ characters, so character count distribution closely matches the word count distribution.

\begin{figure}[h]
\centering
\includegraphics[width=\textwidth]{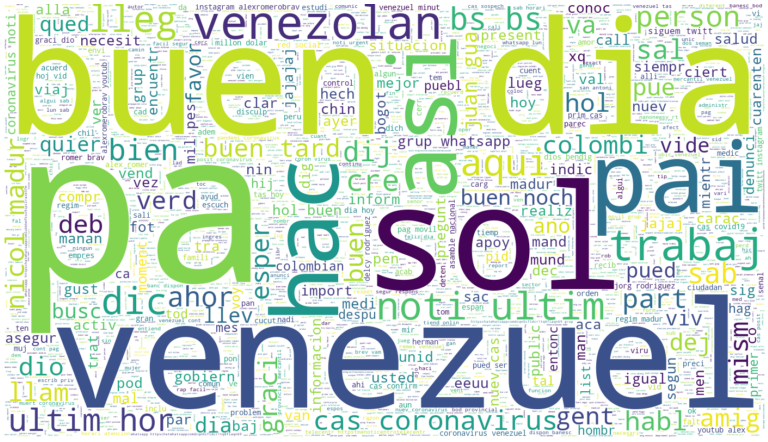}
\caption{Word cloud of WhatsApp text messages, after stemming words and removing commonly used words.}
\label{figure:messages:wordcloud}
\end{figure}

After stemming words (e.g., merging different conjugations of verbs) and removing commonly used words, techniques we describe in much more detail in Chapter \ref{ch:misinformation:labeling} (on labeling misinformation), we obtain the word cloud in figure \ref{figure:messages:wordcloud}. Some common phrases remain, like ``buen día'' (good day), but we can better see themes like Venezuela, the dictator Maduro, the opposition leader Juan Guaidó, coronavirus, news, work, and so on.

Using latent dirichlet allocation---which models documents as random mixtures over hidden (latent) topics, which themselves involve probabilistic distributions of words---we can very nascently parse out topics from text messages in our dataset. Parameterizing this as 10 topics with 10 words each, we obtain the following topics:

\singlespacing\begin{enumerate}
  \item bs hol grup venezuel whatsapp tas pes hoy pag 1
  \item dios senor dia amen vid mund amor mand vide cre
  \item man virus 591 mil pued clar pes sal seman dos
  \item graci pas bien ok buen dia feliz cambi dias grup
  \item coronavirus venezuel cas fuent inform carac nacional covid19 pais servici
  \item venezuel madur gua venezolan pais eeuu nacional gobiern president regim
  \item grup jajaj fals verd vide envi asi notici fot informacion
  \item experient trabaj am jajajaj pm vid mes envi interes priv
  \item coronavirus cas ultim noti covid19 chin hor nuev confirm pais
  \item q buen hac pued pas dias sol sab amig gent
\end{enumerate}\doublespacing

We notice that topics 1, 4, 8, and 10 largely consist in greetings; topic 2 is religious; topic 6 is quite political; and topic 5 centers on the coronavirus.

\subsection{Audio and Video Messages}

\begin{figure}[h]
\centering
\begin{subfigure}{0.45\textwidth}
  \centering
  \includegraphics[width=\textwidth]{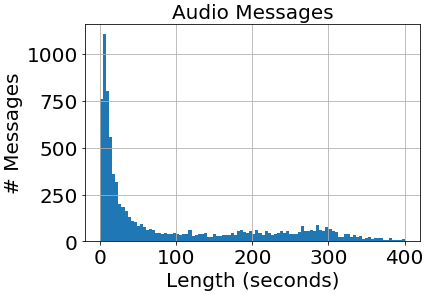}
\end{subfigure}%
\hfill
\begin{subfigure}{0.45\textwidth}
  \centering
  \includegraphics[width=\textwidth]{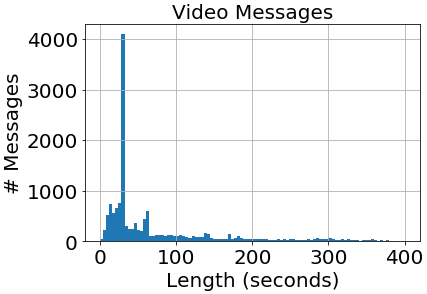}
\end{subfigure}
\caption{Histograms of audio and video message length. We only plot to 400 seconds, but there are much longer audio recordings and videos; 12.1\% of audio messages and 7.5\% of videos are longer than five minutes.}
\label{figure:messages:audio_video_length}
\end{figure}

Figure \ref{figure:messages:audio_video_length} shows histograms of audio and video message lengths. The length of audio messages nearly follows the power law distribution of text message length, though falls much less dramatically---52.9\% of messages are longer than 30 seconds. The tail is fatter than for text messages, with 11.2\% of audio messages between 100-199 seconds, and 14.6\% of audio messages between 200-299 seconds.

The distribution of video message length exhibits an interesting shape, with an unmistakably sharp peak at 30 seconds---nearly 17.5\% of videos are exactly 30 seconds long (compared to 6.3\% that are 29 seconds and 0.7\% that are 31 seconds). A likely explanation for this may be that specialized content creators, like news organizations or propagandists, directly tailor their videos to this length; much of the video content in our groups, then, may be semi-professionally created.

Note that our dataset includes forwarded messages (15.2\% of all messages, but 26.3\% of all audio messages and 40.2\% of videos),\footnote{Here, we only consider ``forwarded'' messages as per the forwarding feature on WhatsApp. Of course, it's possible that users may simply download and re-upload audio/video content, though this is difficult to identify, given the data limitations we discussed in Section \ref{ch:datamethod:data-limitations}.} so the distributions in figure \ref{figure:messages:audio_video_length} are not necessarily indicative of the length of original content. But even when limiting our analysis to non-forwarded messages, the distribution remains basically the same: 20.5\% of non-forwarded videos are 30 seconds long, compared to 7.4\% that are 29 seconds long and 0.8\% that are 31 seconds long.

Given that speaking speed in Spanish is typically between 7-8 syllables per second,\footnote{\url{https://www.transfluent.com/en/2015/07/why-spanish-uses-more-words-than-english-an-analysis-of-expansion-and-contraction/}} it's likely that audio and video messages include substantially more information than text messages on average.\footnote{This has immediate disclaimers: some audio messages may only be music, some videos may not include any spoken words, and so on.} From the perspective of a content creator, say an aid organization attempting to disseminate information, it's likely wise to consider sharing textual content instead as spoken audio or narrated video. Anything over 20 words is an outlier amongst text messages, but 30-second audio recordings and videos aren't; this isn't to say that users necessarily pay less attention to long text messages, but simply that users are more accustomed to content-heavy audio and video.

\subsection{Group Activity}

To measure how active groups are, we use a normalized measure of how many messages they send in our collection time period. This isn't just a simple sum (or dividing that sum by the 53 days we collected data), since we didn't have complete access to every group for the entirety of our time period. Specifically, some groups kicked out the accounts we used to collect data,\footnote{Recall that with six total accounts/smartphones, we joined every group with two different accounts/phones.} which is both unsurprising and inevitable given that we never send any messages, and also join from US phone numbers.\footnote{Non-Colombian/non-Venezuelan phone numbers, especially ones from outside of Latin America, are suspicious in general, though U.S. phone numbers likely attract a disproportionate amount of attention since WhatsApp orders the list of members by ascending country code.}

To account for this discrepancy between groups, we calculated the number of days between the first message collected in each group and the last message (inclusive), and divided the total number of messages we collected in that group by this number of days. This approximates a group activity rate of messages/day, but clearly with large margin of error: we may have collected data from groups on off-days or extremely active days,\footnote{We joined groups around the same time, though, so this seems like a minor issue.} we may bias upwards the activity of very inactive groups,\footnote{Imagine that we never get kicked out a group, but it only has one message on Day 1 and no more messages for Days 2-53; we erroneously record its activity as $1$ instead of $\frac{1}{53}$).} and so on. Still, this is a relatively robust measure for group activity.

\begin{figure}[h]
\centering
\begin{subfigure}{0.45\textwidth}
  \centering
  \includegraphics[width=\textwidth]{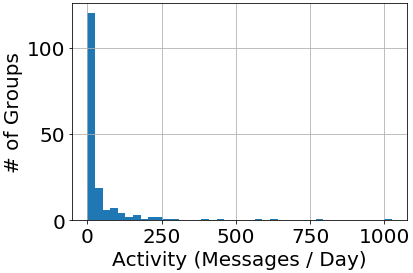}
  \caption{Histogram of our group activity measure.}
  \label{figure:messages:group_activity}
\end{subfigure}%
\hfill
\begin{subfigure}{0.45\textwidth}
  \centering
  \includegraphics[width=\textwidth]{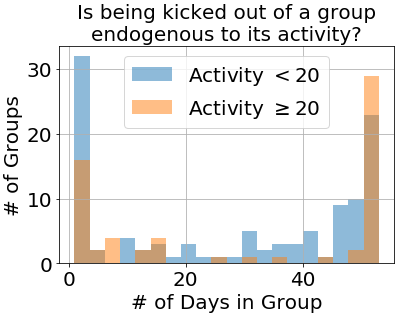}
  \caption{Histogram of how many days we collected data from groups.}
  \label{figure:messages:group_kicked}
\end{subfigure}
\caption{Exploring our measure of group activity.}
\end{figure}

Figure \ref{figure:messages:group_activity} shows a histogram of our group activity measure; figure \ref{figure:messages:group_kicked} examines if being kicked out of groups might be endogeneous to how active they are (this would mean that our activity measurements for active groups are more error-prone/higher variance, since we collect data for fewer days in those groups).

If we set 20 messages/day as the delimiter for inactive/active groups, then there is no statistically significant difference between how long we were able to stay in inactive groups vs. in active groups (we actually stayed longer, on average, in active groups).

Group activity is moderately positively correlated to group size ($r = 0.36$, $p < 0.001$) and group entropy ($r = 0.39$, $p < 0.001$). The OLS estimate for a regression of group activity on size and entropy is presented in table \ref{table:messages:ols_group_activity}.

\begin{table}[h]
\centering
\caption{OLS regression of group activity on group size and group entropy.}
\label{table:messages:ols_group_activity}
\begin{tabular}{p{0.25\textwidth} p{0.35\textwidth} p{0.10\textwidth} p{0.15\textwidth}}
\toprule
 & \textbf{Coefficient (Std. Err.)} & \textbf{$t$} & \textbf{P-Value} \\
\midrule

Intercept & $-18.9513 (14.228)$ & $-1.332$ & $0.185$ \\[0.2em]

Size & $0.4721 (0.130)$ & $3.638$ & $0.000^{\ast}$ \\[0.2em]

Entropy & $62.8113 (14.730)$ & $4.264$ & $0.000^{\ast}$ \\[0.2em]
\midrule
\multicolumn{4}{c}{$n = 174$\,\,(171 d.f.) \quad\quad $R^2 = 0.213$}\\
\bottomrule
\end{tabular}
\end{table}

Unsurprisingly, larger groups are more active---each additional member is linked to an average increase of 0.47 messages/day---and more diverse groups are significantly more active even while controlling for size. There might be some reverse causality in both of these relationships---people, and people from different countries, might be more likely to join more active groups---though there are likely strong effects in both directions. We might imagine that entropy spurs activity in cases like cross-border transactions, cross-border information exchange, and so on. Or we can imagine that cross-border groups have a higher barrier-to-entry (both because they're more difficult to find, and because discussion topics are more limited), so members who do join cross-border groups are more active on average.

\begin{figure}[h]
\centering
\begin{subfigure}{0.45\textwidth}
  \centering
  \includegraphics[width=\textwidth]{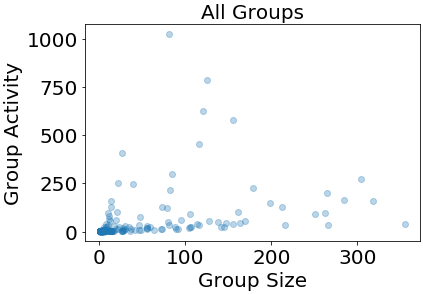}
\end{subfigure}%
\hfill
\begin{subfigure}{0.45\textwidth}
  \centering
  \includegraphics[width=\textwidth]{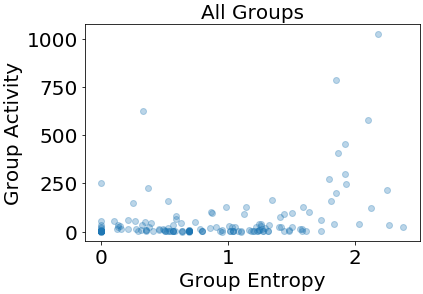}
\end{subfigure}
\caption{Scatter plots of our activity measure against size and entropy.}
\label{figure:messages:scatter_activity}
\end{figure}

Figure \ref{figure:messages:scatter_activity} shows scatter plots of our activity measure against group size and entropy.

\section{Message Concentration and Inequality}

We now present measures of concentration and inequality within groups, as determined by the within-group distributions of how many messages are sent by each user.

A large part of the motivation for these measures is the decentralized nature of the Venezuelan migrant crisis: unlike in other crises where migrants frequently interface with central authorities and institutions---imagine, for example, refugee camps in Greece---Venezuelan migrants have very little interaction with government and aid organizations in Colombia. Much of this stems from the relatively little funding allocated to the crisis by the international community (around \$50 for each Venezuelan migrant to Colombia, compared to around \$50,000 for every Syrian refugee in Europe). However, many of the migrants we interviewed also had family and friends in Colombia, and the transition from Venezuela to Colombia is less overwhelming then, say, from Syria to Western Europe, making turning to aid organizations less necessary.

Within our dataset, some groups are dominated by one or a few users---news groups and dedicated channels for businesses, for example---while others involve much more organic interaction between members. Given the decentralized nature of this crisis, it's worth exploring how concentration in groups can affect how migrants use and share information. More generally, concentration and inequality are important aspects of social networks that impact the relationships and activities of members.

As our principal measure of concentration, we calculate the Herfindahl-Hirschman (H-H) index $\sum_{i = 1}^N s_i^2$, where $s_i$ is the share of messages in the group sent by user $i$, across all $N$ members in the group. Note that this is the same measure as the Simpson index we used to calculate similarity of user countries in groups; since our context---dominance of messages in a group---is closer to industry dominance by firms (the origin of the H-H concentration, an economic concept) than biodiversity, we name it after Mr. Orris C. Herfindahl and Mr. Albert O. Hirschman.

\begin{figure}[h]
\centering
\begin{subfigure}{0.45\textwidth}
  \centering
  \includegraphics[width=\textwidth]{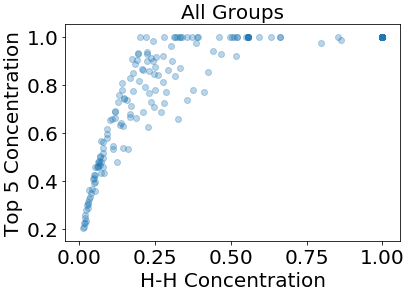}
  \caption{Scatter plot of H-H concentration and top 5 concentration.}
  \label{figure:messages:scatter_hh_top5}
\end{subfigure}%
\hfill
\begin{subfigure}{0.45\textwidth}
  \centering
  \includegraphics[width=0.8\textwidth]{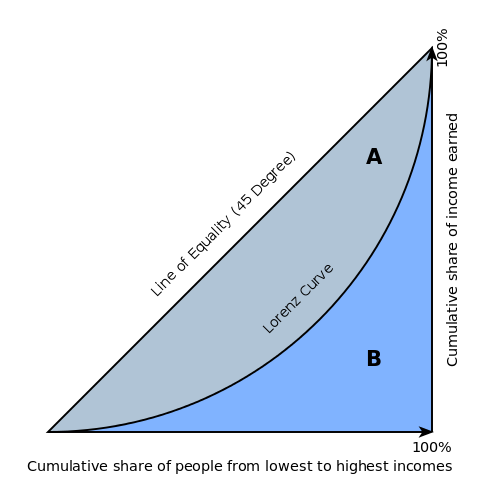}
  \caption{A typical Lorenz curve.}
  \label{figure:messages:lorenz}
\end{subfigure}
\caption{Measuring concentration and inequality within groups.}
\end{figure}

In a group dominated by one user, the H-H concentration is 1, while a perfectly egalitarian group has H-H concentration $\sum (\frac{1}{N})^2 = \frac{1}{N}$. We also calculate the top 5 concentration of each group, simply the proportion of messages sent by the 5 most active members. Figure \ref{figure:messages:scatter_hh_top5} plots these concentration measures against each other.

H-H concentration and top 5 concentration are closely correlated ($r = 0.73$; $p < 0.001$), especially for egalitarian groups (where both concentration measures are low); groups dominated by five (or fewer) members range in concentration. For the rest of this thesis, we default using the H-H concentration, since the top 5 concentration doesn't generalize well between (e.g.) very small and very large groups.

To measure inequality in groups, we calculate the well-known Gini coefficient, which is typically derived from the Lorenz curve, which orders individuals from lowest income to highest income (fewest messages to most messages), and then plots cumulative share of total income (messages) against cumulative share of people \cite{gastwirth-lorenz-1972}. A typical Lorenz curve is shown in figure \ref{figure:messages:lorenz}. The $45^\circ$ line would represent perfect equality, since it integrates a uniform distribution; the Gini coefficient is twice the area of A, the region between the actual Lorenz curve and the perfect equality curve, so A would have no area under perfect equality. Perfect inequality would involve $A = \frac{1}{2}$ and B having no area, since the last person has all of the income (messages).

\begin{figure}[h]
\centering
\begin{subfigure}{0.45\textwidth}
  \centering
  \includegraphics[width=\textwidth]{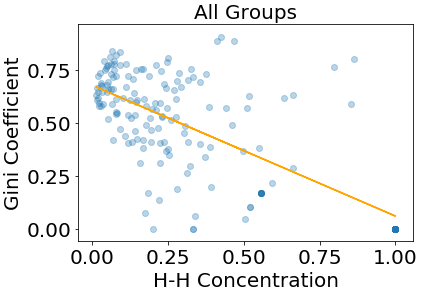}
  \caption{Scatter plot of H-H concentration and Gini coefficient. $r = -0.71$ ($p < 0.001$).}
  \label{figure:messages:scatter_hh_gini}
\end{subfigure}%
\hfill
\begin{subfigure}{0.45\textwidth}
  \centering
  \includegraphics[width=\textwidth]{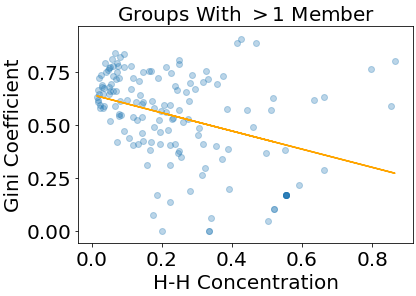}
  \caption{Scatter plot of H-H concentration and Gini coefficient, dropping 1-member groups. $r = -0.37$ ($p < 0.001$).}
  \label{figure:messages:scatter_hh_gini_drop}
\end{subfigure}
\caption{Measuring inequality within groups.}
\end{figure}

We show a scatter plot of H-H concentration against Gini coefficient for each group in figure \ref{figure:messages:scatter_hh_gini}; the same plot without one-person groups, which are perfectly concentrated yet perfectly equal, is shown in figure \ref{figure:messages:scatter_hh_gini_drop}.\footnote{There are 22 one-person groups, so they're not uncommon. These groups aren't as strange as they sound: many businesses restrict their business WhatsApp group so that only they can send messages (imagine a currency exchange operation sending out daily rates).}

It may seems surprising that H-H concentration and Gini/inequality are negatively correlated; we expect that highly concentrated groups are also highly unequal. But the perfectly-concentrated, perfectly-equal one-person groups give us a hint, in that concentration and equality measure different things. Namely, the concentration measure is centered on messages, while the equality measure is centered on users: many ``poor'' users (i.e., users who send one message) joining an active group doesn't affect its concentration, but makes it significantly more unequal.

\subsection{Correlates of Concentration and Inequality}

Amongst the group characteristics we found earlier, concentration is negatively correlated with group size, entropy, degree, and activity; group inequality is positively correlated with these factors. An OLS regression of concentration on these characteristics is shown in table \ref{table:messages:ols_group_concentration}. Regressing Gini coefficient on these characteristics yields nearly the same coefficients in the opposite direction (in particular, $0.0013$ on size and $0.1386$ on entropy, both $p < 0.001$), so we omit that table.

\begin{table}[h]
\centering
\caption{OLS regression of group concentration on group size, group entropy, group activity, and group degree.}
\label{table:messages:ols_group_concentration}
\begin{tabular}{p{0.25\textwidth} p{0.35\textwidth} p{0.10\textwidth} p{0.15\textwidth}}
\toprule
 & \textbf{Coefficient (Std. Err.)} & \textbf{$t$} & \textbf{P-Value} \\
\midrule

Intercept & $0.5460 (0.030)$ & $18.028$ & $0.000^{\ast}$ \\[0.2em]

Size & $-0.0015 (0.000)$ & $-5.021$ & $0.000^{\ast}$ \\[0.2em]

Entropy & $-0.1651 (0.033)$ & $-5.023$ & $0.000^{\ast}$ \\[0.2em]

Degree & $-0.0011 (0.001)$ & $-1.106$ & $0.270$ \\[0.2em]

Activity & $0.0000 (0.000)$ & $0.043$ & $0.966$ \\[0.2em]
\midrule
\multicolumn{4}{c}{$n = 174$\,\,(169 d.f.) \quad\quad $R^2 = 0.375$}\\
\bottomrule
\end{tabular}
\end{table}

\begin{figure}[h]
\centering
\begin{subfigure}{0.45\textwidth}
  \centering
  \includegraphics[width=\textwidth]{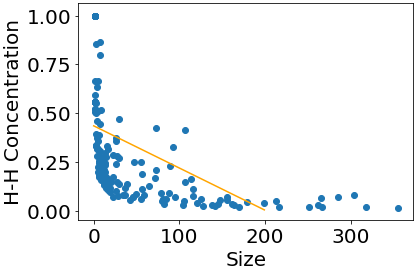}
  \caption{Scatter plot of group size and H-H concentration. $r = -0.50$ ($p < 0.001$).}
  \label{figure:messages:scatter_size_hh}
\end{subfigure}
\hfill
\begin{subfigure}{0.45\textwidth}
  \centering
  \includegraphics[width=\textwidth]{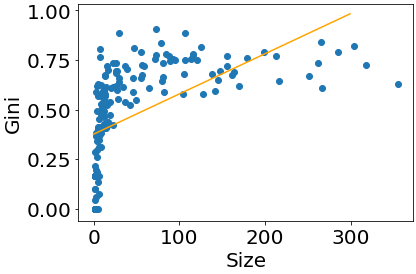}
  \caption{Scatter plot of group size and Gini coefficient. $r = 0.54$ ($p < 0.001$).}
  \label{figure:messages:scatter_size_gini}
\end{subfigure}
\caption{Larger groups are less concentrated but more unequal.}
\end{figure}

We do not rule out reverse causality here, but there are strong explanations for these results. In particular, larger groups are less concentrated, on average, since more members in a group likely means more active participants. Figure \ref{figure:messages:scatter_size_hh} plots H-H concentration against group size; groups of significant size are less concentrated (i.e., in groups with over 100 participants, there's less than a 25\% chance that two randomly selected messages come from the same user).

Larger groups are more unequal, likely because of natural bounds on how many users can truly participate in a WhatsApp conversation. As in many social contexts, WhatsApp groups probably include an ``inner circle,'' while most other members participate very little; the scatter plot in figure \ref{figure:messages:scatter_size_gini} shows that once groups are of size 50 or so, they become quite unequal.

We can frame the negative relationship between entropy and concentration around barrier-to-entry: groups that are more geographically diverse have higher barriers of entry to joining. Users are less likely to find cross-border groups, and if they do, they likely have stronger motivations for joining, whereas users may join news/entertainment groups (which are more likely within national boundaries) with abandon. Conditioned on having scaled the higher barrier to entry, we expect that users will be more active in geographically diverse groups, reducing their concentration.

\begin{figure}[h]
\centering
\begin{subfigure}{0.45\textwidth}
  \centering
  \includegraphics[width=\textwidth]{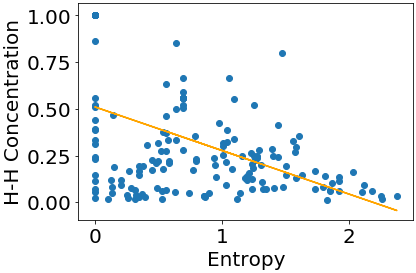}
\end{subfigure}%
\hfill
\begin{subfigure}{0.45\textwidth}
  \centering
  \includegraphics[width=\textwidth]{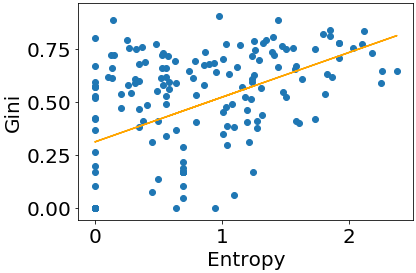}
\end{subfigure}
\caption{Geographically diverse groups are less concentrated and more unequal.}
\label{figure:messages:scatter_entropy_gini_hh}
\end{figure}

The positive relationship between entropy and inequality is more difficult to explain, but we might imagine that geographically diverse (high entropy) groups are more transaction-based. Indeed, imagine currency exchange businesses or transport businesses, or a group where already-crossed migrants (with CO phone numbers) answer questions from crossers (who have VZ phone numbers). Within these groups likely exists a stable element of users (i.e., the business owners, or group administrators), and a plethora of transient users who come and go---a structure which would produce a large Gini coefficient. Scatter plots with entropy are given in figure \ref{figure:messages:scatter_entropy_gini_hh}.

\section{Repeatedly Shared Content\label{ch:messages:repeated}}

In this section, we briefly investigate images, text, and videos that are repeatedly shared in our dataset. As we described in Section \ref{ch:datamethod:data-limitations}, our methodology is limited in only being able to identify content that is shared exactly or near-exactly: we cannot identify images that are slightly altered, or videos that are trimmed and then re-shared. Still, understanding what drives content to be re-shared should inform strategies for disseminating information over public WhatsApp groups, and for elucidating the structures (of users, and of hidden groups) that underlie this network.

\begin{figure}[h]
  \centering
  \includegraphics[width=0.45\textwidth]{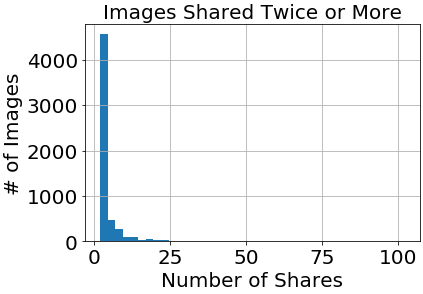}
  \caption{Histogram of number of shares per image.}
  \label{figure:messages:hist_image_nshare}
\end{figure}

We previously identified 38,455 messages with images, and out of these found 23,131 unique images being shared. 75.3\% of these images were only shared once, but in figure \ref{figure:messages:hist_image_nshare} we show the distribution of number of shares, for 5,704 images that were shared multiple times. Most (55.2\%) such images were shared only twice.

In total, 96.3\% of unique images were shared five or fewer times. To better understand what may drive re-sharing, we consider the 850 images that were shared more than five times. Of our original 174 groups, we find 66 groups where these images were first shared (in our dataset), and 96 groups where these images were ever shared. Comparing the set of ``first share'' groups to non-``first share'' groups, we find statistically significant differences in size, entropy, degree, activity, concentration, and inequality (there were no statistically signficant differences in proportion VZ, proportion CO, proportion US, proportion PE, proportion CL, proportion non-CO/non-VZ). Groups where popular images were first shared had average activity 118 messages/day, for example, while non-first share groups had an average of 12 messages/day.

This shouldn't surprise us, since analogously we'd expect early adopters of consumer electronics to be younger, wealthier, and more educated than laggards. Comparing ``any share'' groups to groups where these popular images were never shared yields similar results.

To more accurately understand this dynamic, for our 23,131 images we record certain characteristics of the group where they first appear. An OLS regression of the number of shares for each image on these characteristics is presented in table \ref{table:messages:ols_image_nshare}.

\begin{table}[h]
\centering
\caption{OLS regression of number of shares of each image, on size, entropy, degree, activity, concentration, and inequality of the group where each image first appeared.}
\label{table:messages:ols_image_nshare}
\begin{tabular}{p{0.25\textwidth} p{0.35\textwidth} p{0.10\textwidth} p{0.15\textwidth}}
\toprule
 & \textbf{Coefficient (Std. Err.)} & \textbf{$t$} & \textbf{P-Value} \\
\midrule

Intercept & $1.4880 (0.183)$ & $8.129$ & $0.000^{\ast}$ \\[0.2em]

Size & $-0.0033 (0.000)$ & $-11.599$ & $0.000^{\ast}$ \\[0.2em]

Entropy & $0.3285 (0.036)$ & $9.220$ & $0.000^{\ast}$ \\[0.2em]

Degree & $0.0070 (0.001)$ & $11.993$ & $0.000^{\ast}$ \\[0.2em]

Activity & $-0.0002 (0.000)$ & $-1.449$ & $0.147$ \\[0.2em]

H-H Concentration & $-1.5340 (0.215)$ & $-7.122$ & $0.000^{\ast}$ \\[0.2em]

Gini/Inequality & $0.5078 (0.295)$ & $1.719$ & $0.086$ \\[0.2em]
\midrule
\multicolumn{4}{c}{$n = 23131$\,\,(23124 d.f.) \quad\quad $R^2 = 0.021$}\\
\bottomrule
\end{tabular}
\end{table}

Make no mistake: these coefficients are all small, which comes from the vast majority of images that are only shared once (performing the regression with only the 5,704 images shared twice or more yields larger coefficients in the same directions, with $R^2 = 0.032$).

Still, there are important signals here, in that images shared in more geographically diverse groups are more likely to be re-shared, while images shared in more concentrated groups are less likely to be re-shared. Neither of these relationships is surprising: a nationally diverse member base means expanded conduits for an image, and concentration in a group means content is less likely to arise or spread organically. More unequal groups are linked to more re-shares, which might be because they enjoy a large pool of silent users who mainly consume content.

This is not to imply a causal direction: the reverse direction might be possible, in that images \textit{that are more likely to be re-shared} might simply be shared first in less concentrated, geographically diverse groups. But that would still mean that geographic diversity and low concentration are tied to information spread, in the direction we expect.

\begin{figure}[h]
  \centering
  \includegraphics[width=0.45\textwidth]{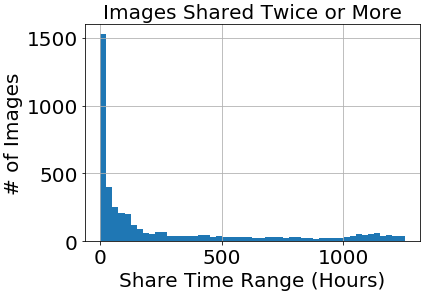}
  \caption{Histogram of the time range for which images (that are shared multiple times) are shared.}
  \label{figure:messages:hist_image_timerange}
\end{figure}

We can also examine the time range for which images are shared, calculated as between when they're first shared in our dataset and when they're last shared (0 for images shared only once); a histogram of these time ranges is shown in figure \ref{figure:messages:hist_image_timerange}.

Regressing the time range that images are shared for on the above variables yields coefficients in the same directions: coefficients on entropy, concentration, and Gini are 37.2854 hours, -160.1983 hours, and -13.5362 hours respectively. Uing only non-zero time ranges (i.e., images shared twice or more), these effects become even stronger, and are shown in table \ref{table:messages:ols_image_timerange_drop}.

\begin{table}[h]
\centering
\caption{OLS regression of time each image was shared for (hours) on size, entropy, degree, activity, concentration, and inequality of the group where images first appeared; we only include images that were shared multiple times.}
\label{table:messages:ols_image_timerange_drop}
\begin{tabular}{p{0.25\textwidth} p{0.35\textwidth} p{0.10\textwidth} p{0.15\textwidth}}
\toprule
 & \textbf{Coefficient (Std. Err.)} & \textbf{$t$} & \textbf{P-Value} \\
\midrule

Intercept & $655.5457 (54.962)$ & $11.927$ & $0.000^{\ast}$ \\[0.2em]

Size & $-0.3593 (0.082)$ & $-4.357$ & $0.000^{\ast}$ \\[0.2em]

Entropy & $120.0194 (10.093)$ & $11.891$ & $0.000^{\ast}$ \\[0.2em]

Degree & $1.3974 (0.154)$ & $9.068$ & $0.000^{\ast}$ \\[0.2em]

Activity & $-0.4606 (0.029)$ & $-15.975$ & $000^{\ast}$ \\[0.2em]

H-H Concentration & $-367.1378 (69.990)$ & $-5.246$ & $0.000^{\ast}$ \\[0.2em]

Gini/Inequality & $-528.2028 (87.003)$ & $-6.071$ & $0.000^{\ast}$ \\[0.2em]
\midrule
\multicolumn{4}{c}{$n = 4462$\,\,(4455 d.f.) \quad\quad $R^2 = 0.108$}\\
\bottomrule
\end{tabular}
\end{table}

It's not clear why the coefficient on Gini is negative (whereas for number of shares the coefficient on Gini was positive); we may try to explain this as that ``poor'' members (users who send few messages) in unequal groups consume and spread more content, but are less invested in re-sharing this content, leading to more shares but for shorter periods. To be clear, this is rather suspect reasoning, but the dynamics of these groups and their memberships are complicated.

Finally, we address the negative coefficients on activity and size: all things being equal, smaller and less active groups mean less crowding out of content and less competing for attention/re-shares.

\subsection{Repeatedly Shared Videos}

We consider analyzing repeatedly shared text, first eliminating any text with fewer than 20 characters, to eliminate trivial messages like ``hola'' and ``gracias.'' But of the remaining 61,159 unique texts, 93.6\% are shared only once, and 98.1\% are shared fewer than two times. This leaves a very small sample to work with, so we instead choose to move on to videos; later, we extensively discuss text-based misinformation---fake news and scams---within our groups, in Chapter \ref{ch:misinformation}.

Of 15,596 video messages, there were 13,733 unique videos in our dataset (we labels videos as identical if they have the same thumbnail and length); 89.6\% of these videos were shared only once. An additional 8.4\% of videos were shared exactly twice.

We proceed as we did for images, recording next to each unique video the properties of the group where it first appeared. Then, only including videos that were shared more than once (the number of videos shared only once is 9x the number of videos shared twice or more; for images, this multipler was 3x, so it makes sense now to limit our sample), we regress number of shares on the same aforementioned group characteristics. The only significant coefficients are a slight negative coefficient on entropy (-0.1765) and a slight positive coefficient on degree (0.0012).

When we regress the time range that videos (that were shared multiple times) were shared for, we obtain the estimates in table \ref{table:messages:ols_video_timerange_drop}. We again see the significant positive coefficient we've come to expect on entropy, and the significant negative coefficient we expect on concentration.

\begin{table}[h]
\centering
\caption{OLS regression of how long videos were shared for (hours), for videos that were shared multiple times, on size, entropy, degree, activity, concentration, and inequality of the group where the video first appeared.}
\label{table:messages:ols_video_timerange_drop}
\begin{tabular}{p{0.25\textwidth} p{0.35\textwidth} p{0.10\textwidth} p{0.15\textwidth}}
\toprule
 & \textbf{Coefficient (Std. Err.)} & \textbf{$t$} & \textbf{P-Value} \\
\midrule

Intercept & $73.5493 (42.349)$ & $1.737$ & $0.083$ \\[0.2em]

Size & $-0.1421 (0.065)$ & $-2.190$ & $0.029^{\ast}$ \\[0.2em]

Entropy & $31.4391 (8.263)$ & $3.805$ & $0.000^{\ast}$ \\[0.2em]

Degree & $0.4011 (0.096)$ & $4.189$ & $0.000^{\ast}$ \\[0.2em]

Activity & $-0.0725 (0.032)$ & $-2.235$ & $0.026^{\ast}$ \\[0.2em]

H-H Concentration & $-105.5255 (50.429)$ & $-2.093$ & $0.037^{\ast}$ \\[0.2em]

Gini/Inequality & $-35.2518 (63.275)$ & $-0.557$ & $0.578$ \\[0.2em]
\midrule
\multicolumn{4}{c}{$n = 1435$\,\,(1428 d.f.) \quad\quad $R^2 = 0.045$}\\
\bottomrule
\end{tabular}
\end{table}

\chapter{Reply Cascades\label{ch:replycascades}}

By design, WhatsApp provides scant information about how messages are received and regarded by other users. Unlike Facebook and Twitter, users cannot ``like'' or share\footnote{Users can forward messages to other groups, but WhatsApp doesn't link forwards of a message to the original message.} messages. In \cite{caetano-attention-2019}, Caetano et al. call WhatsApp ``an unsophisticated platform'' since ``information is shared through a very loosely structured interface.'' Yet we derive critically important information about messages from features like ``likes'' and ``shares''; they illustrate how others perceive certain content---or if they saw the content at all. When Instagram removed ``likes,'' for instance, it ``completely changed the platform'' \cite{independent-20191116}.

WhatsApp, however, allows users to send replies to other messages (and the original message is explicitly mentioned in the reply). Caetano et al. analyze replies in \cite{caetano-attention-2019}, tying their investigation to ``information overload'' within groups, arguing that messages compete for ``cognitive resources'' and form an ``economy of attention.''\footnote{The concept of an attention economy was first theorized by Herbert Simon \cite{ciampaglia-attention-2015}, who had said that, ``A wealth of information creates a poverty of attention'' \cite{simon-attention-1971}.} Per Caetano, since replies include a snippet of the original message, they bring renewed attention to that message, valuable in the scarce attention economy.

\begin{figure}[h]
\centering
\includegraphics[width=0.6\textwidth]{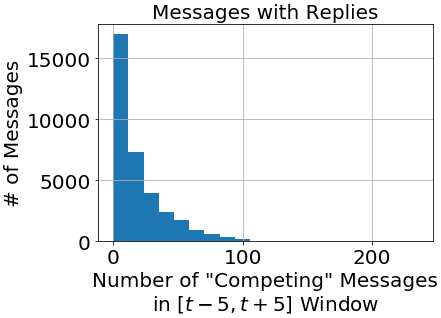}
\caption{In our groups, most messages aren't ``competing'' for attention.}
\label{figure:replycascades:competing_msg}
\end{figure}

Within our groups, however, this ``information overload'' effect doesn't seem as drastic. For message $m$ in group $i$ at time $t$ (minutes), we define ``competing'' messages as other messages in group $i$ in the time interval $[t - 5, t + 5]$ (minutes). Figure \ref{figure:replycascades:competing_msg} shows a histogram, for messages with replies, of how many messages they compete against. Of messages with replies, we find that 44\% are ``competing'' with fewer than 10 other messages, and that 64.9\% are competing with fewer than 20 other messages. But even 20 messages in a single 11-minute interval ($[t-5, t+5]$) isn't excessive, and shouldn't suggest to us any kind of ``information overload'' as Caetano et al. assert.

Nevertheless, replies are important because they do characterize, to some (limited) degree, user attention and interaction. Moreover, on WhatsApp, replying to messages and forwarding messages involve the same two-step process of highlighting the message and clicking on a new toolbar that appears. Whether users click on the ``reply'' button may give us the best approximation of whether users click on the ``forward'' button, which is important for official actors---governments and aid organizations---seeking to disseminate information over WhatsApp.

In the following sections, we discuss the prevalence of replies in our dataset, and pushback on directly understanding replies as popularity or engagement measures, as suggested in Caetano et al. \cite{caetano-attention-2019}. We move on to understanding what might drive replies within groups, including concentration and geographic entropy of each group. Later, we analyze the structural characteristics of reply cascades by interpreting them as graphs, constructing a ``virality'' metric similar to (but not the same as) that used by Caetano et al. in \cite{caetano-attention-2019}. We show how this measure of structural virality better accounts for different content types, and is still correlated to various group characteristics.

\section{Overview of Replies\label{ch:replycascades:replies}}

Out of 171,634 messages in our dataset, 49,212 messages (28.7\%) were replies. We only managed to fully trace 43,912 (89.2\%) of these replies to their source, for various reasons: some replies were to messages sent before we joined the group (though we gave a 24-hour buffer after we joined groups before recording messages), some original messages were deleted before we could capture them, and so on.

Of the 171,634 messages in our dataset, 34,444 (20.1\%) were replied to. In Figure \ref{figure:replycascades:hist_replies_n} below, we plot a histogram of the distribution of how many replies each of these messages received, but an overwhelming majority---29,478 messages (85.6\%)---received fewer than five replies; 16,959 messages (49.2\%) received only one reply, and 6,944 (20.2\%) received exactly two replies.

\begin{figure}[h]
  \centering
  \includegraphics[width=0.45\textwidth]{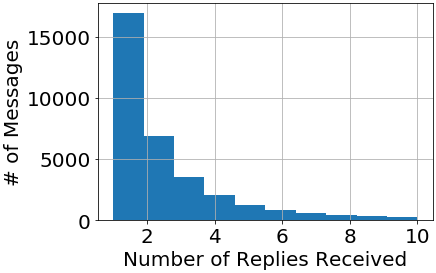}
  \caption{Histogram of number of replies received. We cut the graph off at 10, but some messages received many more than 10 replies.}
  \label{figure:replycascades:hist_replies_n}
\end{figure}

In Table \ref{table:replycascades:replies_breakdown_by_type} below, we break down messages by content type, and calculate what proportion of each type received replies.

\begin{table}[h]
\centering
\caption{Breakdown by content type, showing proportion of each type that receives replies. Note that number of messages don't add up to anything meaningful, since messages can contain more than one content type (or not contain any).}
\label{table:replycascades:replies_breakdown_by_type}
\begin{tabular}{p{0.15\textwidth} p{0.20\textwidth} p{0.20\textwidth} p{0.20\textwidth}}
\toprule
\textbf{Content} & \textbf{\# Messages} & \textbf{\# with Replies} & \textbf{\% with Replies} \\
\midrule

Text & 101,414 & 25,191 & 24.8\% \\[0.2em]

Image & 38,455 & 6,768 & 17.6\% \\[0.2em]

Video & 15,596 & 1,872 & 12.0\% \\[0.2em]

Emojis & 23,886 & 4,305 & 14.9\% \\[0.2em]

Audio & 8,918 & 2,053 & 23.0\% \\[0.2em]

Forwarded & 26,168 & 1,354 & 5.2\% \\[0.2em]

\bottomrule
\end{tabular}
\end{table}

By far, text messages are most likely to receive replies (24.8\% of text messages receive replies, compared to 17.6\% of images and 12.0\% of videos); this should \textit{instantly} give us pause in how we understand and analyze replies. On other media platforms like Facebook, images and videos are by far the most popular content, and also have the highest levels of engagement; one report estimates that the average video post on Facebook reaches 12.05\% of page audience, the average image reaches 11.63\% of page audience, and text updates only reach 4.56\% \cite{buzzsumo-20170823}. The newspaper The Guardian sees users engage most with text articles on its own website, but video content on social media platforms \cite{buzzsumo-20170823}.

Though we don't have access to true engagement/view data from WhatsApp, there's no reason to not expect this trend to also hold in WhatsApp groups. So if images and videos are actually the most popular and engaging content, what does it mean for text messages to receive replies at much higher rates? \textit{Across categories, replies are not a good measurement of a message's popularity or engagement.} For whatever reason, users may find it unnatural to reply to photos (akin to quoting a photo, we might say); alternatively, it's possible that images/videos shared in our groups are forwarded from other sources (as opposed to text more likely being original content), so users are less likely to respond to such forwarded content.\footnote{By forwarded, we don't only mean forwarded through WhatsApp: only 14.2\% of image content is ``forwarded'' through WhatsApp (from one conversation to the other, using the forward feature in WhatsApp). Many images, for example, are downloaded by users and re-uploaded, though we have no way of determining this.} This latter point is shown in our table; forwarded content, by far, is less likely to receive replies, with only 5.2\% of forwarded messages receiving replies.

What does this mean for us? First and most importantly, that across content categories, we cannot use replies as an accurate metric for popularity, engagement, etc. (within categories, this metric is significantly less suspect). But this also means that when comparing replies across groups, we must either restrict or normalize content type, since a highly-interactive group where only videos are shared could result in many fewer replies than an inactive group where only texts are shared. Finally, this means that the understanding in \cite{caetano-attention-2019} of replies as attention is outright misleading; they had written that, ```We say that...messages in the cascade \textit{caught the attention} of a group member, motivating her to interact.''

\subsubsection{Comparing Messages by Sender Country}

With the discussion above, before we compare messages sent by Colombian numbers vs. those sent by Venezuelan numbers, we first compare content type distributions. We found that of messages sent by Colombian numbers and Venezuelan numbers, there were roughly equal proportions of text messages (62.6\% of messages by Colombian numbers, and 60.3\% of messages by Venezuelan numbers), images (20.1\% and 24.2\% respectively), and video messages (8.2\% and 7.3\% respectively).

17.9\% of messages from Venezuelan numbers received replies, while 20.7\% of messages from Colombian numbers received replies. This small difference is somewhat accounted for by the subtle differences in content type distribution---Venezuelan users send more images and fewer text messages.

\subsubsection{Comparing Messages by Time of Day}

\begin{figure}[h]
  \centering
  \includegraphics[width=0.6\textwidth]{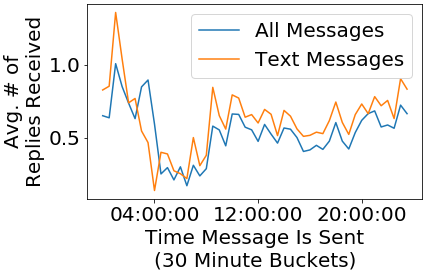}
  \caption{Messages in the wee morning hours are barely replied to; starting in mid-morning, messages start to receive more replies, and this pattern rises through the night before peaking at 12-1 AM.}
  \label{figure:replycascades:time_of_day}
\end{figure}

Figure \ref{figure:replycascades:time_of_day} shows the average number of replies to each message, by time of day. This mirrors what we expect, though we might be surprised to see messages receiving many more replies in late-night hours. This might be the result of more active/serious users being on at that time, or conversations turning more personal, and so on.

\begin{figure}[h]
  \centering
  \includegraphics[width=0.6\textwidth]{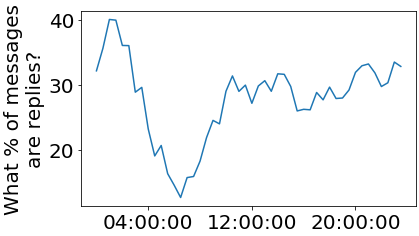}
  \caption{Few of the messages sent at 4-5 AM are replies, but in mid-day through evening nearly 30\% of messages are replies, peaking to 40\% of messages at 12-1 AM.}
  \label{figure:replycascades:time_of_day_areReplies}
\end{figure}

Graphing the proportion of all messages that \textit{are} replies in figure \ref{figure:replycascades:time_of_day_areReplies}, these explanations seem plausible.

\subsubsection{Replies Within Groups}

\begin{figure}[h]
  \centering
  \includegraphics[width=0.45\textwidth]{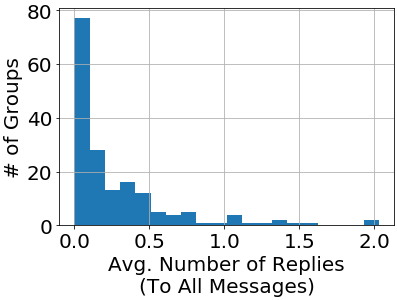}
  \caption{Histogram of average number of replies to messages within each group.}
  \label{figure:replycascades:hist_group_replies_n}
\end{figure}

Given the patterns we've seen with repeatedly shared images and video (specifically, that number and timespan of re-shares is positively correlated with geographic diversity, and negatively correlated with group concentration), we might wonder if similar patterns of interaction take place with replies. For each group, we calculate the average number of replies to all messages; this is plotted in the histogram in figure \ref{figure:replycascades:hist_group_replies_n} (in 50 groups, no replies are recorded).

 Before proceeding, we re-emphasize that comparing average number of replies across groups is suspect, since different content types are replied to at different rates; this was the discussion in Section \ref{ch:replycascades:replies}. Later, in Section \ref{ch:replycascades:virality}, we redo the following analysis using an alternative measure robust to content types. Still, it's interesting to compare the average number of replies across groups, as is.

The average number of replies to messages in each group is correlated with group size, entropy, activity, degree, concentration, and Gini. Previously, we saw that the rate at which (and timespan for which) images are re-shared is linked to the entropy, concentration, and inequality (Gini) of the group where they're first shared. With this, we decide to regress the average number of replies within groups on these three group characteristics; the results are shown in table \ref{table:replycascades:ols_group_replies_n}.

\begin{table}[h]
\centering
\caption{OLS regression of average number of replies for messages within a group, on entropy, concentration, and inequality of the group. When regressing only on groups with replies ($n = 124$), the effects are stronger and remain statistically significant.}
\label{table:replycascades:ols_group_replies_n}
\begin{tabular}{p{0.25\textwidth} p{0.35\textwidth} p{0.10\textwidth} p{0.15\textwidth}}
\toprule
 & \textbf{Coefficient (Std. Err.)} & \textbf{$t$} & \textbf{P-Value} \\
\midrule

Intercept & $0.2548 (0.100)$ & $2.551$ & $0.012^{\ast}$ \\[0.2em]

Entropy & $0.1328 (0.046)$ & $2.876$ & $0.005^{\ast}$ \\[0.2em]

H-H Concentration & $-0.3739 (0.117)$ & $-3.195$ & $0.002^{\ast}$ \\[0.2em]

Gini/Inequality & $0.0815 (0.136)$ & $0.601$ & $0.549$ \\[0.2em]
\midrule
\multicolumn{4}{c}{$n = 174$\,\,(170 d.f.) \quad\quad $R^2 = 0.253$}\\
\bottomrule
\end{tabular}
\end{table}

\begin{figure}[h]
\centering
\begin{subfigure}{0.45\textwidth}
  \centering
  \includegraphics[width=\textwidth]{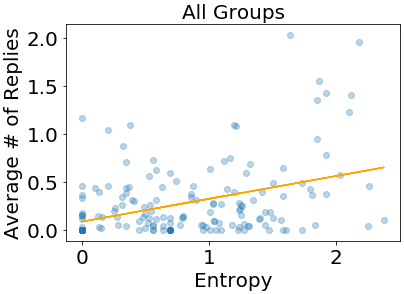}
\end{subfigure}%
\hfill
\begin{subfigure}{0.45\textwidth}
  \centering
  \includegraphics[width=\textwidth]{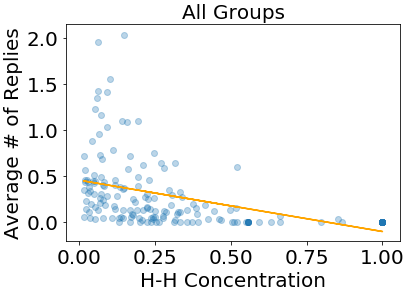}
\end{subfigure}
\caption{There are more replies in more geographically diverse groups, and fewer replies in highly concentrated groups.}
\label{figure:replycascades:scatter_replies_n}
\end{figure}

We see the familiar pattern: the average of number of replies is higher in more geographically diverse groups, and lower in more concentrated groups. Restating the hypotheses discussed in Chapter \ref{ch:messages:repeated}, concentrated groups are less fertile ground for organic interaction with and spread of content. More replies in cross-border groups could again arise from the purpose-directed nature of these groups: imagine currency exchange operations, or groups where potential migrants ask questions of Venezuelans already settled in Colombia. Scatter plots of entropy and concentration with average number of replies are shown in figure \ref{figure:replycascades:scatter_replies_n}.

\section{Construction of Reply Graphs}

We now proceed to investigate the structural characteristics of replies and reply cascades (chains).

Define a reply cascade by all messages that terminate their reply chains at the same root. Similar to \cite{caetano-attention-2019},\footnote{In \cite{caetano-attention-2019}, Caetano et al. (incorrectly) construct directed acyclic graphs, where characteristics like average distance are not well-defined since paths between nodes may not exist.} we construct for each reply cascade an undirected graph where there is an edge between messages $X, Y$ if $X$ is a reply to $Y$ or vice versa.

Within each cascade/graph, we calculate both the average shortest-path distance between nodes, and the maximum shortest-path distance between any pair of nodes; we obtain both with the canonical technique of breadth-first search from each node in a reply cascade. The average distance between nodes represents, to some effect, the ``virality'' of a message, where virality is not only a measure of some content's popularity, but also how much of that popularity was driven by peer-to-peer sharing \cite{goel-strucvir-2015}. Contrast this to ``broadcast'' content, whose sharing is less peer-to-peer than driven by some central source. Bad Superbowl ads still reach many people, but they never go viral.

If we represent replies as graphs, high virality implies a certain decentralization, with a larger average distance between nodes. In \cite{goel-strucvir-2015}, Goel et al. define \textit{structural virality} as exactly this measure. Specifically, in graph $G$ with $n$ nodes, structural virality $v(G) = \frac{1}{n(n-1)} \sum_{i, j} d_{ij}$ where $d_{ij}$ is the length of the shortest path between nodes $i$ and $j$.

\begin{figure}[h]
\centering
\begin{subfigure}{0.5\textwidth}
  \centering
  \includegraphics[width=\textwidth]{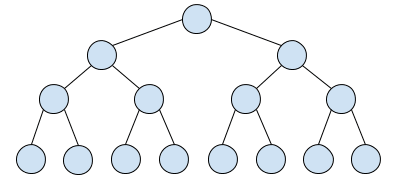}
  \caption{A (perfect) binary tree.}
\end{subfigure}%
\hfill
\begin{subfigure}{0.4\textwidth}
  \centering
  \includegraphics[width=\textwidth]{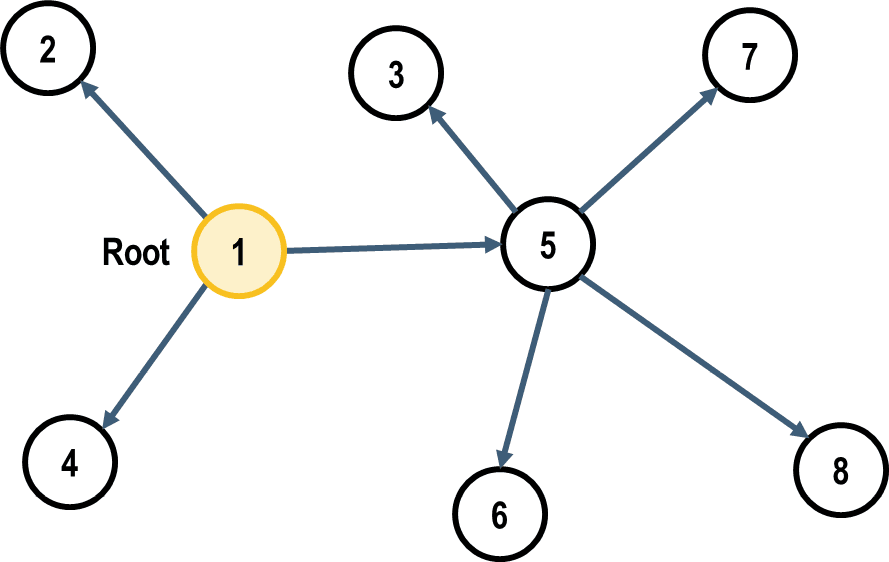}
  \caption{A ``broadcast'' graph.}
\end{subfigure}
\caption{The binary tree has a much higher average distance between nodes (virality) than the broadcast graph.}
\label{figure:replycascades:binary_vs_broad}
\end{figure}

\pagebreak
Consider, for example, the two graphs in figure \ref{figure:replycascades:binary_vs_broad}. In the binary tree, each user receives a message and shares it with to two others, while in the broadcast graph, most of the sharing is driven by two central users; most people who receive the message do not later go on to share it. The structural virality of the first graph is much higher than the structural virality of the second graph, since the second graph is highly centralized, so all nodes are close to some central nodes (and thus, to each other).

\begin{figure}[h]
\centering
\includegraphics[width=0.6\textwidth]{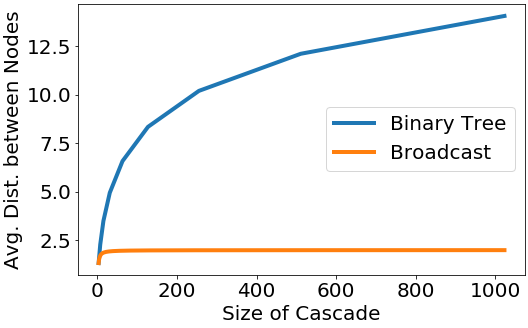}
\caption{The virality, per Goel et al., of perfect binary trees and perfect broadcast graphs. The virality of the broadcast graph is bound by 2 (specifically, its virality is $\frac{1 \times (n-1) + (n-1) \times [1 + 2(n-2)]}{n(n-1)} = \frac{2(n-1)^2}{n(n-1)} = \frac{2(n-1)}{n} \rightarrow 2$), no matter its size, while the virality of the binary tree is unbounded.}
\label{figure:replycascades:binary_vs_broadcast}
\end{figure}

To better illustrate virality, we generate perfect ``broadcast graphs'' (one central node connected to all other nodes, which are only connected to the central node) and perfect binary trees of various sizes, and plot their viralities in figure \ref{figure:replycascades:binary_vs_broadcast}.

Note that Goel et al., as well as Caetano et al. from UFMG, compute structural virality as the average distance over all pairs of distinct nodes. We argue for instead using a different measure of structural virality, where we define $v(G) = \frac{1}{n^2} \sum_{i, j} d_{ij}$. Instead of averaging over all distinct pairs of nodes, we simply average over all pairs of nodes (i.e., include distances from nodes to themself). Effectively, we scale down the Goel et al. measure by $\frac{n-1}{n}$. With large $n$, this difference is clearly insignificant, but we argue for its importance on small graphs.

Consider 2-node, 3-node, and 4-node chains. Using our virality measure, where we compute over all pairs of nodes (instead of all distinct pairs of nodes), the virality of a 2-node chain is $0.5$ (since each node is connected to itself with distance 0 and the other node with distance 1), while the Goel et al. structural virality for such a graph is 1. Now consider a 4-node chain $\circ - \circ - \circ - \circ$. The distance matrix is given by $\begingroup
\renewcommand*{\arraystretch}{0.5} \begin{bmatrix} 0 & 1 & 2 & 3 \\ 1 & 0 & 1 & 2 \\ 2 & 1 & 0 & 1 \\ 3 & 2 & 1 & 0 \end{bmatrix} \endgroup$, so the Goel et al. structural virality yields $\frac{20}{12} \approx 1.67$. On the other hand, our measure of structural virality is $\frac{20}{16} = 1.25$. These viralities, as well as those of a 3-node chain, are shown in table \ref{table:replycascades:viralitymeasure}.

\begin{table}[htbp]
\centering
\caption[Virality Measure]{Our measure of structural virality vs. Goel et al.}
\label{table:replycascades:viralitymeasure}
\begin{tabular}{p{0.2\textwidth} p{0.3\textwidth} p{0.4\textwidth}}
\toprule
\textbf{Graph} & \textbf{Our Virality $\frac{1}{n^2} \sum d_{ij}$} & \textbf{Goel et al. Virality $\frac{1}{n(n-1)} \sum d_{ij}$} \\
\midrule
$\circ - \circ$
& $\frac{1}{2} = 0.5$ & 1 \\[0.2em]

$\circ - \circ - \circ$
& $\frac{8}{9} \approx 0.89$ & $1 \frac{1}{3} \approx 1.33$ \\[0.2em]

$\circ - \circ - \circ - \circ$
& $1 \frac{1}{4} = 1.25$ & $1 \frac{2}{3} \approx 1.67$ \\[0.2em]
\bottomrule
\end{tabular}
\end{table}

A 2-node chain---a message with one reply---is \textit{much} less viral than a 4-node chain, where a message is replied to, its reply also replied to, and that second reply \textit{also} replied to. Yet the Goel et al. measure puts the virality of a 2-node chain at 60\% of the virality of a 4-node chain, while our measure puts it at 40\% of the 4-node chain's virality.

Our virality measure makes more sense when considering the 3-node chain as well. A 2-node chain (a message with one reply) is substantially less viral than a 3-node chain; the Goel et al. measure puts a 2-node chain at 75\% of the virality of a 3-node chain, while our measure puts it at around 56\%.

In short, using our virality measure instead of that by Goel et al. allows us to much better compare viralities when we include 2-node chains (i.e., messages with one reply). Since most messages with replies in our dataset (and likely in general) \textit{are} 2-node chains (given the power-law distribution of number of replies), our measure allows us to more robustly investigate virality.

\begin{figure}[h]
\centering
\includegraphics[width=0.75\textwidth]{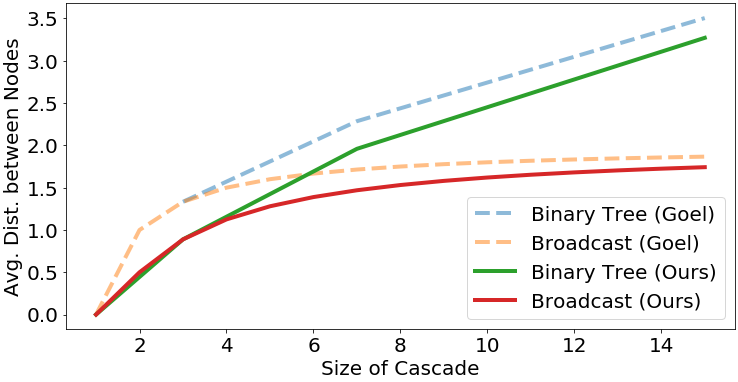}
\caption{Our virality converges to the Goel et al. virality, but starts out less steeply. We argue that this makes the most sense, since a message with one reply shouldn't be considered as very ``viral.''}
\label{figure:replycascades:chang_vs_goel}
\end{figure}

For good measure, in figure \ref{figure:replycascades:chang_vs_goel}, we plot our measure of structural virality vs. the Goel et al. measure, for the aforementioned perfect binary trees and perfect broadcast graphs. Both measures quickly converge as $n$ increases. But our measure increases less steeply for smaller cascades, which we argues makes sense, since a 2-node cascade (a message with one reply) really isn't that viral.

\section{Virality\label{ch:replycascades:virality}}

\begin{figure}[h]
\centering
\begin{subfigure}{0.45\textwidth}
  \centering
  \includegraphics[width=\textwidth]{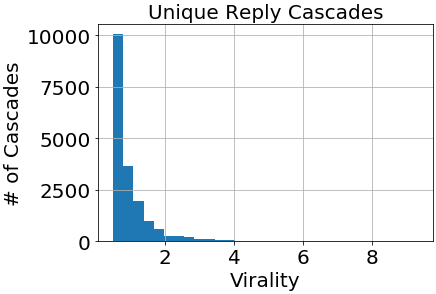}
  \caption{Histogram of virality across reply cascades.}
  \label{figure:replycascades:hist_virality}
\end{subfigure}%
\hfill
\begin{subfigure}{0.45\textwidth}
  \centering
  \includegraphics[width=\textwidth]{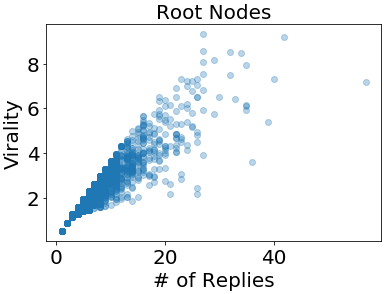}
  \caption{Scatter plot of how many replies each root is connected to, compared to its virality.}
  \label{figure:replycascades:scatter_root_repliesn_virality}
\end{subfigure}
\caption{Virality in our dataset.}
\end{figure}

Figure \ref{figure:replycascades:hist_virality} shows a histogram of virality for each reply cascade (only counting each cascade once, regardless of how many messages it includes), and the scatter plot in figure \ref{figure:replycascades:scatter_root_repliesn_virality} plots the virality of each root node against how many replies it receives.

For messages within reply cascades (including the root), we set their virality as the virality of the reply cascade they're in. The most important motivation for this measure is that it allows us to compare reply cascades across content types, since we no longer focus on the prevalence of reply cascades, but on properties \textit{within} reply cascades.

In particular, we previously saw that text is replied to at much higher rates than images, even though we know images to typically be more ``viral.'' Now, the average virality across all images in reply cascades is 1.71, which is \textit{14\% higher} than the average virality of text in reply cascades, which is 1.50. Similarly, we saw that messages from Venezuelan users received replies at a lower rate than messages from Colombian users. When examining messages from Venezuelans that are part of reply cascades, compared to messages from Colombians in reply cascades, it turns out that Venezuelans' messages are more viral (1.53 vs. 1.45, $p < 0.001$).

\subsubsection{Diameter}

\begin{figure}[h]
  \centering
  \includegraphics[width=0.45\textwidth]{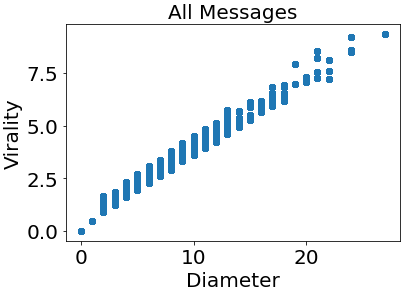}
  \caption{Virality (average distance between nodes in a reply cascade) and diameter (maximum distance) are nearly the same measure.}
  \label{figure:replycascades:scatter_diameter_virality}
\end{figure}

Instead of computing virality, the average distance between nodes in a reply cascade, we might consider diameter, the maximum distance between nodes. Letting the diameter of each message in a reply cascade being the diameter of the reply cascade, figure \ref{figure:replycascades:scatter_diameter_virality} reveals that these are nearly the same measure.

\pagebreak
\subsection{Virality Across Groups}

That virality can be compared across content types means we can also compare virality across groups. We define each group's virality as the average virality of all messages in that group that are part of reply cascades;\footnote{If we defined virality as the average virality of all messages (including messages that are not part of reply cascades), that quickly yields a near-identical measure to the average number of replies measure we used in Section \ref{ch:replycascades:replies}.} for groups without any replies, we imputed their virality as 0.

\begin{figure}[h]
\centering
\begin{subfigure}{0.45\textwidth}
  \centering
  \includegraphics[width=\textwidth]{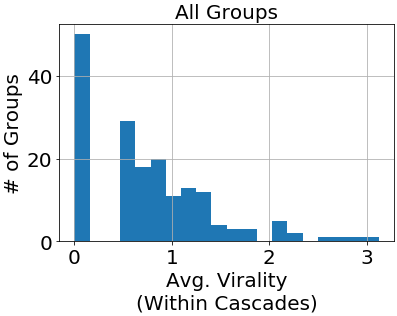}
  \caption{Histogram of average virality (within reply cascades) across groups.}
  \label{figure:replycascades:hist_group_virality}
\end{subfigure}%
\hfill
\begin{subfigure}{0.45\textwidth}
  \centering
  \includegraphics[width=\textwidth]{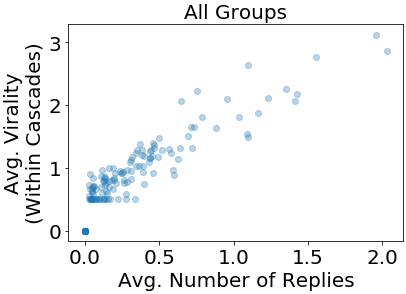}
  \caption{Scatter plot of virality in each group, which can be more accurately generalized across content type, with average number of replies in each group.}
  \label{figure:replycascades:scatter_group_repliesn_virality}
\end{subfigure}
\caption{Virality in groups.}
\end{figure}

Figure \ref{figure:replycascades:hist_group_virality} shows a histogram of virality across groups, and the scatter plot in figure \ref{figure:replycascades:scatter_group_repliesn_virality} plots virality against the average number of replies to each message across groups. The two are closely correlated (Pearson $r = 0.90$), though from here we default to using virality, since, as we mentioned, it allows us to better compare reply cascades across content types (and, consequently, groups with different content types).

Before, we had found that the average number of replies in each group was linked to the group's entropy, concentration, and inequality. Here, we perform the same regressions with virality as the dependent variable, yieling the results in table \ref{table:replycascades:ols_group_virality}. A separate regression using only groups with replies (recall that 50 groups have no replies, likely either they're highly inactive or dedicated business channels where only group administrators can send messages) is shown in table \ref{table:replycascades:ols_group_virality_drop}.

\begin{table}[h]
\centering
\caption{OLS regression of virality in a group, on entropy, concentration, and inequality of the group.}
\label{table:replycascades:ols_group_virality}
\begin{tabular}{p{0.25\textwidth} p{0.35\textwidth} p{0.10\textwidth} p{0.15\textwidth}}
\toprule
 & \textbf{Coefficient (Std. Err.)} & \textbf{$t$} & \textbf{P-Value} \\
\midrule

Intercept & $0.5648 (0.147)$ & $3.830$ & $0.000^{\ast}$ \\[0.2em]

Entropy & $0.1381 (0.068)$ & $2.027$ & $0.044^{\ast}$ \\[0.2em]

H-H Concentration & $-0.7948 (0.173)$ & $-4.601$ & $0.000^{\ast}$ \\[0.2em]

Gini/Inequality & $0.7055 (0.200)$ & $3.526$ & $0.001^{\ast}$ \\[0.2em]
\midrule
\multicolumn{4}{c}{$n = 174$\,\,(170 d.f.) \quad\quad $R^2 = 0.474$}\\
\bottomrule
\end{tabular}
\end{table}

We see patterns we're all too familiar with by now: concentration in groups means that reply cascades are less viral (which is completely unsurprising, since messages are more centralized); entropy in groups is linked to more virality and decentralization (we can imagine, for example, a reply cascade splitting into separate chains amongst Venezuelan and Colombian members in the group). Inequality in groups is linked to more virality, which might come from ``poor'' group members (users who send few messages) breaking off into separate discussion.

\begin{table}[h]
\centering
\caption{OLS regression of virality in a group, on entropy, concentration, and inequality of the group; we only include groups with replies.}
\label{table:replycascades:ols_group_virality_drop}
\begin{tabular}{p{0.25\textwidth} p{0.35\textwidth} p{0.10\textwidth} p{0.15\textwidth}}
\toprule
 & \textbf{Coefficient (Std. Err.)} & \textbf{$t$} & \textbf{P-Value} \\
\midrule

Intercept & $0.7623 (0.183)$ & $4.177$ & $0.000^{\ast}$ \\[0.2em]

Entropy & $0.2393 (0.073)$ & $3.257$ & $0.001^{\ast}$ \\[0.2em]

H-H Concentration & $-1.1033 (0.311)$ & $-3.543$ & $0.001^{\ast}$ \\[0.2em]

Gini/Inequality & $0.4438 (0.262)$ & $1.694$ & $0.093$ \\[0.2em]
\midrule
\multicolumn{4}{c}{$n = 124$\,\,(120 d.f.) \quad\quad $R^2 = 0.246$}\\
\bottomrule
\end{tabular}
\end{table}

To re-iterate, when using virality as a measure (and especially dropping groups without replies), our analysis no longer involves the prevalence of reply cascades, but simply the dynamics \textit{within} reply cascades. That these patterns retain significance means that even controlling for the fact that there are more replies in unconcentrated and geographically diverse groups, replies in those groups are \textit{still} more viral.

\subsection{Temporal Characteristics of Reply Cascades}

Virality is a \textit{structural} characteristic of reply cascades; in \cite{caetano-attention-2019}, Caetano et al. also focus on cascade duration (defined as the time between the message time of the root node, and when the last reply is sent), which they term the ``main \textit{temporal} attribute'' (emphasis ours) of reply cascades. Amongst other findings, Caetano et al. report that ``political cascades last longer than non-political ones...A possible explanation is that political cascades stir more debate among the participants of the group'' \cite{caetano-attention-2019}.

Clearly, Caetano et al. associate cascade duration with the amount of participation in each cascade; 12-hour reply cascades involve much more back-and-forth discussion than 6-hour reply cascades. This might be true if both groups were equally active, but that's not how public WhatsApp groups work (unless Brazilian groups are somehow staggeringly different from Colombian groups). Some groups are highly active, but many aren't, making cascade duration a blatantly flawed and deceptive measure. Just imagine a highly-inactive group where someone replies to messages a few days later, on average, with no other messages in between; ``cascades'' in that group last much longer than cascades in a higly-active group with many people participating (and many more messages being sent). What could anyone possibly say about cascade duration given that these circumstances do exist?

\begin{figure}[h]
  \centering
  \includegraphics[width=\textwidth]{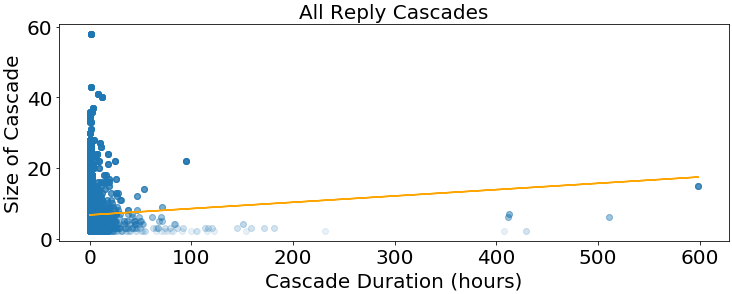}
  \includegraphics[width=\textwidth]{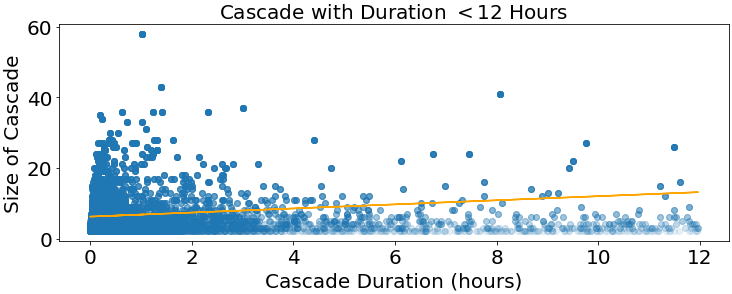}
  \caption{Scatter plots of the size of reply cascades against their duration. The top plot includes all cascades; the bottom plot only includes cascades with duration less than 12 hours.}
  \label{figure:replycascades:temporal_size}
\end{figure}

\pagebreak
In figure \ref{figure:replycascades:temporal_size}, we plot the size of reply cascades against their duration, for all cascades and for cascades lasting less than 12 hours. Most cascades of any significant size have short durations---\textit{which is what we'd expect}, since viral/popular cascades likely take place in highly active groups where attention soon turns to new topics. The slope of the best fit line in the top picture is 0.018; in the bottom picture, it's 0.579. Even 0.579 is miniscule, telling us that for each additional hour of a cascade, there are 0.579 more messages in that cascade, on average. So cascade duration and cascade size aren't even moderately correlated (Pearson $r = 0.04$ between duration and size across all cascades; Pearson $r = 0.15$ for $<12$ hour cascades).

\begin{figure}[h]
  \centering
  \includegraphics[width=\textwidth]{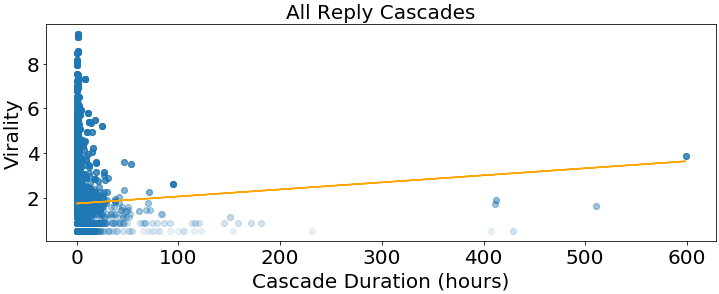}
  \includegraphics[width=\textwidth]{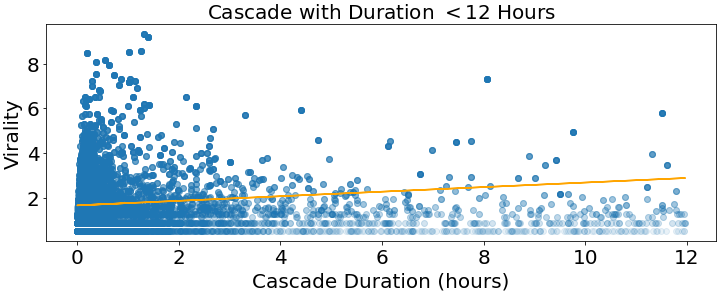}
  \caption{Scatter plots of the virality of reply cascades against their duration. The top plot includes all cascades; the bottom plot only includes cascades with duration less than 12 hours.}
  \label{figure:replycascades:temporal_virality}
\end{figure}

Figure \ref{figure:replycascades:temporal_virality} shows scatter plots of the virality of reply cascades against their duration. Longer cascades are very, very slightly more viral ($r = 0.03$ across all cascades; $r = 0.10$ in cascades lasting less than 12 hours), but in general, cascade duration says little about how active/involving/popular cascades are.

With this discussion, we completely discard temporal characteristics of reply cascades; structural characteristics like virality are clearly much more useful measures of the activity that drives reply cascades.

\chapter{Misinformation: Fake News and Scams\label{ch:misinformation}}

During our fieldwork, we found that most migrants with knowledge of public WhatsApp groups also had significant concerns about misinformation within groups. Migrants were most concerned with employment-related scams, like job offers where employers would skip town before payday, or unscrupulous ``domestic'' positions for women that would turn out to be sex work (or worse). A second major area of concern involved fraudulent service providers, like exchange/remittance services that steal customers' funds.

While these kinds of criminality are certainly present in our network, they're extremely difficult to investigate with the data we have. Nothing necessarily distinguishes legitimate employment offers from deceitful ones, and the same advertisements are used by both legitimate service providers and fraudulent ones.

But more obvious internet-based scams are also present in these groups, and they're easily identifable. Consider the following message, for example, which offers free food cards:\footnote{Here, ``food card'' refers to the cash transfers implemented by (e.g.) the World Food Programme, which are distributed as prepaid debit cards.}

\singlespacing\begin{lstlisting}
BONOS NUEVA TARJETA ALIMENTARIA 2020-2025:
-Madres solteras
-Madre trabajadora
-Joven estudiante
-Vives con tus padres
-Madre separada
-Madres extranjeras
-Estos y Mas...
TARJETA ALIMENTARIA DE:
 $6550.65 (BONO).
NUEVA TARJETA ALIMENTARIA EN:
  https://tarjetas.global/solicitud-tarjeta-alimentaria-aquii/
Clic en el link encima para enviar su solicitud y preparar su entrevista
INICIO IMEDIATO!
Fonte: G1 - O portal de noticias del Globo
\end{lstlisting}\doublespacing

In the 53 days we collected messages, this scam was shared 34 times, in 15 unique groups by 30 different individuals (these numbers don't even include the numerous variants that are also shared). Or consider the following scam that purports to offer free mobile data due to the coronavirus quarantine:

\singlespacing\begin{lstlisting}
100 GB de datos de Internet sin ninguna recarga Por Motivo de CUARENTENA (CORONA VIRUS)
Obtenga 100 GB de datos de Internet gratis en cualquier red movil durante 60 dias.
Consiguelo ahora AQUI
https://appcutt.link/netfree
\end{lstlisting}\doublespacing

While internet-based scams like these are certainly different from offline scams (like unscrupulous employment offers or fraudulent service providers), and less a cause of concern for migrants, we argue that online scams are meaningfully related to offline scams. Groups with more internet-based scams likely contain more offline scams; users who fall prey to internet-based scams will more likely be victims of serious employment-related scams. To be clear, this is exactly broken windows theory, but online and offline scams are probably more connected than turnstile jumping and homicide.

These online scams might be easily identifiable to our eyes (or to an algorithm), but many users fall prey to them. In section \ref{ch:misinformation:scams}, we show that such (easily identifiable) scams are nonetheless widespread across groups and users. Either there are a lot of dedicated scammers, or many users have had accounts taken over by scammers, or both; in any case, the concerted effort to spread easily-identifiable internet scams in these groups makes clear that they \textit{do} trick users.

One Good Samaritan group administrator puts it best. In an English translation of their message:

\singlespacing\begin{lstlisting}
* ATTENTION GROUP To help us from home
THERE are the links that circulate in Whatssap groups and other networks
There are no free megabytes.
No free bonuses.
No national or international company is looking for employees with astronomical salaries.
Coca Cola is NOT giving away fridges.
Pepsi is NOT giving away bonuses.
Nike is NOT giving away shoes
Zara is NOT giving away bonuses.
Netflix is NOT giving away accounts.
WhatsApp is free, you do NOT need to forward anything.
It will not change the color of WhatsApp nor will they verify your account.
No one is giving away wheelchairs or medicines.
What will happen to you is that when you click on those links that are viruses, they will hack you and lose your data, they will steal your social media accounts, impersonate You to scam with dollars, your cell phone collapse and many other things.
Does not exist!
Talk to your friends and family so they don't fall into the trap.
\end{lstlisting}\doublespacing

Some users even accuse the Venezuelan and Cuban intelligence services of being behind common scams (we share an English translation):

\singlespacing\begin{lstlisting}
URGENT:
Don't even think about opening the link to Exito right now [referring to free coupons for the Exito chain of grocery stores], and of course tell your contacts not to open an email or message that says ``free internet'' and another that says free pensions. It is a virus that SEBIN [the Venezuelan intelligence service] and the Cuban G2 [intelligence service] are promoting to harm cell phones and computers to block social networks.
Tell all the contacts on your list not to accept a video called Ministeriodetrabajo. It is a virus that erases your mobile. Be careful, it's very dangerous. Pass it on to your list as people open it thinking it's a joke. They are broadcasting it today on the radio. Pass it to whoever you can.
\end{lstlisting}\doublespacing

More generally, other misinformation---particularly fake news---is also widespread on social media networks in Venezuela and Colombia, and popularly disseminated through WhatsApp \cite{cigi-20180613} \cite{publica-20170360}.

In the historic 2018 Colombian presidential election, Oscar Palma, professor of political science at Rosario University in Bogotá, declared that ``WhatsApp chains are [the] worst thing to have happened in this election'' \cite{cigi-20180613}, after many Colombians engaged with false content concerning the candidates. In Venezuela, Carlos Correa, director of free expression NGO Espacio Público, called his country ``an information desert'' \cite{poynter-venezuela-20191107}. ``Counterclaims, fake news, and outright lies'' are spread by both the Venezuelan government and opposition supporters, through both official media outlets and social media \cite{publica-20170360}.

In this chapter, we analyze fake news and online scams within our dataset, and how their prevalences vary across groups and users. We begin in chapter \ref{ch:misinformation:labeling} by discussing our methodology for labeling fake news and scams, centering on text pre-processing techniques and the cosine similarity metric. In chapter \ref{ch:misinformation:fakenews}, we move on to analyzing fake news, in its message properties, as well as user and group dynamics; we then do the same for scams in chapter \ref{ch:misinformation:scams}. Finally, in chapter \ref{ch:misinformation:detecting}, we employ various machine learning classifiers in an attempt to classify scam messages, using the content of the messages and some additional features, based on the results in section \ref{ch:misinformation:scams}.

In most contexts, scams receive comparatively less attention from researchers than fake news and other forms of misinformation. As of late March 2020, for instance, Google Scholar returns around 68,500 articles for the query ``Facebook fake news,'' but only 15,400 for the query ``Facebook scam(s)'' (the corresponding numbers for ``WhatsApp fake news'' and ``WhatsApp scam(s)'' are 11,600 and 1,550).

This disparity may be largely because online users typically treat social media networks as sources of entertainment and news, making fake news particularly appealing and troublesome. But this doesn't hold in our context of public WhatsApp groups related to the Venezuelan refugee crisis. Indeed, we found in field work that migrants treat these groups as sources of employment, resources, assistance, and day-to-day information. Political misinformation likely matters much less in this context, but economic scams matter much more, since they may be particularly appealing to migrants desperate to cross the border or longing for employment.

\section{Labeling Misinformation\label{ch:misinformation:labeling}}

Fact-checking sources are prevalent in Colombia and even Venezuela---in November 2019, the Poynter Institute, a highly-acclaimed American nonprofit journalism research institute, published an article titled, ``Against all odds, fact-checking is flourishing in Venezuela'' \cite{poynter-venezuela-20191107}. Colombian and Venezuelan fact-checking websites center on content shared by official sources, including Venezuelan president Nicolás Maduro and Venezuelan opposition representatives, as well as viral content shared over social media like Facebook and WhatsApp.

Using two Colombian sources---\textit{La Silla Vacía} and \textit{ColombiaCheck} (the two Colombian fact-checkers recognized by Poynter)---we construct a repository of fake news corpuses. We also manually inspect ``popular'' content shared in our groups (any messages that are at least 20 characters long and identically shared thrice or more, by any user/in any group), and include fake news corpuses obtained that way.

We then apply the canonical methods for processing text and detecting text similarity \cite{huang-similarity-2008}. Specifically, put our WhatsApp messages as $w_1, \dots, w_n$ and our fake news corpuses as $f_1, \dots, f_m$ so that $D = \{w_1, \dots, w_n, f_1, \dots, f_m\}$ is our set of WhatsApp messages and fake news corpuses. Put $T = \{t_1, \dots, t_p\}$ as the set of distinct terms in $D$. We do not include stop words, commonly used words that don't significantly alter the meaning of a document. In English, stop words are words like ``what,'' ``their,'' and so on; we obtain stop words from the Python \textit{Natural Language Toolkit (NLTK)} package.\footnote{The NLTK package directly provides stop words in Spanish. They include: de, la, que, el, en, y, a, los, del, se, las, por, un, para, con, no, una, su, al, lo, como, más, pero, sus, le, ya, o, este, sí, porque, esta, entre, cuando, muy, sin, sobre, también, me, hasta, hay, donde, quien, desde, todo, nos, durante, todos, uno, les, ni, contra, otros, ese, eso, ante, ellos, e, esto, mí, antes, algunos, qué, unos, yo, otro, otras, otra, él, tanto, esa, estos, mucho, quienes, nada, muchos, cual, poco, ella, estar, estas, algunas, algo...
}

After removing stopwords, we replace punctuation with spaces.\footnote{Commonly, pre-processing for text similarity involves directly removing punctuation (as opposed to replacing it with spaces); ``anti-communist,'' for example, is probably better represented as ``anticommunist'' rather than ``anti'' and ``communist,'' since the latter will pickup similarities with ``communist.'' In our circumstances, however, many scams involve hyperlinks, where it makes more sense to separately tokenize the domain names and post-domain parts of the URL. One scam purporting to offer free coupons for the \textit{Plaza}/\textit{Vea} chain of grocery stores, for example, uses the URL http://bit.ly/plazavea-cupon. Future iterations of the scam may involve variations on the url, such as http://tinyurl.com/plazavea-cupon or http://bit.ly/something-else. Splitting strings by punctuation allows us to detect both of these variants (since the original URL is tokenized as [``bit'', ``ly'', ``plazavea'', ``cupon'']), while simply removing punctuation would tokenize the original URL as [``httpbitlyplazaveacupon''] (a single token) and fail to match future variants.} We also ``stem'' words, using the NLTK Snowball algorithm (Spanish) \cite{snowball-spanish}, which maps similar words with different endings to the same root. For example, \textit{chico} (meaning boy or small) and \textit{chica} (meaning girl or small) both map to \textit{chic}, while \textit{chicago} maps to \textit{chicag} and no further.

Our next step is to vectorize each text as a $p \times 1$ feature vector, whose $i$-th term is the count of how many times $t_i$ appears in the text. But instead of just using counts directly, we normalize the counts by inverse document frequency, or $\log \frac{|D|}{|\{d \in D: t_i \in d\}|}$. This gives us the well-known statistic TF-IDF (term frequency-inverse document frequency) \cite{ramos-tfidf-2003}, which measures how uniquely relevant each term is in a document. Consider, for example, the set of articles written about Princeton: ``university'' would likely appear in almost all of them,\footnote{We do not automatically remove ``university'' as a stop word, since it's not so commonly used across English language texts in general.} and quite frequently in each, so the frequency of ``university'' in an article isn't informative about the article. A word like ``Eisgruber'' might appear in relatively few documents, on the other hand, so using TF-IDF would weight its appearances more, and this is meaningful to us in helping differentiate the articles. An article that mentions Eisgruber 13 times is likely much more similar to an article that mentions Eisgruber 11 times, than an article that mentions ``university'' 13 times is to one that mentions ``university'' 11 times.

In deciding a distance/similarity measure to compare feature vectors with, we might consider Euclidean distance, letting distance $d(x, y) = \|x-y\|$. But such a measure is not robust to document size; in particular, it might dictate that the abstract of this thesis is more similar to the abstract of a thesis about giraffes, than it is to the body of this thesis.\footnote{Let the abstracts be, for example, ``whatsapp venezuela'' and ``giraffe neck'' repeated 3 times, and let the thesis body be ``whatsapp venezuela'' repeated 100 times. Then if $T = \{\textrm{whatsapp}, \textrm{venezuela}, \textrm{giraffe}, \textrm{neck}\}$, the feature vectors are (without slight TF-IDF adjustments) $(3, 3, 0, 0), (0, 0, 3, 3)$ for the abstracts and $(100, 100, 0, 0)$ for the thesis body; clearly, the abstracts are closer.}

Instinctively, we de-norm our feature vectors, so $d(x, y) = \| \frac{x}{\|x\|} - \frac{y}{\|y\|} \|$. This immediately motivates cosine similarity, $s(x, y) = \frac{\langle x, y \rangle}{\|x\| \|y\|}$, as a similarity measure since $\| \frac{x}{\|x\|} - \frac{y}{\|y\|} \| = \sqrt{\frac{x^T x}{\|x\|^2} - 2\frac{x^T y}{\|x\| \|y\|} + \frac{y^T y}{\|y\|^2}} = \sqrt{2 - 2 \cdot \frac{\langle x, y \rangle}{\|x\| \|y\|}}$. Cauchy-Schwarz gives us $0 \leq s(x,y) = \frac{\langle x, y \rangle}{\|x\| \|y\|} \leq 1$, and this measure is 1 iff $x$ and $y$ are parallel. Particularly relevant, cosine similarity is independent of document length.

We collected 56 fake news corpuses from the two fact-checking sources; all 56 were present in our data. Additionally, from our manual inspection of popular messages in our dataset, we extracted 64 further fake news corpuses; these 64 corpuses involved 291 different messages.

To exclude messages like ``hola,'' ``gracias,'' and so on, and to reduce the probability of false positives, we only considered messages as possible fake news if their tokenization was at least five words long. Ultimately, after removing the 291 messages that we already (manually) flagged as fake news, this resulted in 43,734 candidates.

For each of the 43,734 candidates, we calculated their maximum cosine similarity to any known fake news corpus, and manually inspected 497 messages with maximum cosine similarity over 0.3. In \cite{resende-textual-2019}, Resende et al. only considered cosine similarity over 0.4 (apparently, in a small manual sample they found no fake news matches when cosine similarity was below 0.4), but we show below that a decent number of fake news corpuses had maximum cosine similarity (to any known fake news) less than 0.4. Ultimately, we found 181 true positives (36.4\%).

\begin{figure}[h]
\centering
\includegraphics[width=0.6\textwidth]{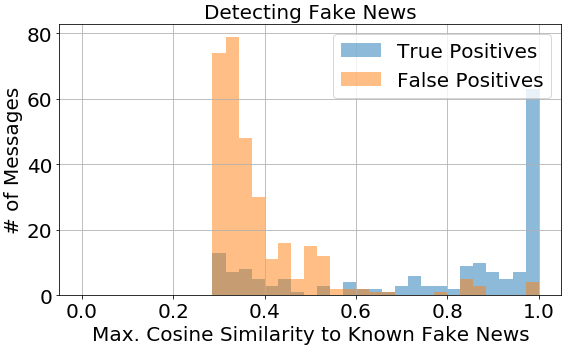}
\caption{This histogram includes all messages with cosine similarity over 0.3 to known fake news. The cosine similarity of true positives peaks at 1.0 and quickly diminishes; the distribution of cosine similarity of false positives increases exponentially at cosine similarity less than 0.4. True positives had mean cosine similarity 0.775, while false positives had mean cosine similarity 0.400 ($p < 0.001$).}
\label{figure:misinformation:detecting_fakeNews_cs}
\end{figure}

Figure \ref{figure:misinformation:detecting_fakeNews_cs} presents a histogram comparing the maximum cosine similarity of true positives and the maximum cosine similarity of false positives. The former distribution peaks at 1 but continues even at cosine similarity less than 0.4.

Clearly, limiting our manual inspection to messages with cosine similarity over 0.4, as \cite{resende-textual-2019} did, would leave those messages unlabeled as fake news.\footnote{Resende et al. did not necessarily do anything wrong, since the range of cosine similarities can depend on the pre-processing, the token structure, and the actual text involved. But it's clear that the cosine similarities of true positives are a continuous distribution on $[0,1]$, so imposing a bound of 0.4 based on a small manual sample is rather suspect.} In fact, there almost certainly exists fake news with cosine similarity less than 0.3, but the histogram makes clear that false positives grow exponentially by that point, really making true positives needles in a haystack.

To label scams, as no ``scam-checking'' sources exist in our context, we constructed a repository of 84 known scams through the same process of manually inspecting popular content. These 84 scams were found in 663 different messages. We pre-processed and tokenized as we did for fake news, and ultimately filtered 43,362 candidate messages to check (as for fake news, these were messages whose tokenization was $\geq 5$ words, and that were not already known to be scams).

\begin{figure}[h]
\centering
\includegraphics[width=0.6\textwidth]{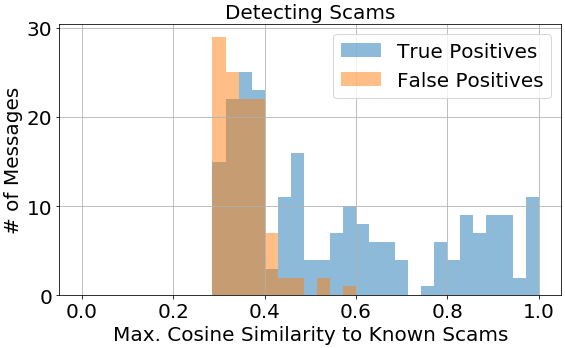}
\caption{This histogram includes all messages with cosine similarity over 0.3 to known scams. The cosine similarity of true positives peaks is more scattered than before; again, the distribution of false positives' cosine similarity exponentially grows at cosine similarity less than 0.4. True positives had mean cosine similarity 0.563, while false positives had mean cosine similarity 0.354 ($p < 0.001$).}
\label{figure:misinformation:detecting_scam_cs}
\end{figure}

We manually inspected 335 messages with cosine similarity over 0.3 to a known scam, and out of these found 223 true positives (66.6\%). True positives had mean cosine similarity 0.563, while false positives had mean cosine similarity 0.354 ($p < 0.001$). Figure \ref{figure:misinformation:detecting_scam_cs} presents a histogram comparing the maximum cosine similarity of true positive scams and the maximum cosine similarity of false positive scams.

Note that the cosine similarity of true positives is much more scattered than it was for fake news: this likely arises because scams are much shorter (in sections \ref{ch:misinformation:fakenews:message} and \ref{ch:misinformation:scams:message}, we show that fake news are much longer than other messages, while scams are somewhat shorter than other messages). With scams being shorter, there is less for cosine similarity to pick up on, so we lose the cosine similarity peak at 1.0 that we saw for fake news. We might also imagine that those responsible for scams are also more intentionally and more frequently altering messages, resulting in lower cosine similarity to known scams.

Again, limiting our manual inspection to messages with cosine similarity over 0.4, as \cite{resende-textual-2019} did, would leave a \textit{lot} of scams unlabeled. There's a stronger case to be made here for manually inspecting messages even below cosine similarity 0.3, but we skip this because of time constraints, since the number of messages to inspect grows exponentially as we drop cosine similarity.

\section{Analyzing Fake News\label{ch:misinformation:fakenews}}

With the labeling methodology described in section \ref{ch:misinformation:labeling}, we ended up labeling 472 messages as fake news. To better characterize the prevalence of fake news, we filtered our original 171,634 messages to 44,025 messages whose text tokenizations were at least five words (i.e., text messages with some greater meaning than greetings, etc.). For the rest of this chapter, we call these ``meaningful text messages.'' Remarkably, the proportion we found of fake news within meaningful text messages (1.1\%) is nearly the same as in \cite{resende-textual-2019}, which found 578 fake news amongst 59,979 textual messages (1.0\%).

\subsection{Message Dynamics\label{ch:misinformation:fakenews:message}}

On average, fake news messages received fewer replies, as compared to other ``meaningful text messages'': 0.0805 versus 0.5659, a stark difference ($p < 0.001$). This effect was slightly weaker when comparing fake news messages and all other messages (which on average received 0.5415 replies), though as we described in Chapter \ref{ch:replycascades}, comparing replies across content types is suspect. Because the distribution of replies is so skewed (where most messages receive no replies), we can also examine differences at the 95th percentile of fake news and other meaningful text messages, in terms of how many replies they receive. It turns out that only 4.9\% of fake news receive \textit{any} replies, while the 95th quantile of non-fake news text messages receives 3 replies.

Fake news messages were also significantly less viral, based on the structural virality metric defined in Chapter \ref{ch:replycascades:virality}; we re-emphasize that virality is only calculated across messages \textit{with} replies (i.e., we are controlling for the fact that fake news messages receive significantly fewer replies).

The average virality of fake news in reply cascades was 0.7667, compared to 1.3565 for other meaningful text messages, and 1.5022 for all non-fake news messages ($p = 0.01$). The 95th quantile of fake news in reply cascades has virality 1.41, while for other meaningful text messages this virality is 3.55.

Fake news messages were also longer (here, we only compare to other meaningful text messages, for obvious reasons). On average, fake news was 1384 characters long, compared to 318 characters long for other meaningful text messages, and involved 233 words compared to 49 words ($p < 0.001$ for both).

As with text messages in general, we can use latent dirichlet allocation to parse out topics underlying fake news messages. Setting parameters of 10 topics with 10 words each, we obtain the following topics:

\singlespacing\begin{enumerate}
  \item virus chin mund salud pais cas egipt limon pas merc
  \item virus dias pulmon tom vias agu chin evit pais sol
  \item alert hij inform nin pas compart ser pais segur escuel
  \item limon tom agu pued calient cuerp celul cuid alcalin sustanci
  \item agu tom inclu ibuprofen sintom sal favor salv ajo virus
  \item contact virus pasal urgent celular mensaj llam dil vide murcielag
  \item chin accion telon mund coronavirus mundial virus empres compr tod
  \item virus pued calient sol agu coronavirus man hor beb hac
  \item 40 dios dias person jesus mand mensaj pued despu famili
  \item chin virus caf wuh mund coron quimic pacient km beijing
\end{enumerate}\doublespacing

Unsurprisingly, eight of these ten topics involve the coronavirus (topic 3 is a fear-mongering news alert about organ-trafficking mafias, and topic 9 is a religious chain message). Topics 6, 7, and 10 center on current events related to the coronavirus; topics 1, 2, 4, 5, and 8 include fake scientific and medical information.

\subsection{User Dynamics\label{ch:misinformation:user}}

\begin{figure}[h]
\centering
\begin{subfigure}{0.45\textwidth}
  \centering
  \includegraphics[width=\textwidth]{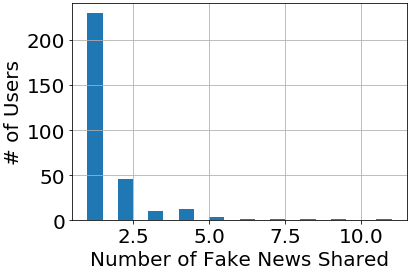}
  \caption{A histogram of the number of fake news messages sent by users who've shared fake news.}
  \label{figure:misinformation:hist_fakenews_user_frequency}
\end{subfigure}
\hfill
\begin{subfigure}{0.45\textwidth}
  \centering
  \includegraphics[width=\textwidth]{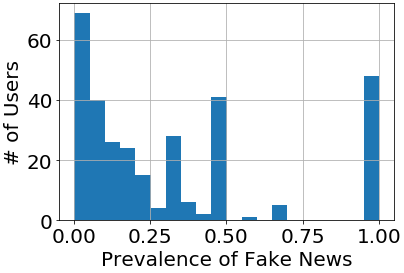}
  \caption{A histogram of the prevalence of fake news amongst fake news sharers' meaningful text messages.}
  \label{figure:misinformation:hist_fakenews_user_prevalence}
\end{subfigure}%
\caption{Histograms with users who've shared fake news.}
\end{figure}

In total, 309 unique users shared fake news (3.9\% of 7,860 active members). Figure \ref{figure:misinformation:hist_fakenews_user_frequency} plots a histogram of the number of times users shared fake news; 74.4\% of sharers only shared fake news once, and 14.9\% of sharers only shared fake news twice.

We move on to analyzing the prevalence of fake news, which we define as the proportion of fake news amongst all meaningful text messages sent by a user. Of 4,645 users who sent meaningful text messages, 93.3\% never shared fake news. But, as seen in figure \ref{figure:misinformation:hist_fakenews_user_prevalence}, of users who've shared fake news, many haven't shared much other meaningful text content.

\subsubsection{Comparing Across Countries}

\begin{figure}[h]
\centering
\includegraphics[width=0.45\textwidth]{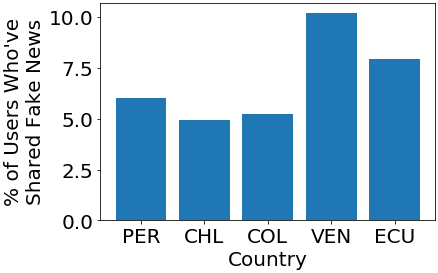}
\caption{Comparing fake news senders by country. (ANOVA $p < 0.001$; $t$-tests between VEN and PER/CHL/COL $p < 0.05$; $t$-tests between COL and PER/CHL/ECU and between VEN and ECU not significant.)}
\label{figure:misinformation:bar_fakenews_country_whoshared}
\end{figure}

An extremely interesting disparity we found was that Venezuelan users were twice as likely to be fake news sharers as Colombian users: 10.2\% of Venezuelans have shared fake news, compared to 5.2\% of Colombians ($p < 0.001$). Figure \ref{figure:misinformation:bar_fakenews_country_whoshared} graphs the percentage of users who've shared fake news from our five principal countries; around the same proportion of Peruvian, Chilean, and Colombian users have shared fake news, while somewhat more Ecuadorians and many more Venezuelans have.

\begin{figure}[h]
\centering
\begin{subfigure}{0.45\textwidth}
  \centering
  \includegraphics[width=\textwidth]{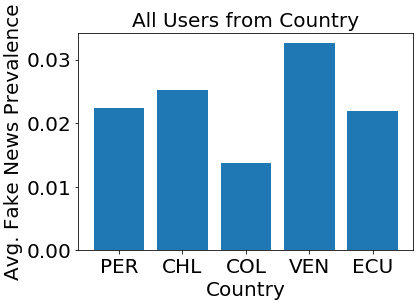}
  \caption{(ANOVA $p < 0.001$; only the $t$-test between VEN and COL was significant.)}
  \label{figure:misinformation:bar_fakenews_country_prevalence}
\end{subfigure}%
\hfill
\begin{subfigure}{0.45\textwidth}
  \centering
  \includegraphics[width=\textwidth]{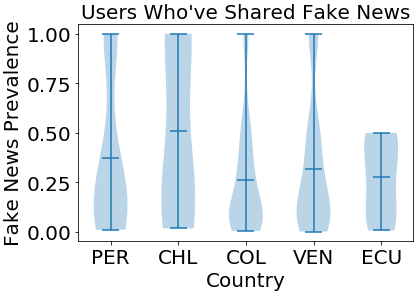}
  \caption{(Only the $t$-test between COL and CHL was statistically significant.)}
  \label{figure:misinformation:violin_fakenews_country_prevalence_sharers}
\end{subfigure}
\caption{Cross-country comparisons of average prevalence of fake news.}
\end{figure}

The same disparity holds when examining the average prevalence of fake news of users from each country, shown in figure \ref{figure:misinformation:bar_fakenews_country_prevalence}.

On average, 3.3\% of each Venezuelan's user's text content was fake news, compared to 1.4\% for each Colombian user.

The difference in average prevalence of fake news across users, however, is almost entirely explained by the fact that a mucher higher percentage of Venezuelans have shared fake news compared to Colombian users. Figure \ref{figure:misinformation:violin_fakenews_country_prevalence_sharers} shows a violin plot of fake news prevalence across users who've shared fake news, with the horizontal line at each country's mean, and makes clear that of users who've shared fake news, average fake news prevalence is roughly equal across countries (in particular, the difference between Venezuelan users and Colombian users is not statistically significant).

\begin{figure}[h]
\centering
\begin{subfigure}{0.45\textwidth}
  \centering
  \includegraphics[width=\textwidth]{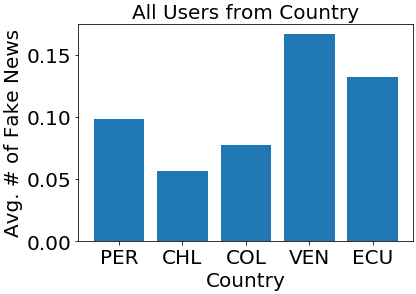}
  \caption{(ANOVA $p < 0.001$; $t$-tests between VEN and PER/COL $p < 0.05$.)}
  \label{figure:misinformation:bar_fakenews_country_frequency}
\end{subfigure}%
\hfill
\begin{subfigure}{0.45\textwidth}
  \centering
  \includegraphics[width=\textwidth]{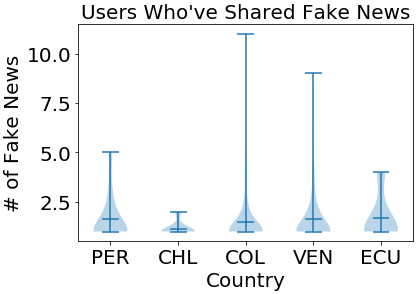}
  \caption{(No statistical significance.)}
  \label{figure:misinformation:violin_fakenews_country_frequency_sharers}
\end{subfigure}
\caption{Cross-country comparisons of average frequency of fake news.}
\end{figure}

Even when looking at the raw number of fake news shared, Venezuelan users shared significantly more fake news. On average, each Venezuelan user shared 0.167 fake news messages, compared to 0.078 for Colombian users ($p < 0.001$). Again, this difference was mostly accounted for by more Venezuelans having shared fake news: when only considering users who've shared fake news, users from each of the countries shared 1.0-1.5 fake news messages on average. The graph in figures \ref{figure:misinformation:bar_fakenews_country_frequency} plots the averages, and the violin plot in figure \ref{figure:misinformation:violin_fakenews_country_frequency_sharers} shows per-country distributions of how many fake news messages have been shared, by sharers of fake news (the horizontal line indicates each country's mean).

Even though we've found that Venezuelans are twice as likely to have shared fake news, we argue that this is actually an underestimate. Recall that the two fact-checking sources we used were both from Colombia, making our methodology inherently more likely to catch fake news that is relevant to Colombians; that even with this bias we found such a discrepancy suggests the actual effect is even stronger.

Why are Venezuelans so much more likely to share fake news? Generally, anyone with even passing knowledge of Latin America would point to the country's political environment, and the ``information desert'' in Venezuela that we described in the introduction to this chapter.\footnote{We emphasize that Venezuelans are particularly susceptible to fake news because of their ``post-truth'' environment, not because of differences in intelligence. In field interviews, most Venezuelan migrants, even those living in informal settlements without electricity or running water, clearly had high levels of education.} Misinformation has been spread by the Maduro regime over both official state channels and social media; a 2018 article from The Guardian puts it bluntly by describing fake news as one of the dictator's ``weapons'' \cite{guardian-fakenews-20180125}. Unsurprisingly, opposition forces under Juan Guaido have responded with the same strategies.\footnote{In a more personal encounter, one of the reasons Princeton's travel oversight staff gave for not approving our proposed field travel to Cucuta was that, ``anti-Maduro groups send people over the border to use their phones to send messages and information to a wider network over WhatsApp or Telegram.''}

In our case, however, one particular circumstance may best explain Venezuelan users' disposition to fake news. Our data collection period included late March and early April 2020, when the coronavirus pandemic arrived and exponentially worsened in Colombia and Venezuela. Most fake news messages in our dataset were coronavirus-related, and a significant portion related to home cures against the virus, like lemon water or leaving clothes in the sun. In a country with a collapsed health system, promised cures to a devastating illness may be especially appealing.\footnote{Of course, not everyone is fooled. Within our groups, rebuttals to these false cures include ``Now if we screwed up, the eighth plague of Egypt arrived'' (in response to fake news announcing that Chinese doctors had cured the coronavirus with an Egyptian serum), as well as, ``What a mess, we will end up drinking garlic water.''}

\subsection{Group Dynamics\label{ch:misinformation:fakenews:group}}

For each group, we construct two measures for the prevalence of fake news within that group: first, the proportion of meaningful text messages in that group that involve fake news---which we call the ``message prevalence'' of fake news---and second, the proportion of users in that group who've shared fake news (in the same group), which we call the ``user prevalence'' of fake news.

\begin{figure}[h]
\centering
\begin{subfigure}{0.45\textwidth}
  \centering
  \includegraphics[width=\textwidth]{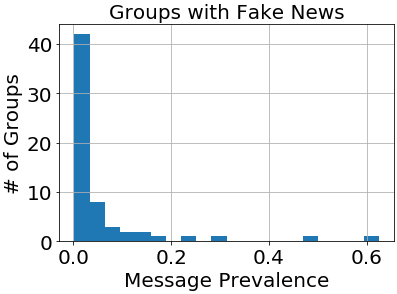}
  \caption{Histogram of message prevalence of fake news.}
  \label{figure:misinformation:hist_fakenews_group_prevalence}
\end{subfigure}%
\hfill
\begin{subfigure}{0.45\textwidth}
  \centering
  \includegraphics[width=\textwidth]{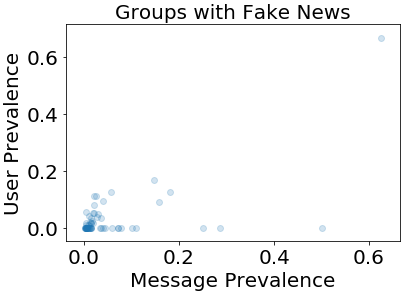}
  \caption{Scatter plot of message prevalence and user prevalence of fake news.}
  \label{figure:misinformation:scatter_fakenews_group}
\end{subfigure}
\caption{Plots involving groups with fake news.}
\end{figure}

Of 174 groups, over 64\% (112 groups) did not have any messages flagged as fake news. The histogram in figure \ref{figure:misinformation:hist_fakenews_group_prevalence} reveals that across even groups where fake news was shared, message prevalence was low; in 54 of the 62 groups were fake news was shared, fake news made up less than 10\% of the group's meaningful text messages. The scatter plot in figure \ref{figure:misinformation:scatter_fakenews_group} reveals that message prevalence and user prevalence were both typically low. The two groups where fake news made up over 40\% of meaningful text messages were both quite inactive (one had only two messages in our months-long collection period, and the other had 24).

Across all groups (including the 112 groups where no fake news was shared), the proportion of users who shared fake news was correlated, unsurprisingly, with the Venezuelan user proportion of groups. On average, an increase in the Venezuelan user proportion by 10\% increased the proportion of users who shared fake news by 0.45\% ($p < 0.01$).

\begin{figure}[h]
\centering
\includegraphics[width=0.45\textwidth]{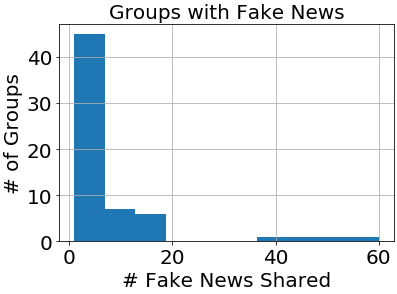}
\caption{A histogram of the number of fake news messages sent in groups, for groups where fake news was shared.}
\label{figure:misinformation:hist_fakenews_group_frequency}
\end{figure}

Finally, figure \ref{figure:misinformation:hist_fakenews_group_frequency} shows a histogram of the raw number of fake news messages in groups with fake news.

To better understand group dynamics, we move on to looking only at groups where fake news was shared. In these groups, the message prevalence of fake news was weakly negatively correlated with group size and activity, moderately positively correlated with group concentration, moderately negatively correlated with virality, and strongly negatively correlated with Gini (group inequality). An OLS regression of fake news message prevalence on these factors is shown in table \ref{table:misinformation:ols_group_fakenews}.

\begin{table}[h]
\centering
\caption{OLS regression of fake news message prevalence on group size, activity, concentration, virality, and inequality; we only include groups with fake news.}
\label{table:misinformation:ols_group_fakenews}
\begin{tabular}{p{0.25\textwidth} p{0.35\textwidth} p{0.10\textwidth} p{0.15\textwidth}}
\toprule
 & \textbf{Coefficient (Std. Err.)} & \textbf{$t$} & \textbf{P-Value} \\
\midrule

Intercept & $0.2825 (0.051)$ & $5.496$ & $0.000^{\ast}$ \\[0.2em]

Size & $0.0000 (0.000)$ & $0.509$ & $0.612$ \\[0.2em]

Activity & $0.0001 (0.000)$ & $1.047$ & $0.299$ \\[0.2em]

(H-H) Concentration & $0.1592 (0.070)$ & $2.276$ & $0.027^{\ast}$ \\[0.2em]

Gini/Inequality & $-0.3662 (0.082)$ & $-4.451$ & $0.000^{\ast}$ \\[0.2em]

Virality & $-0.0348 (0.037)$ & $-0.932$ & $0.356$ \\[0.2em]
\midrule
\multicolumn{4}{c}{$n = 62$\,\,(56 d.f.) \quad\quad $R^2 = 0.506$}\\
\bottomrule
\end{tabular}
\end{table}

Yet because of the power-law distribution of message prevalence across groups, where fake news made up less than 10\% of text content in 54 of 62 groups where fake news was shared, any regression is strongly affected by the two aforementioned outlier groups, where fake news made up over half of textual content. We perform another OLS regression dropping those outliers, obtaining the coefficients in table \ref{table:misinformation:ols_group_fakenews_drop}.

\begin{table}[h]
\centering
\caption{OLS regression of fake news message prevalence on group size, activity, concentration, virality, and inequality; we only include groups with fake news, and dropped groups with fake news message prevalence over 0.4.}
\label{table:misinformation:ols_group_fakenews_drop}
\begin{tabular}{p{0.25\textwidth} p{0.35\textwidth} p{0.10\textwidth} p{0.15\textwidth}}
\toprule
 & \textbf{Coefficient (Std. Err.)} & \textbf{$t$} & \textbf{P-Value} \\
\midrule

Intercept & $0.1653 (0.034)$ & $4.804$ & $0.000^{\ast}$ \\[0.2em]

Size & $0.0000 (0.000)$ & $-0.172$ & $0.864$ \\[0.2em]

Activity & $0.0000 (0.000)$ & $0.077$ & $0.939$ \\[0.2em]

(H-H) Concentration & $0.0932 (0.055)$ & $1.681$ & $0.098$ \\[0.2em]

Gini/Inequality & $-0.2159 (0.061)$ & $-3.536$ & $0.001^{\ast}$ \\[0.2em]

Virality & $0.0027 (0.024)$ & $0.114$ & $0.910$ \\[0.2em]
\midrule
\multicolumn{4}{c}{$n = 60$\,\,(54 d.f.) \quad\quad $R^2 = 0.337$}\\
\bottomrule
\end{tabular}
\end{table}

From these regressions, more concentrated groups are linked to greater fake news prevalence, while more unequal groups are strongly linked to less fake news prevalence, even while controlling for group size, activity, and virality. That these coefficients are in opposite directions should not surprise us; we previously saw that group concentration and group inequality are distinct measures (in particular, group inequality increases significantly when there are many ``poor'' individuals with few messages, though they barely affect concentration\footnote{Imagine wealth in New York City: 8 million poor individuals arriving in the city would significantly increase inequality measures, but have little impact on concentration of wealth at the top.}).

There are clear hypotheses for why more concentrated groups might have higher fake news prevalence: ``echo chambers'' on social media networks, where like-minded individuals are insulted from diverse and alternative perspectives, have been well studied, especially since the 2016 U.S. presidential election \cite{allcott-fakenews-2017} \cite{guess-selective-2018}. We might also hypothesize that more concentrated groups feel more familiar to at least the frequent users, since a small subset of the group dominates conversation, so they may pass on information more inattentively (in a ``Forwards from Grandma'' kind of manner\footnote{\url{https://knowyourmeme.com/memes/forwards-from-grandma}}).

This second point might explain why fake news message prevalence decreases as group inequality rises: highly unequal groups may appear to include many strangers (i.e., members who send few messages). Users may fear getting called out for sharing fake news, or may simply pay more attention to messages they forward along. More directly, these message-poor members may occasionally chime in with alternate perspectives.

\subsection{Variants of Fake News\label{ch:misinformation:fakenews:variants}}

Above, we looked at fake news in aggregate; now we identify unique pieces of fake news. Because fake news is slightly altered as users pass it on, whether insiduously or not,\footnote{We can imagine malicious users altering content slightly to avoid spam filters, for example, but also innocent users rewording false medical advice to be more credible and context-specific. For example, false medical advice in our dataset usually cites an invented doctor, but the nationality of this doctor changes based on the group it's sent to.} we must aggregate together fake news messages with subtle differences.

Take, for example, the following message, which provides false medical advice about the coronavirus (that the sun kills the coronavirus\footnote{At the time of publication, this is unsupported by medical experts.}):

\singlespacing\begin{lstlisting}
*Consejo del Dr. Yuri Ortega Sotelo +51987453411
El coronavirus es de gran tamano con un diametro celular de 400-500 micras, por lo que cualquier mascara impide su entrada, por lo que no es necesario explotar a los farmaceuticos para comerciar con bozales.
El virus no se instala en el aire, sino en el suelo, por lo que no se transmite por el aire.
El virus, cuando cae sobre una superficie de metal, vivira durante 12 horas, por lo que lavarse bien las manos con agua y jabon sera suficiente.
El virus cuando cae sobre las telas permanece durante 9 horas, por lo que lavar la ropa o exponerla al sol durante dos horas es suficiente para matarlo.
El virus vive en las manos durante 10 minutos, por lo que llevar un desinfectante con alcohol en el bolsillo y aplicar es suficiente para prevenirlo.
Si el virus se expone a una temperatura de 26-27 C, se matara, no vive en areas calientes. Tambien es suficiente beber agua caliente y exponerse al sol. Mantenerse alejado del helado y la comida fria es importante.
Hacer gargaras con agua tibia y sal mata el virus en las amigdalas y evita que se filtren a los pulmones.
Cumplir con estas instrucciones es suficiente para prevenir el virus.
Dr. Yuri Ortega Sotelo
\end{lstlisting}\doublespacing

The following variant is a shorter snippet of the first message, and also changes the supposed medical source from a doctor to UNICEF.

\singlespacing\begin{lstlisting}
Consejos de la Unicef
El coronavirus es de gran tamano con un diametro celular de 400-500 micras, por lo que cualquier mascara impide su entrada, por lo que no es necesario explotar a los farmaceuticos para comerciar con bozales.
El virus no se instala en el aire, sino en el suelo, por lo que no se transmite por el aire.
El virus, cuando cae sobre una superficie de metal, vivira durante 12 horas, por lo que lavarse bien las manos con agua y jabon sera suficiente.
El virus cuando cae sobre las telas permanece durante 9 horas, por lo que lavar la ropa o exponerla al sol durante dos horas es suficiente para matarlo.
El virus vive en las manos durante 10 minutos, por lo que llevar un desinfectante con alcohol en el bolsillo y aplicar es suficiente para prev...
\end{lstlisting}\doublespacing

To detect altered messages, we again use cosine similarity, purely within the set of fake news messages and with a manually-tuned baseline of 0.8. Out of 214 different fake news texts, we identified 98 that were variants of other messages, leaving 116 unique fake news messages. Two particularly viral messages involved seven variants each; both promised cures to the coronavirus.

\begin{figure}[h]
\centering
\includegraphics[width=0.45\textwidth]{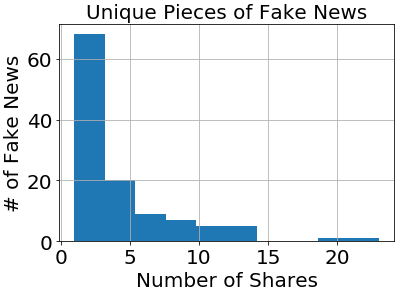}
\caption{Histogram of the number of times each unique fake news was shared.}
\label{figure:misinformation:hist_fakenews_unique_frequency}
\end{figure}

\pagebreak
Figure \ref{figure:misinformation:hist_fakenews_unique_frequency} depicts a histogram of how many times each of the 116 unique fake news pieces was shared; the two messages that were shared most frequently were about a child-kidnapping organ-trafficking mafia (with 23 shares) and another advising that the Chinese cured coronavirus with hot liquids and gargling with saltwater (19 shares).

\begin{figure}[h]
\centering
\begin{subfigure}{0.45\textwidth}
  \centering
  \includegraphics[width=\textwidth]{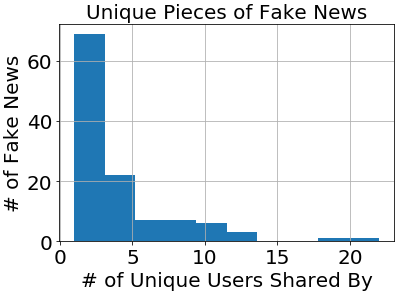}
\end{subfigure}%
\hfill
\begin{subfigure}{0.45\textwidth}
  \centering
  \includegraphics[width=\textwidth]{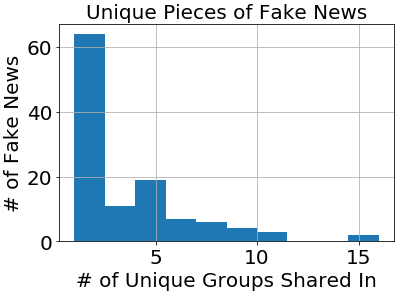}
\end{subfigure}
\caption{Histograms of how many different users shared each fake news piece, and in how many different groups each fake news piece was shared.}
\label{figure:misinformation:hist_fakenews_unique_user_group}
\end{figure}

Finally, the histograms in figure \ref{figure:misinformation:hist_fakenews_unique_user_group} show that the vast majority of fake news were shared by five or fewer users and in five or fewer groups. For each unique fake news, we calculated the average number of shares per user (who shared the message), and the average number of shares per group (where the message was shared). Across the 116 fake news, these averaged to 1.06 and 1.28 respectively, showing that users typically only shared each fake news message once, and that fake news weren't repeatedly shared in the groups they reached.

\section{Analyzing Scams\label{ch:misinformation:scams}}

In analyzing scams, we proceed nearly identically to the previous section (though we expect, and proceed to show, completely different results).

With our labeling methodology, we identified 886 messages as scams (88\% more messages than fake news messages), and again separated out 44,025 ``meaningful'' text messages (i.e., text messages with at least five words in their tokenization).

\subsection{Message Dynamics\label{ch:misinformation:scams:message}}

Like fake news messages, scam messages received starkly fewer replies: 0.1219 replies on average, compared to 0.5697 for other meaningful text messages, and 0.5424 across all non-scam messages ($p < 0.001$). The 95th quantile of scams received only one reply, while the 95th quantile of non-scam text messages received three replies.

As with fake news, scams in even reply cascades went significantly less viral, with average virality 0.6695, compared to 1.3590 across other meaningful text messages, and 1.5029 across all non-scam messages ($p < 0.001$). The 95th quantile of scams and non-scam meaningful text messages had viralities 1.28 and 3.55, respectively.

While fake news messages were much longer than other text messages (435\% as long), scams were actually slightly shorter than other meaningful text messages, at 297 vs. 330 characters on average (n.s.), and 41 vs. 51 words long ($p < 0.01$).

A latent dirichlet allocation parameterized with 10 topics of 10 words each yields the following topics within scam messages:

\singlespacing\begin{enumerate}
  \item bon pais prestam http com cupon diner hol exit 000
  \item internet gb 100 dat gratis obteng ahor https consiguel cualqui
  \item grup vide bienven prestam https siguient pas va voy javi
  \item ayud resib us 77 alimentari onu earn 00 invest clic
  \item prest 000 prestam personal 3 eur tas plaz whatsapp interes
  \item https whatsapp and oscur com ly bit of to activ
  \item https 000 sisb netflix period aislamient cupon com entra rap
  \item netflix period aislamient https pandemi dand gratis deb coronavirus mund
  \item https com chat diplom whatsapp ayud c z l grup
  \item tarjet alimentari madr cp https nuev solicitud bon to crypto
\end{enumerate}\doublespacing

The topics of scams are a bit more varied than topics in fake news messages. Topics 7 and 8 involve free Netflix accounts during the coronavirus quarantine; topics 4 and 10 offer financial assistance from the government; and topic 2 purports to offer free internet. Topics 1, 3, and 5 are fake loan offers, while topic 6 involves WhatsApp (i.e., ``Change the color of your WhatsApp!'').

\subsection{User Dynamics\label{ch:misinformation:scams:user}}

\begin{figure}[h]
\centering
\begin{subfigure}{0.45\textwidth}
  \centering
  \includegraphics[width=\textwidth]{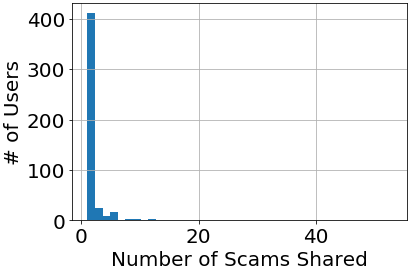}
\end{subfigure}
\hfill
\begin{subfigure}{0.45\textwidth}
  \centering
  \includegraphics[width=\textwidth]{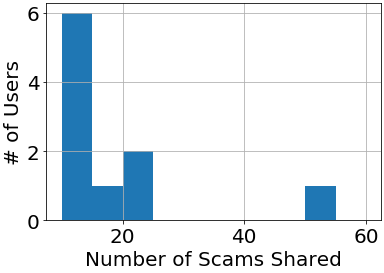}
\end{subfigure}
\caption{The left includes all users who shared scams; the right only includes frequent scam-sharers.}
\label{figure:misinformation:hist_scam_user_frequency}
\end{figure}

Scams were shared by 473 users, 6.0\% of the 7,860 users in total. Like with fake news, most users who shared scams only shared them once (70.8\%, compared to 74.4\% of fake news sharers), and 16.1\% shared scams exactly twice. But as figure \ref{figure:misinformation:hist_scam_user_frequency} shows, the tail of this distribution is a lot higher than for fake news.

Users who shared fake news shared a maximum of 11 fake news messages, but the right graph in figure \ref{figure:misinformation:hist_scam_user_frequency} makes clear that some users are frequent scam-sharers. This is expected: even setting aside intentional troublemakers, we may imagine that scam victims have had their accounts commandeered to bulk-send scams.

\begin{figure}[h]
\centering
\begin{subfigure}{0.45\textwidth}
  \includegraphics[width=\textwidth]{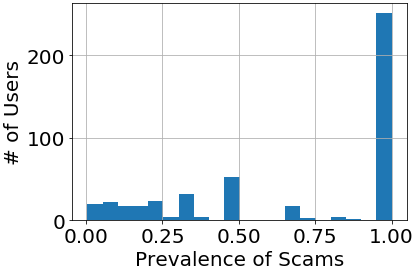}
  \caption{Most users who share scams \textit{only} share scams.}
  \label{figure:misinformation:hist_scam_user_prevalence}
\end{subfigure}
\hfill
\begin{subfigure}{0.45\textwidth}
  \includegraphics[width=\textwidth]{ch-misinformation/images/hist_fakenews_user_prevalence.png}
  \caption{Most users who share fake news share other meaningful text content.}
  \label{figure:misinformation:hist_fakenews_user_prevalence_v2}
\end{subfigure}
\caption{Prevalence of scams amongst sharers, compared to prevalence of fake news amongst sharers.}
\end{figure}

We move on to analyzing the prevalence of scams, which we similarly define as the proportion of scams amongst all meaningful text messages sent by a user. Of 4,645 users who sent meaningful text messages, 89.8\% never shared scams. But, as seen in figure \ref{figure:misinformation:hist_scam_user_prevalence}, of 473 users who've shared scams, 251 users (53.1\%) have only sent scam messages and no other meaningful text messages!

Figure \ref{figure:misinformation:hist_fakenews_user_prevalence_v2} plots again the relevant histogram for fake news sharers; these graphs are \textit{starkly} different. Comparing these two distributions allows us to better characterize scam sharers, but also provide remedies. Whereas banning fake news sharers would prevent them from sharing other meaningful content, most scam sharers don't share any other meaningful content!

Across users, sharing fake news and sharing scams were very weakly positively correlated ($r = 0.03$, $p = 0.01$). 9.4\% of users who've shared fake news also shared scams, compared to 5.9\% of users who haven't shared fake news; similarly, 6.1\% of users who've shared scams also have shared fake news, compared to 3.8\% of users who haven't shared scams.

The prevalence of fake news and scams for each user were not correlated (or, rather, very weakly negatively correlated with no statistical significance), which might be due to crowding-out effects between fake news and scams. Of users who've shared fake news, scams on average made up 0.8\% of users' messages, compared to 2.7\% of messages from users who've never shared fake news; fake news prevalence was 0.5\% amongst scam sharers, and 0.7\% for non-scam sharers.

\subsubsection{Comparing Across Countries}

\begin{figure}[h]
\centering
\includegraphics[width=0.45\textwidth]{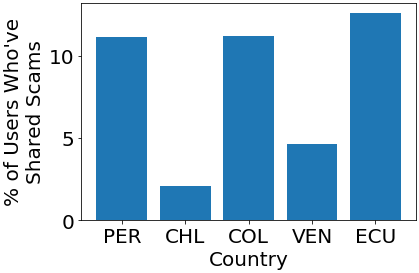}
\caption{Comparing scam senders by country. (ANOVA $p < 0.001$; $t$-tests between VEN and PER/COL/ECU, and between COL and CHL, $p < 0.001$.)}
\label{figure:misinformation:bar_scam_country_whoshared}
\end{figure}

Before, we had seen that Venezuelan users were more likely to have shared fake news. It turns out that this disparity is flipped on its head for scam-sharers: Colombian users were 240\% as likely to share scams, compared to Venezuelans! Specifically, 11.2\% of Colombian users have shared scams, while only 4.6\% of Venezuelan users have. The bar graph in figure \ref{figure:misinformation:bar_scam_country_whoshared} shows that Chileans and Venezuelans are significantly less likely to have shared scams than users from Peru, Colombia, and Ecuador.

This presents an interesting juxtaposition to the crime narrative-based xenophobia against Venezuelan migrants that is so present in Colombia (even in field interviews with Venezuelan migrants, they also shared narratives where Venezuelan migrants were disproportionately responsible for criminality). In our collection of public WhatsApp groups, Colombian users, not Venezuelans, are significantly more likely to be the ones sharing scams! Of course, this comes with numerous disclaimers---many (or most) Colombian users may be Venezuelan migrants, users who share scams may be doing so unintentionally (perhaps as victims themselves), and so on---but it's certainly interesting that only knowing a user's country code, we should be much more wary of messages from Colombian users.

\begin{figure}[h]
\centering
\begin{subfigure}{0.45\textwidth}
  \centering
  \includegraphics[width=\textwidth]{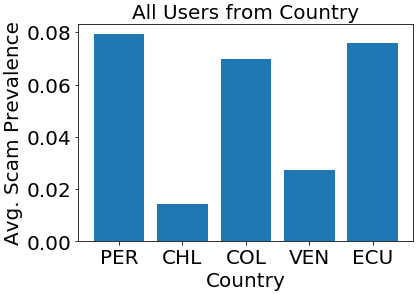}
  \caption{(ANOVA $p < 0.001$; $t$-tests between VEN and PER/COL/ECU, and between COL and CHL, $p < 0.001$.)}
  \label{figure:misinformation:bar_scam_country_prevalence}
\end{subfigure}%
\hfill
\begin{subfigure}{0.45\textwidth}
  \centering
  \includegraphics[width=\textwidth]{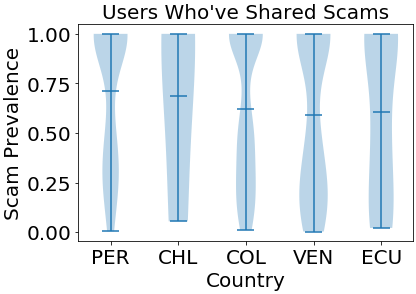}
  \caption{(No statistical significance.)}
  \label{figure:misinformation:violin_scam_country_prevalence_sharers}
\end{subfigure}
\caption{Cross-country comparisons of average prevalence of scams.}
\end{figure}

The same disparity exists when examining the average prevalence of scams for users from each country. On average, 7.0\% of each Colombian user's text content involves scams, compared to 2.7\% for Venezuelan users. Figure \ref{figure:misinformation:bar_scam_country_prevalence} plots these proportions by country; for the average Venezuelan user, fewer of their text messages are scams, compared to the average Peruvian, Colombian, and Ecuadorian users.

This difference, however, is almost entirely explained by the fact that a higher percentage of Colombians (and Peruvians/Ecuadorians) have shared scams compared to Venezuelans. The violin plot in figure \ref{figure:misinformation:violin_scam_country_prevalence_sharers} only includes users who've shared scams; for these users, scams make up roughly 60\% of their meaningful text content regardless of what country they're from.

\begin{figure}[h]
\centering
\begin{subfigure}{0.45\textwidth}
  \centering
  \includegraphics[width=\textwidth]{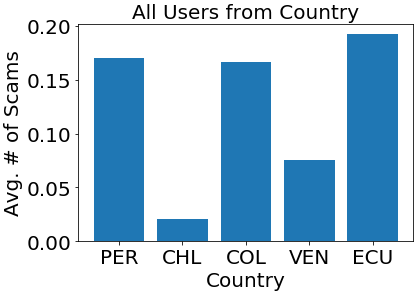}
  \caption{(ANOVA $p < 0.001$; $t$-tests between VEN and PER/COL, and between COL and CHL, $p < 0.01$.)}
  \label{figure:misinformation:bar_scam_country_frequency}
\end{subfigure}%
\hfill
\begin{subfigure}{0.45\textwidth}
  \centering
  \includegraphics[width=\textwidth]{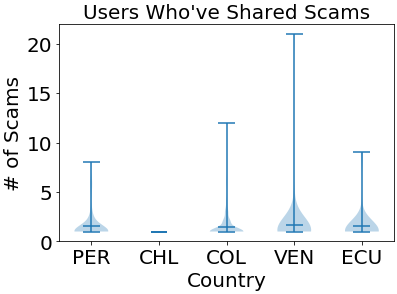}
  \caption{(No statistical significance.)}
  \label{figure:misinformation:violin_scam_country_frequency_sharers}
\end{subfigure}
\caption{Cross-country comparisons of average frequency of scams.}
\end{figure}

Finally, when looking at raw number of scams shared, Colombian users on average have shared 0.17 scams, compared to 0.08 scams for Venezuelan users ($p < 0.001$). Again, this difference was mostly accounted for by more Colombians (and Peruvians/Ecuadorians) having shared scams: when only considering users who've shared scams, users from each of the countries all shared 1.0-1.5 scams on average. The bar graph in figure \ref{figure:misinformation:bar_scam_country_frequency} and the violin plot in figure \ref{figure:misinformation:violin_scam_country_frequency_sharers} show these frequencies.

\subsection{Group Dynamics\label{ch:misinformation:scams:group}}

For each group, we again construct two measures for the prevalence of scams within that group: first, the proportion of meaningful text messages in that group that involve scams---message prevalence---and second, the proportion of users in that group who've shared scams---user prevalence.

\begin{figure}[h]
\centering
\begin{subfigure}{0.45\textwidth}
  \centering
  \includegraphics[width=\textwidth]{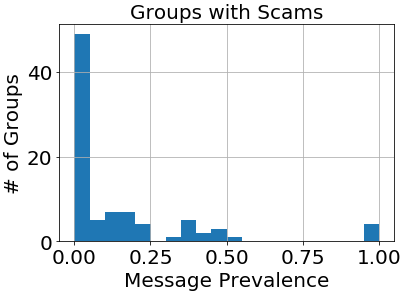}
  \caption{Histogram of message prevalence of scams.}
  \label{figure:misinformation:hist_scam_group_prevalence}
\end{subfigure}%
\hfill
\begin{subfigure}{0.45\textwidth}
  \centering
  \includegraphics[width=\textwidth]{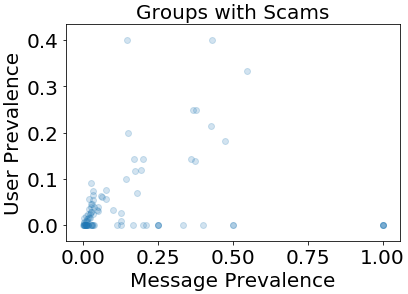}
  \caption{Scatter plot of message prevalence and user prevalence of scams.}
  \label{figure:misinformation:scatter_scam_group}
\end{subfigure}
\caption{Plots involving groups with scams.}
\end{figure}

Of 174 groups, only 49.4\% (86 groups) did not have any messages flagged as scams; this was significantly lower than the 112 groups where no fake news was shared. In the 88 groups where scams were shared, however, the message prevalence of scams was low: in 54 of these groups, less than 10\% of text messages consisted of scams; the histogram in figure \ref{figure:misinformation:hist_scam_group_prevalence} looks extremely similar to the histogram before of message prevalence of fake news. Considering both message prevalence and user prevalence, the prevalence of scams wasn't correlated to fake news prevalence within groups.

The scatter plot in figure \ref{figure:misinformation:scatter_scam_group} reveals a weak correlation between user prevalence and message prevalence of scams in groups (Pearson $r = 0.35$, $p < 0.001$). Of the seven groups where scams made up over 50\% of group text messages (i.e., message prevalence over 0.5), six were mostly inactive (with few messages over our months-long collection period), but one was a highly active internet money-making group that, as the topic suggests, was filled with scams.

\begin{figure}[h]
\centering
\includegraphics[width=0.45\textwidth]{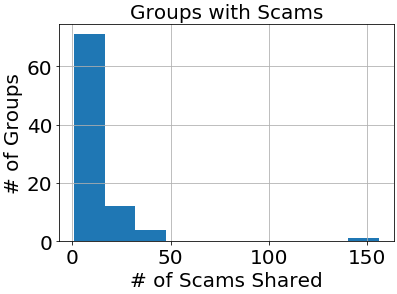}
\caption{A histogram of the number of scams sent in groups, for the 88 groups where scams were shared.}
\label{figure:misinformation:hist_scam_group_frequency}
\end{figure}

Finally, figure \ref{figure:misinformation:hist_scam_group_frequency} shows a histogram of the raw number of scams across groups with scams. The scam-filled group seen to the graph's right was a very active gaming group, where 156 scams were part of 1751 text messages.

Across all groups (including the 86 groups where no messages were flagged as scams), the message prevalence of scams was correlated with group entropy (geographic heterogeneity) and average group virality. An OLS regression of message prevalence on those factors is presented in table \ref{table:misinformation:ols_group_scam_message}.

\begin{table}[h]
\centering
\caption{OLS regression of scam message prevalence on group entropy and group virality; we include all groups, including groups where no scams were shared.}
\label{table:misinformation:ols_group_scam_message}
\begin{tabular}{p{0.25\textwidth} p{0.35\textwidth} p{0.10\textwidth} p{0.15\textwidth}}
\toprule
 & \textbf{Coefficient (Std. Err.)} & \textbf{$t$} & \textbf{P-Value} \\
\midrule

Intercept & $0.0750 (0.022)$ & $3.345$ & $0.001^{\ast}$ \\[0.2em]

Entropy & $0.0867 (0.023)$ & $3.726$ & $0.000^{\ast}$ \\[0.2em]

Virality & $-0.0863 (0.022)$ & $-3.855$ & $0.000^{\ast}$ \\[0.2em]
\midrule
\multicolumn{4}{c}{$n = 174$\,\,(171 d.f.) \quad\quad $R^2 = 0.104$}\\
\bottomrule
\end{tabular}
\end{table}

Higher entropy in groups (more geographic diversity) is weakly linked to higher message prevalence, which could be for a number of reasons: users may be less familiar with each other, or these groups are more centered on general/online themes (as opposed to groups for specific locations in Colombia, etc.), and so on. Higher scam message prevalence is also linked to lower virality within groups. This could be a result of less interaction in the group---scammers may be less afraid of being called out---though we previously showed that scam messages are generally much less viral.

Scam user prevalence was weakly negatively linked to Venezuelan user proportion (which should be clear from what we've already discussed), and weakly positively related to group inequality. The OLS coefficients are shown in table \ref{table:misinformation:ols_group_scam_user}.

\begin{table}[h]
\centering
\caption{OLS regression of groups' scam user prevalence on group proportion VZ and group Gini.}
\label{table:misinformation:ols_group_scam_user}
\begin{tabular}{p{0.25\textwidth} p{0.35\textwidth} p{0.10\textwidth} p{0.15\textwidth}}
\toprule
 & \textbf{Coefficient (Std. Err.)} & \textbf{$t$} & \textbf{P-Value} \\
\midrule

Intercept & $0.0012 (0.010)$ & $0.120$ & $0.904$ \\[0.2em]

Proportion VZ & $-0.0481 (0.019)$ & $-2.555$ & $0.011^{\ast}$ \\[0.2em]

Inequality/Gini & $0.0699 (0.018)$ & $3.950$ & $0.000^{\ast}$ \\[0.2em]
\midrule
\multicolumn{4}{c}{$n = 174$\,\,(171 d.f.) \quad\quad $R^2 = 0.104$}\\
\bottomrule
\end{tabular}
\end{table}

The relationship with Gini/inequality is opposite that found with fake news \textit{message} prevalence (where message prevalence of fake news decreased with inequality), but message prevalence and user prevalance are different characteristics.

In particular, higher inequality means more strangers in the group: these strangers can make fake news messages less likely (say, by potentially calling out fake news, or by bringing in alternate perspectives), but can also mean that more users are sharing scams (perhaps these very strangers are sharing scams). Still, these relationships are weak across both fake news and scams, so we don't discount the possibility of these simply being spurious coefficients.

We now only examine groups where scams were shared to obtain stronger effects. Message prevalence of scams was correlated with size, activity, degree, concentration, inequality, and virality; results from our kitchen sink regression are shown in table \ref{table:misinformation:ols_group_scam_drop_message}. Remarkably, these results are exceedingly similar to those for fake news message prevalence: again, concentration and group inequality are significant, concentration in the positive direction, and inequality in the negative direction.

\begin{table}[h]
\centering
\caption{OLS regression of groups' scam message prevalence on group size, group activity, group degree, group concentration, group virality, and group inequality. Only across groups where scams were shared.}
\label{table:misinformation:ols_group_scam_drop_message}
\begin{tabular}{p{0.25\textwidth} p{0.35\textwidth} p{0.10\textwidth} p{0.15\textwidth}}
\toprule
 & \textbf{Coefficient (Std. Err.)} & \textbf{$t$} & \textbf{P-Value} \\
\midrule

Intercept & $0.5147 (0.083)$ & $6.232$ & $0.000^{\ast}$ \\[0.2em]

Size & $0.0003 (0.000)$ & $1.105$ & $0.272$ \\[0.2em]

Activity & $0.0000 (0.000)$ & $0.459$ & $0.647$ \\[0.2em]

Degree & $0.0003 (0.001)$ & $0.379$ & $0.706$ \\[0.2em]

(H-H) Concentration & $0.4887 (0.156)$ & $3.140$ & $0.002^{\ast}$ \\[0.2em]

Inequality/Gini & $-0.7290 (0.137)$ & $-5.307$ & $0.000^{\ast}$ \\[0.2em]

Virality & $-0.0363 (0.070)$ & $-0.520$ & $0.605$ \\[0.2em]
\midrule
\multicolumn{4}{c}{$n = 88$\,\,(81 d.f.) \quad\quad $R^2 = 0.489$}\\
\bottomrule
\end{tabular}
\end{table}

The high positive coefficient on concentration is a bit surprising, since ``echo chambers'' are less applicable in this case. We might imagine, however, that if users who mostly share scams are behind the concentration---as they are in the aforementioned internet money-making group, and perhaps other groups---concentration breeds greater scam prevalence. As before, higher inequality likely means that ``strangers'' (message-poor users) will call out messages, or that users will pay more attention before forwarding on scams.

\subsection{Variants of Scams\label{ch:misinformation:scams:variants}}

Like with fake news, scams are altered as they're shared, though likely more insiduously than fake news. Consider, for example the following message, which purports to offer a loan:

\singlespacing\begin{lstlisting}
Buenos dias .  Para todos aquellos que necesitan prestamos de dinero, el servicio de prestamos lo ayudara al ayudarlo en varias areas de prestamos de dinero.  Para la comunicacion
whatsapp: +229 636 963 16
\end{lstlisting}\doublespacing

and:

\singlespacing\begin{lstlisting}
Buenos dias .  Para todos aquellos que necesitan prestamos de dinero, el servicio de prestamos lo ayudara al ayudarlo en varias areas de prestamos de dinero.  Para la comunicacion
whatsapp: +22 963 696 316
\end{lstlisting}\doublespacing

The formatting of the number has been slightly changed in the second message, likely to disguise the country code (the country code +229 is from Benin, in West Africa, while a +22 country code---which doesn't exist, since country codes must be instaneous $\Leftrightarrow$ no code is a prefix of some other code---might seem European).

Or take the following message, also a fishy loan offer:

\singlespacing\begin{lstlisting}
OFERTA DE PReSTAMO DE DINERO
 Somos una empresa que ofrece prestamos para la vivienda, prestamos de inversion, prestamos para automoviles, prestamos personales que van desde  4,000 [Euros] a  1,000,000 [Euros] con una tasa de interes del 3% sobre capital a corto y largo plazo. Si estas interesado contactanos por whatsapp: +33752534155
\end{lstlisting}\doublespacing

In one variant, the heading has been slightly modified (from ``LOAN OFFER'' to ``We offer the loan''), amounts in the messages are different (and the currency was even changed from Euros to Kuwaiti dinars?!), and an additional sentence was added:

\singlespacing\begin{lstlisting}
Ofrecer el prestamo
  Somos una empresa que ofrece prestamos para vivienda, prestamos de inversion, prestamos para automoviles, prestamos personales que van desde 5,000 hasta 1,000,000 de dinares kuwaities con una tasa de interes del 3% sobre capital a corto y largo plazo.
  Con este prestamo, puede restaurar completamente su hogar, pagar sus impuestos y contribuir a sus necesidades personales y familiares.  Si esta interesado, contactenos a traves de WhatsApp: +33752534155
\end{lstlisting}\doublespacing

With 247 unique scam messages, we again use cosine similarity (with manually-tuned limit again 0.8) to find 105 that are variants on other scams. After merging, we end up with 142 unique scams; all but one were shared fewer than 50 times. But this message, which purports to offer free mobile data, involved 26 variants (!), which were shared 116 times (!) by 77 unique users (!) across 40 unique groups (!):

\singlespacing\begin{lstlisting}
100 GB de datos de Internet sin ninguna recarga
Obtenga 100 GB de datos de Internet gratis en cualquier red movil durante 60 dias.
Consiguelo ahora \nhttps://internet4goffers.com/es
\end{lstlisting}\doublespacing

Another message involved 29 variants (!), and purports to offer free cash transfers (from the UN and an unnamed ``government''):

\singlespacing\begin{lstlisting}
La OMS y el Gobierno han destinado un BONO de dinero para todos los paises por Motivo de CUARENTENA (CORONA VIRUS)
Obtenga su BONO gratis en cualquier pais.
Consiguelo ahora AQUI
https://bit.ly/Bono-Comida-8
\end{lstlisting}\doublespacing

\begin{figure}[h]
\centering
\includegraphics[width=0.45\textwidth]{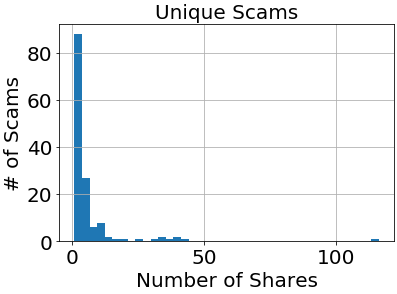}
\caption{Histogram of the number of times each unique scam was shared.}
\label{figure:misinformation:hist_scam_unique_frequency}
\end{figure}

\pagebreak
Figure \ref{figure:misinformation:hist_scam_unique_frequency} gives a histogram of how many times each of the 142 unique scams were shared.

\begin{figure}[h]
\centering
\begin{subfigure}{0.45\textwidth}
  \centering
  \includegraphics[width=\textwidth]{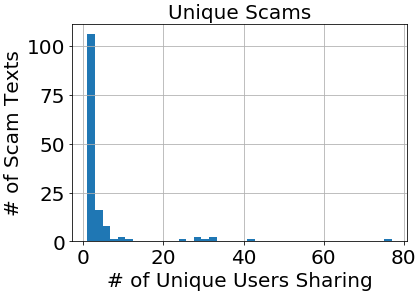}
\end{subfigure}%
\hfill
\begin{subfigure}{0.45\textwidth}
  \centering
  \includegraphics[width=\textwidth]{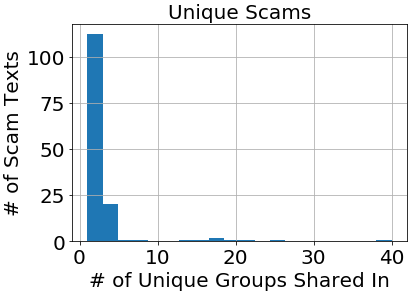}
\end{subfigure}
\caption{Histograms of how many different users shared each unique scam, and in how many different groups each unique scam was shared.}
\label{figure:misinformation:hist_scam_unique_user_group}
\end{figure}

Finally, the histograms in figure \ref{figure:misinformation:hist_scam_unique_user_group} show that of all unique scams, the vast majority were shared by five or fewer users, and in five or fewer groups (just like with fake news).

For each unique scam, we calculated the average number of shares per user, and the average number of shares in each group. Across all scams, these averaged to 2.52 and 2.64 respectively, significantly higher than we found for fake news. Clearly, there seems to be greater intent and greater maliciousness behind the sharing of scams.

\section{Detecting Scams with Machine Learning\label{ch:misinformation:detecting}}

Given that many scams rely on the same nuances (like offering something for free or conveying a sense of urgency), it may be possible to automatically flag scam messages. In this section, we put various machine learning classifiers to this task, using certain characteristics of scams based on our results from Chapter \ref{ch:misinformation:scams}:
\begin{enumerate}
  \item Text only (tokens)
  \item Tokens and message length
  \item Tokens and user country code
  \item Tokens and group dynamics (concentration and inequality)
\end{enumerate}

We proceed with the same labeled data as before, and only work with messages with 5-word tokens or longer, leaving us 44,025 messages with 886 labeled scams. Immediately, we perform an 80-20 train-test chronological split, giving us a size 35,225 overall training set, and a 8,800 sample overall test set.\footnote{Note that our methodology does not perfectly exclude test data from processing: previously, to find and label scams, we had manually verified messages that were shared identically thrice or more, including during the test set period. If a message was only shared twice in the training time period but twice more in the test set period, we still manually reviewed and labeled it. This is an extremely minor violation, since we could've avoided it with simply more manual labor.}

Within our overall training set, we create four cross-validation folds by ``forward chaining,'' which ensures that in each fold we never train on data after the beginning of that fold's test. Specifically, we chronologically split our training set into five equal sets, then: train on $\{1\}$ and test on 2 (fold 1), train on $\{1, 2\}$ and test on 3 (fold 2), train on $\{1, 2, 3\}$ and test on 4 (fold 3), and train on $\{1, 2, 3, 4\}$ and test on 5 (fold 4). Clearly, we should give higher credence to performance in folds 3 and 4, since training set size in those folds nears the actual training set size.

Given that 98\% of our data are true negatives, overall accuracy is a poor measure of performance here; indeed, any metric with true negatives in the denominator will be uselessly close to 0 or 1 (in particular, just classifying everything as ``not-scam'' results in an outstanding 98\% accuracy with an amazing 0\% false positive rate). We settle for recall, $\frac{TP}{TP + FN}$, which measures our detection rate of actual scams, and precision, $\frac{TP}{TP + FP}$, which measures how precise our positive (scam) prediction is.

\subsection{Text Only}

We proceed with five well-known classifiers: logistic regression, SVM, nearest neighbors, decision trees, and random forest. Although applications of Naive Bayes classifiers to the spam detection problem are well-known, here we discard those classifiers because they assume strong independence between features. In our dataset, the relationships between tokens \textit{do matter}---``free'' and ``http'' each become much more suspicious in combination.

Given our extremely high-dimensional feature space (with 42,904 tokens), we test each classifier with various regularization parameters. The first two classifiers are both linear, which may be a setback given our circumstances: some tokens are likely to be red flags for scams regardless of context (e.g., ``free'' or ``loan''), but some tokens only become red flags in combination. Consider, for example, the tokens ``United Nations,'' ``assistance'' and ``http'': none of these tokens by themself is very suspicious (indeed, most messages about the UN are probably news-related), but in combination, these immediately signal the assistance/cash-transfer scams we discussed earlier.

\begin{figure}[h]
\centering
\includegraphics[width=\textwidth]{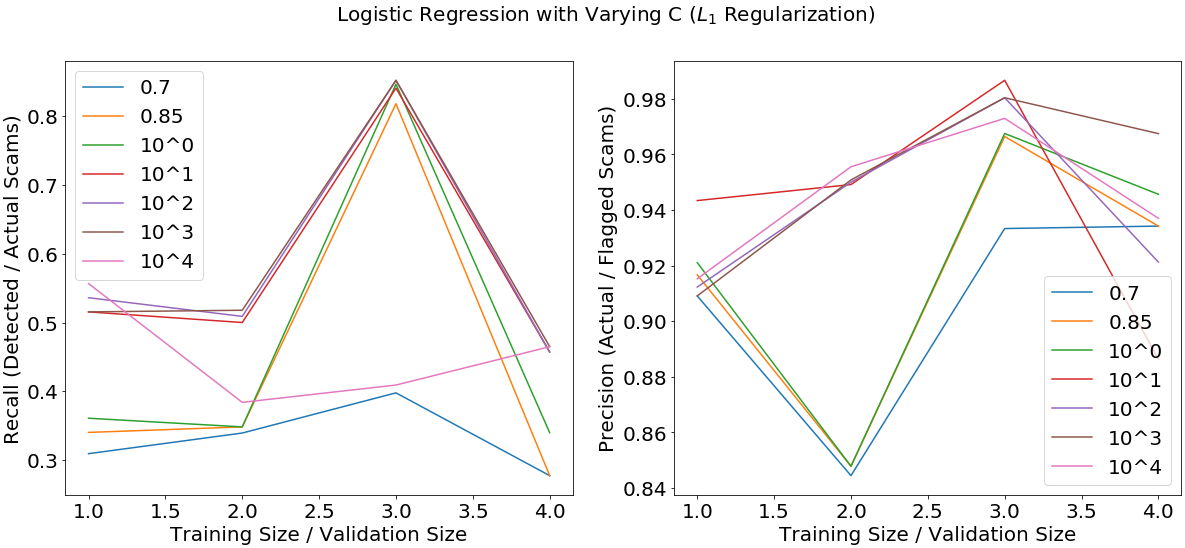}
\caption{Performance of $L_1$-penalty logistic regressions.}
\label{figure:misinformation:ml_log_l1}
\end{figure}

\begin{figure}[h]
\centering
\includegraphics[width=\textwidth]{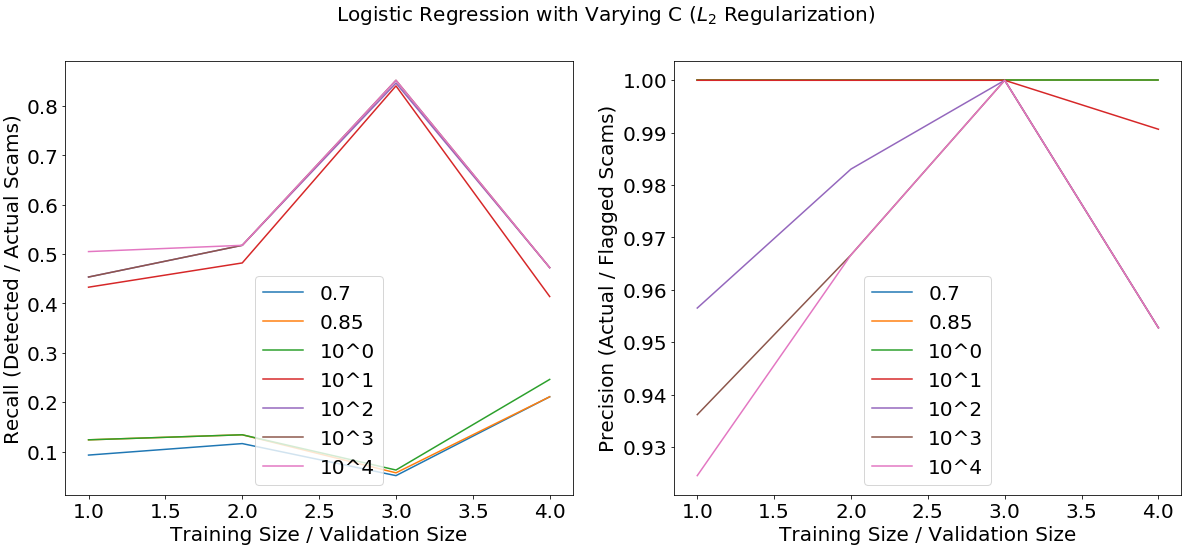}
\caption{Performance of $L_2$-penalty logistic regressions.}
\label{figure:misinformation:ml_log}
\end{figure}

\pagebreak
The plots in figures \ref{figure:misinformation:ml_log_l1} and \ref{figure:misinformation:ml_log} show the performance of $L_1$ and $L_2$-regularized logistic regression, respectively, with recall (detection rate of actual scams) in the left graph and precision (quality of positive predictions) in the right graph. In each plot, the different lines represent different regularization parameters $C$ ($C$ is the inverse of the conventional regularization penalty $\lambda$, so a low $C$ means stronger regularization).

In the $L_2$ regularization, we notice that $C \leq 1$ quickly brings coefficients in the logistic regression to 0, resulting in very low recall (but near-perfect precision, since we're not flagging anything).

$L_1$-regularization, which more likely yields sparse solutions given the shape of the 1-norm unit ball, results in around the same recall as the $L_2$ penalty, but worse precision: more of the predicted scams turn out to be false positives. We might imagine that this arises simply from having to consider fewer tokens. Imagine, for example, that hyperlinks are generally suspicious, but links ending in ``.gov'' are generally legitimate: if there are few .gov links, the $L_1$ penalty might only assign a positive coefficient to the token ``http'', while the $L_2$ penalty could assign a positive coefficient on ``http'' and a negative coefficient ``gov'' for the same cost.

\begin{figure}[h]
\centering
\includegraphics[width=\textwidth]{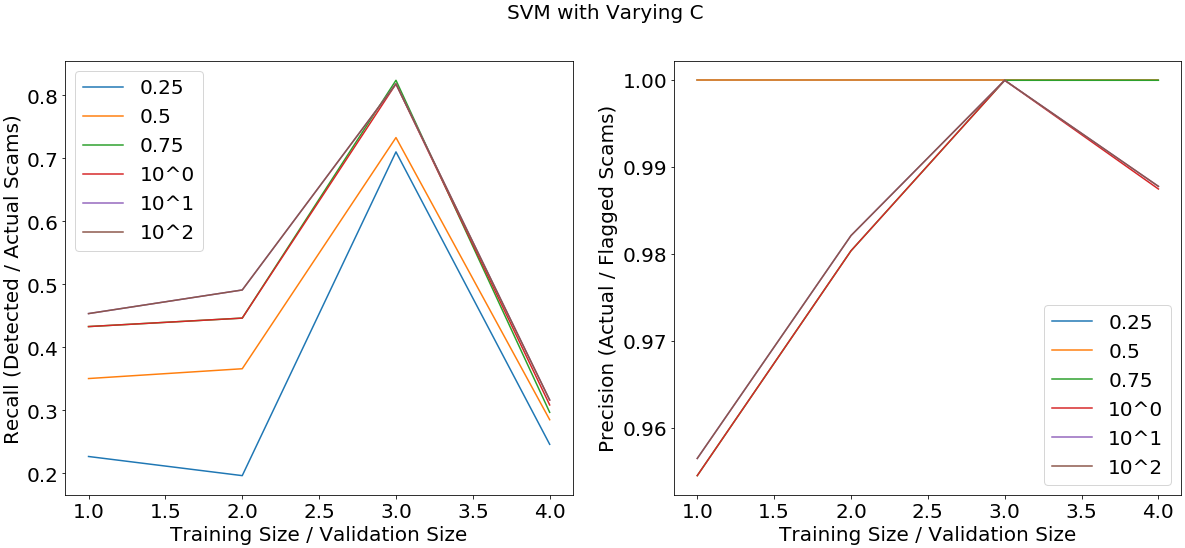}
\caption{Performance of SVM classifiers.}
\label{figure:misinformation:ml_svm}
\end{figure}

Next, in figure \ref{figure:misinformation:ml_svm}, we show recall and performance from SVM classifiers with varying regularization. With $C \geq 0.75$, the classifiers begin to converge, with quite similar performance, SVM the classifiers have worse recall than logistic regression.

Precision of SVMs is generally better than in logistic regression, but this is simply the recall-precision trade off (in predicting fewer scams overall). Given our context, it's more important to prioritize recall---being able to identify more scams---even at the cost of false positives (especially since the false positive rate is still exceedingly low in general, given that 98\% of messages aren't scams).

\begin{figure}[h]
\centering
\includegraphics[width=\textwidth]{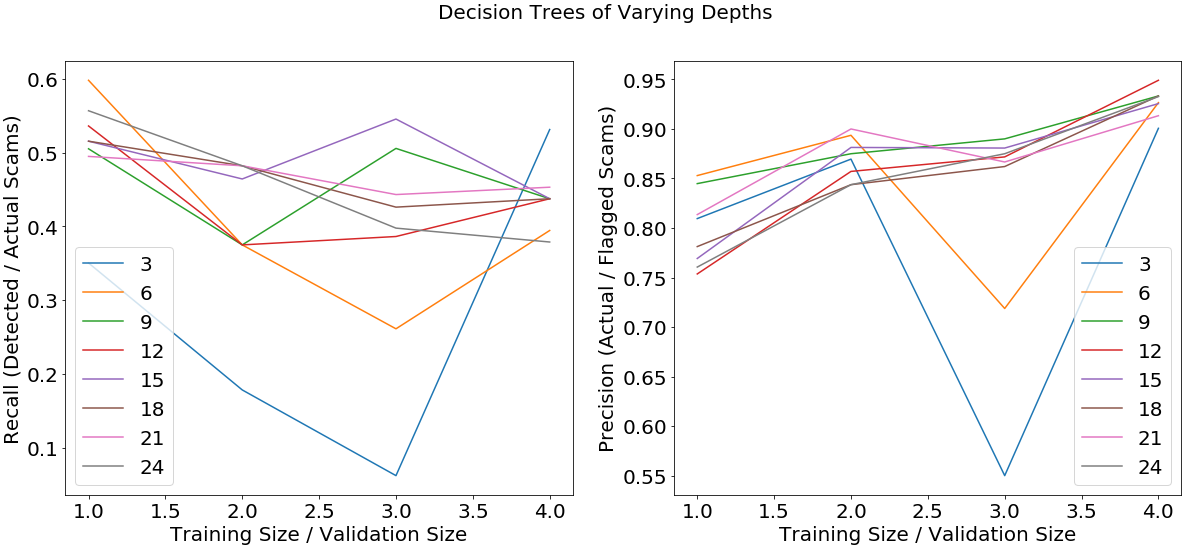}
\caption{Performance of decision tree classifiers.}
\label{figure:misinformation:ml_dt}
\end{figure}

Figure \ref{figure:misinformation:ml_dt} plots performance for decision trees of varying depths. There seem to be little gains, and substantial risk of overfitting, once the depth of a decision tree is 12 or so; these trees seem to perform equally well as $L_2$-penalized logistic regression, though noise with our small sample size makes comparison difficult.

\begin{figure}[h]
\centering
\includegraphics[width=\textwidth]{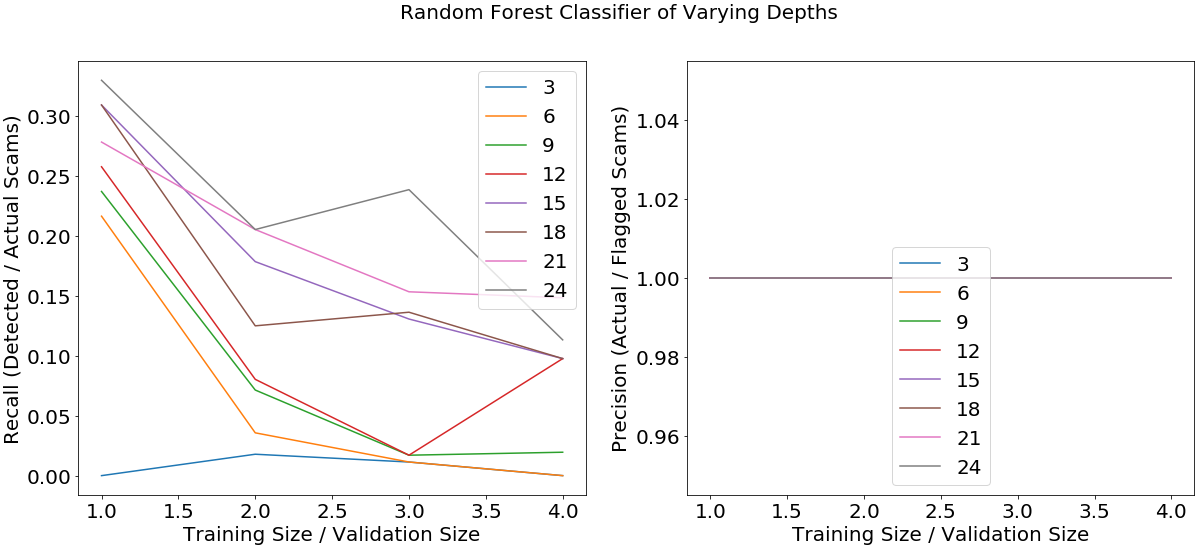}
\caption{Performance of random forest classifiers (100 trees).}
\label{figure:misinformation:ml_rf}
\end{figure}

Figure \ref{figure:misinformation:ml_rf} plots performance for random forests of varying depth.

Random forests have much worse performance decision trees (in particular, the 1.00 precision signals that our classifier flags very few scams), and we might attribute this to the bootstrap sampling involved in fitting each decision tree within the random forest. Because there are relatively few positives (scams) in our dataset, comprising only 2\% of messages, bootstrap sampling is likely to leave out important training examples altogether, and consistently do this across trees.

\begin{figure}[h]
\centering
\includegraphics[width=\textwidth]{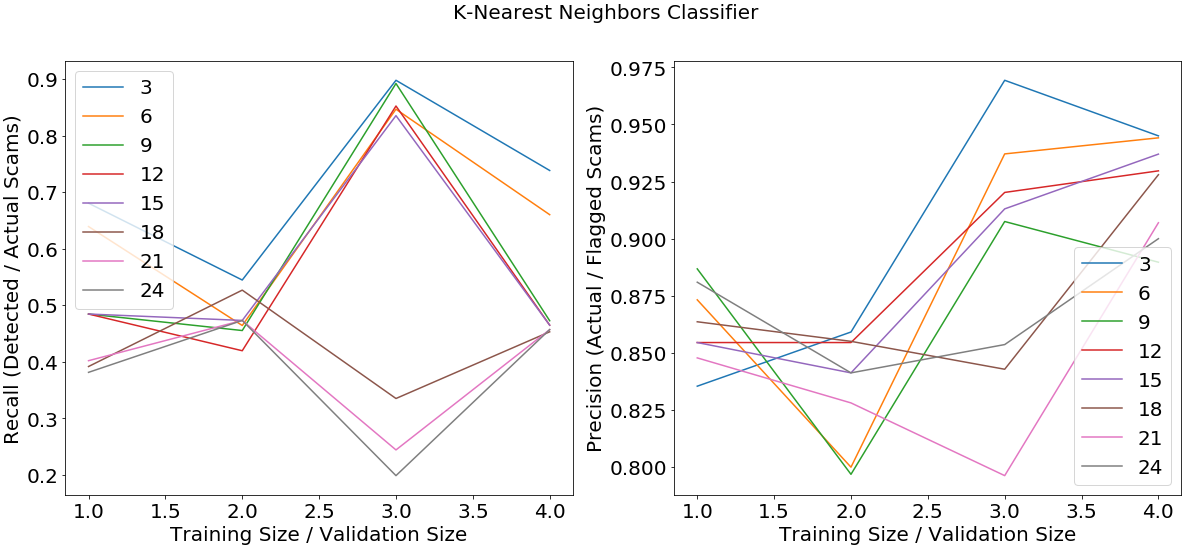}
\caption{Performance of $k$-nearest neighbors classifiers.}
\label{figure:misinformation:ml_knn}
\end{figure}

Finally, in figure \ref{figure:misinformation:ml_knn} we plot the performance of $k$-nearest neighbors classifiers, which seem to do substantially better than our other classifiers. This shouldn't surprise us, at all: because scams develop so many variants over time, a close match on some tokens to a known scam should be an immediate red flag (this, after all, was our motivation for using cosine similarity to label scams and then to merge variants).

\subsection{Text and Message Dynamics}

Given our results in Section \ref{ch:misinformation:scams:message} (where we studied the message properties of scams), we add text length (in number of words, normalized) as a feature. Scams involved nearly 20\% fewer words on average than other meaningful text messages. Given poor performance in the previous section of random forests and $L_1$ logistic regression, we only focus in this section on $L_2$ logistic regression, SVM, decision tree, and $k$-nearest neighbors classifiers.

\begin{figure}[h]
\centering
\includegraphics[width=\textwidth]{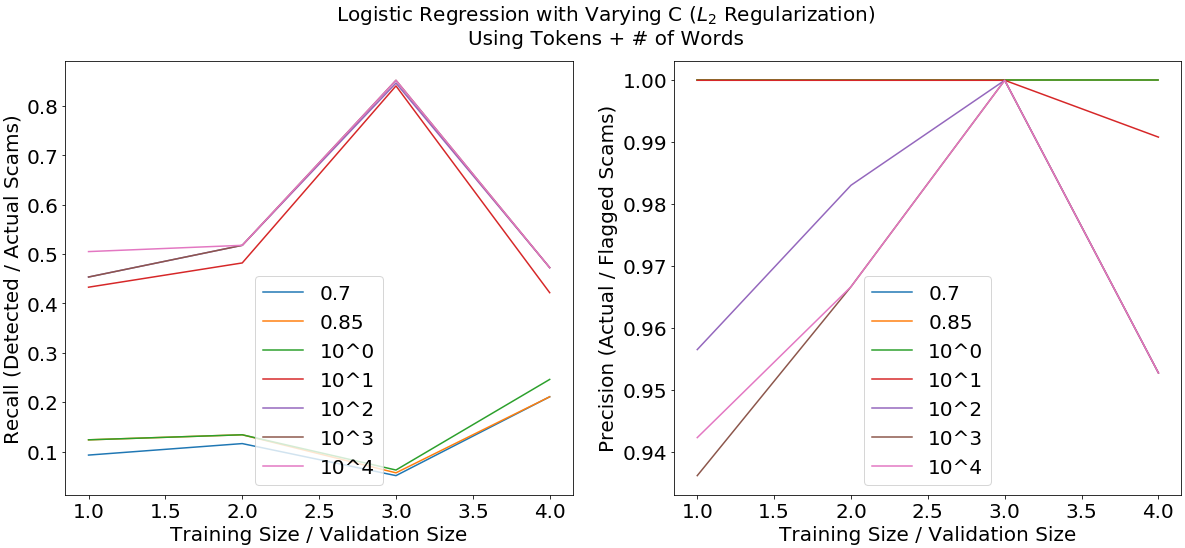}
\caption{Performance of $L_2$-penalty logistic regressions, using message length.}
\label{figure:misinformation:ml_log_wl}
\end{figure}

The performance of $L_2$-regularized logistic regression (hereafter, just ``logistic regression'') is presented in figure \ref{figure:misinformation:ml_log_wl}. There appears to be no difference in performance from before, which makes sense since the importance of message length differs based on the token context. Under specific scenarios, say when receiving a message that includes tokens about the United Nations, message length might be extremely important---shorter messages are likely scams, while longer messages about the UN are likely news. But because logistic regression is linear in the feature space, it can't incorporate these non-linear nuances. SVM with word length also performs similarly as SVM with only tokens (figure omitted).

\begin{figure}[h]
\centering
\includegraphics[width=\textwidth]{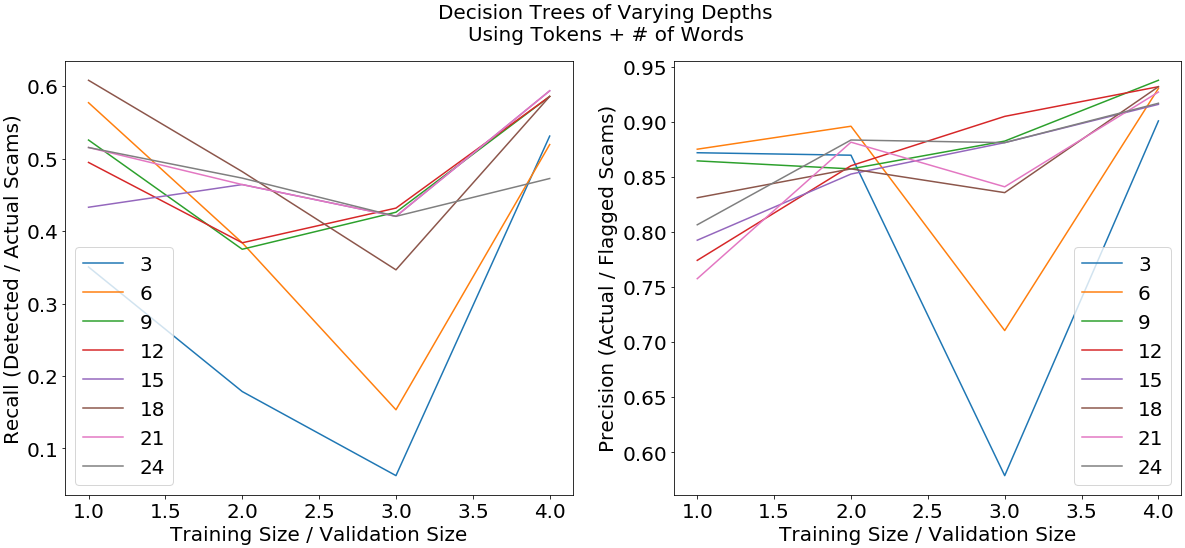}
\caption{Performance of decision tree classifiers, using message length.}
\label{figure:misinformation:ml_dt_wl}
\end{figure}

\pagebreak
Decision trees perform substantially better when incorporating word length, likely for the reason we just discussed, where message length becomes important in certain scenarios. In figure \ref{figure:misinformation:ml_dt_wl}, recall is around 10-20\% higher than before, on average! This improvement comes without any significant loss in precision.

\begin{figure}[h]
\centering
\includegraphics[width=\textwidth]{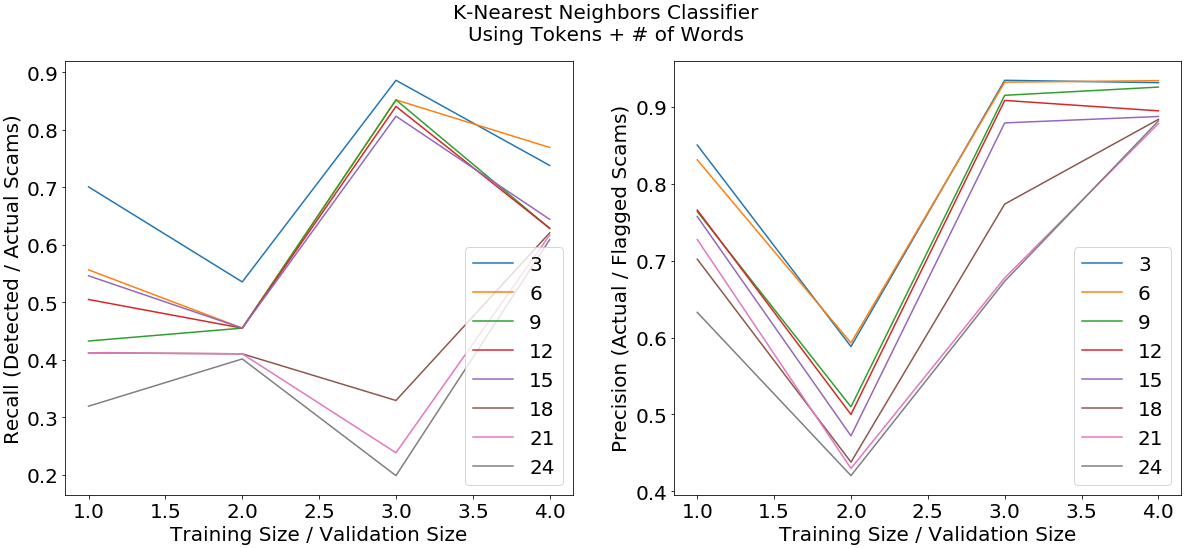}
\caption{Performance of $k$-nearest neighbors classifiers, using message length.}
\label{figure:misinformation:ml_knn_wl}
\end{figure}

Performance for $k$-neighbors classifiers using message length is shown in figure \ref{figure:misinformation:ml_knn_wl}. There is little improvement, likely because word counts only matter in specific contexts---and those contexts are already accounted for by the neighbors matching.

\subsection{Text and User Dynamics}

Now, alongside tokens, we incorporate information about senders, but only whether they have a VZ or CO country code (or neither). A more sophisticated approach would surely take into account specific telephone numbers, but that would quickly converge towards simply blacklisting certain users, given what we saw in section \ref{ch:misinformation:scams:user}; we leave this strategy out of our analysis, since it would lead to overestimates of our scam detection performance in more general contexts.

\begin{figure}[h]
\centering
\includegraphics[width=\textwidth]{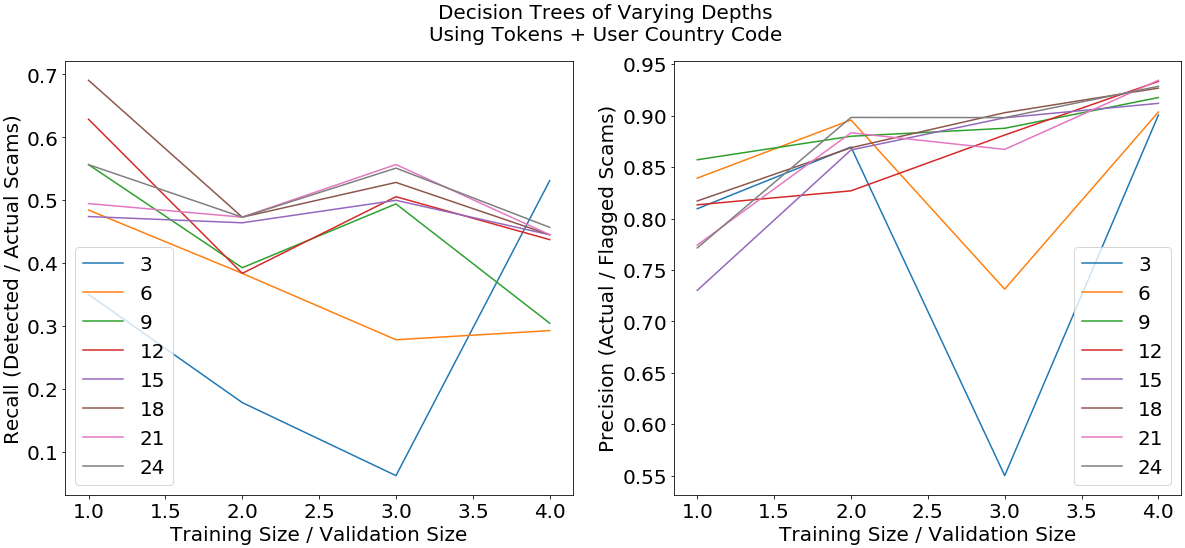}
\caption{Performance of decision tree classifiers, using user country code.}
\label{figure:misinformation:ml_dt_cc}
\end{figure}

As with incorporating message length, there are no substantial improvements in either logistic regression or SVM, because of what we discussed earlier. Performance of decision trees with message length, shown in figure \ref{figure:misinformation:ml_dt_cc}, is slightly better than decision trees on tokens only (and around the same as decision trees with message length). With user country code, the $k$-nearest neighbors classifier performs the same, if not worse (figure omitted).

\subsection{Text and Group Dynamics}

Finally, we incorporate measures of group concentration and inequality,\footnote{There is a bit of cheating from the test set here, since in chapter \ref{ch:misinformation:scams:group} we used the entire dataset to determine that group concentration and inequality were linked with scam prevalence. But that's an extremely minor violation, since it's not like we borrow coefficients or anything.\\\\A larger issue might be that we use group concentration and inequality as calculated across our entire dataset (for both the training feature vectors and the test feature vectors). This is still a relatively small violation, and can be completely ignored if we assume that these are permanent underlying characteristics of a group (that can be perfectly sampled), which seems fine.} which we found to have statistically significant relations (in opposite directions) with scam message prevalence. As with users, blacklisting/whitelisting certain groups might make more sense for an actual scam detection algorithm, but here we attempt to determine if more abstract group characteristics can be applied towards scam detection.

\begin{figure}[h]
\centering
\includegraphics[width=\textwidth]{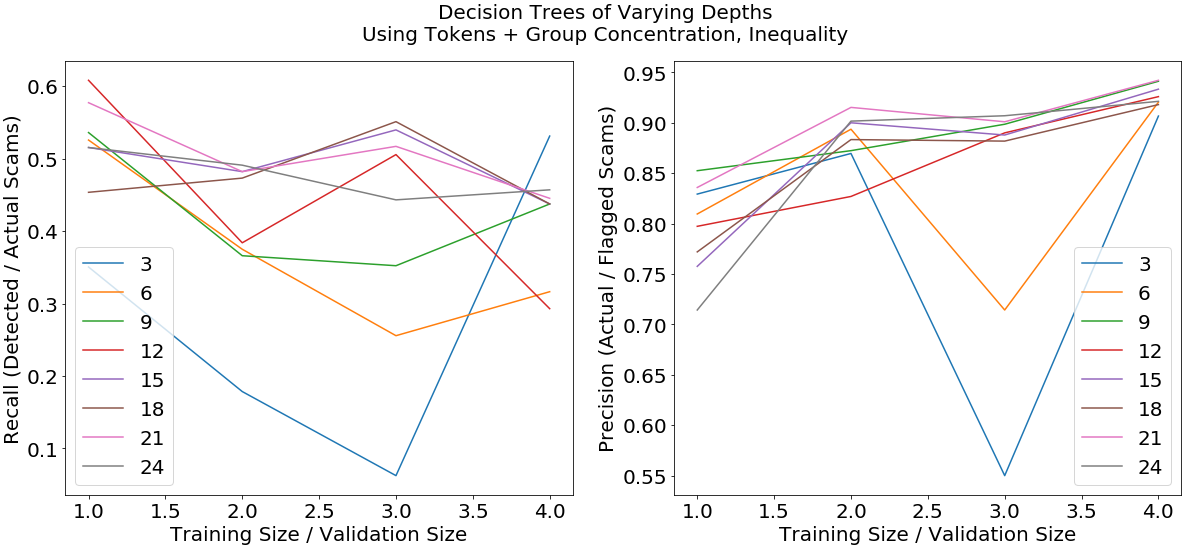}
\caption{Performance of decision tree classifiers, using group concentration and inequality.}
\label{figure:misinformation:ml_dt_group}
\end{figure}

Unsurprisingly, logistic regression and SVM again perform no better than using only tokens (though this conclusion isn't necessarily obvious, because group dynamics may linearly interact with message content, whereas message length likely has non-linear interactions). Performance of decision trees is slightly better than the baseline decision tree classifier, as shown in figure \ref{figure:misinformation:ml_dt_group}. The performance of the $k$-nearest neighbors classifiers remains unchanged (figure omitted).

\subsection{Test Results}

Given these findings, we lean towards using a $k$-nearest neighbors classifier with three neighbors and only text tokens. We compare this with another decent classifer, a 12-level decision tree that incorporates text tokens, user country code (more specifically, whether they're CO or VZ), and message length (in words).

The confusion matrix for the 3-nearest neighbors classifier is $\begin{bmatrix} TP = 103 & FP = 11 \\ FN = 49 & TN = 8637 \end{bmatrix}$, giving recall 67.8\% (not bad!) and precision 90.4\%.

\begin{figure}[h]
\centering
\includegraphics[width=\textwidth]{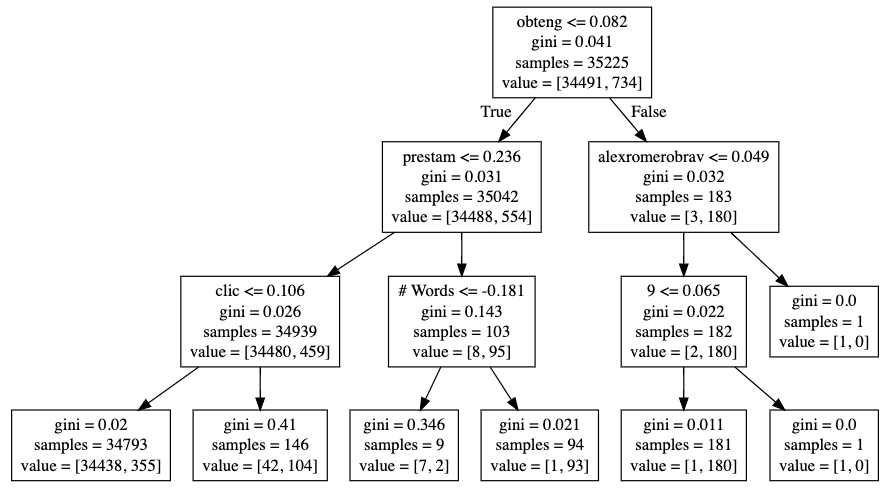}
\caption{3-level decision tree fit on training data.}
\label{figure:misinformation:tree}
\end{figure}

The confusion matrix for our decision tree is $\begin{bmatrix} TP = 58 & FP = 7 \\ FN = 94 & TN = 8641 \end{bmatrix}$, which is clearly much worse, at recall 38.2\% and precision 89.2\%. To briefly provide some interpretability to our decision tree, in figure \ref{figure:misinformation:tree} we show a 3-level decision tree (which, to be clear, has much worse performance).

The feature for number of words appears in the second level; important tokens in the decision tree include ``obteng'' (obtain), ``prestam'' (loan), and ``click'' (click here).

\chapter{Coronavirus\label{ch:coronavirus}}

\begin{figure}[h]
  \centering
  \includegraphics[width=0.8\textwidth]{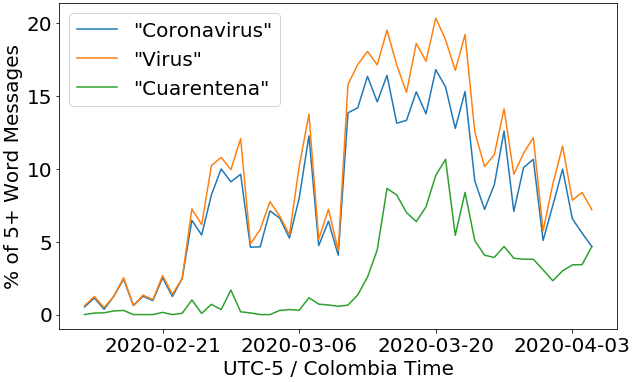}
  \caption{Plot of keyword popularity amongst all text messages with 5+ words.}
  \label{figure:coronavirus:5word_proportion}
\end{figure}

The 2019-20 coronavirus (COVID-19) pandemic began in December 2019, in Wuhan, China, but spread worldwide in the early months of 2020; the World Health Organization recognized it as a pandemic on March 11, 2020. Cases began appearing in Colombia in early March \cite{minsalud-20200306}, while Venezuela reported its first two cases on March 13 \cite{reuters-confirm-20200313}. The effects of the coronavirus pandemic are as visible in public WhatsApp groups as they are on Bogotá's empty streets. Figure \ref{figure:coronavirus:5word_proportion} shows the daily proportion of all 5+ word text messages that mention coronavirus, virus, and quarantine.

No matter what topics groups were originally cented on, discussions came to include the virus itself, its worldwide spread, and its socioeconomic effects in Colombia and Venezuela. In Chapter \ref{ch:misinformation}, we've already seen how widespread coronavirus-related misinformation is: fake news about cures, fake news about cases, and internet scams offering food aid or free mobile data in this especially difficult time. Even religious chain messages have come to center on the coronavirus---take the following snippet, a Biblical allusion passed around in multiple groups (we share the English translation):

\singlespacing\begin{lstlisting}
"WHAT WOULD YOU DO IN THIS CASE? [...]
Everyone hears the news: 2 women have died in New York. Within hours, the disease seems to be invading everyone. Scientists are still working to find the antidote, but nothing works. And suddenly, the expected news comes: The DNA code of the virus has been deciphered. You can make the antidote. It is going to require the blood of someone who has not been infected and in fact throughout the country the word is spread that everyone should go to the nearest hospital for a blood test. [...]
Suddenly the doctor comes out yelling a name he has read in the registry. The smallest of your children is next to you, grabs your jacket and says: Daddy? That's my name! [...] The older doctor comes up to you and says: Thank you, sir, your son's blood is perfect, it is clean and pure, you can make the antidote to this disease ... The news spreads everywhere, people are praying and crying with happiness.
Then the doctor approaches you and your wife and says: Can we talk for a moment? We did not know that the donor would be a child and we need you to sign this form to give us permission to use their blood. When you are reading the document you realize that they are not putting the amount they will need and you ask: How much blood? ... The doctor's smile disappears and he replies: We did not think it would be a child. We were not prepared. We need it all ... You can't believe it and you try to answer: 'But, but ... '. The doctor keeps insisting on you, 'You don't understand, we are talking about the cure for everyone. Please sign, we need it ... "
\end{lstlisting}\doublespacing

In this chapter, we analyze two aspects of the coronavirus pandemic that we're uniquely poised to investigate: interest in \textit{trochas}, illegal border crossings which surged in popularity after Colombia shut down its borders, and the effects of Colombia's mandatory nationwide quarantine.

\section{Proliferation of \textit{Trocha} Crossings}

On Friday, March 13, Venezuela reported its first two cases of coronavirus; the same day, President Iván Duque of Colombia began restricting entry for visitors from Europe and Asia, and announced a closing of all border crossings with Venezuela \cite{eltiempo-cerrar-20200314}.

Many have criticized the Colombian border shutdown, given the vulnerabilities of Venezuelans amidst their country's collapsed health system \cite{devex-20200317}. Moreover, near the two largest crossings, in Cúcuta and Maicao, are hundreds to thousands of \textit{trochas}, irregular border crossings controlled by criminal organizations and paramilitaries, who require payment for passage, and often rob and/or assault migrants. The belief of regional experts, and many along the border, is that shutting down the official border directly means increasing \textit{trocha} crossings \cite{devex-20200317}.

\begin{figure}[h]
  \centering
  \includegraphics[width=\textwidth]{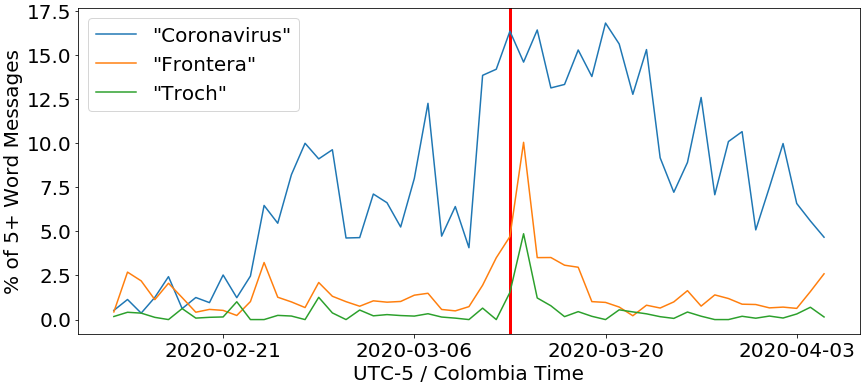}
  \caption{Plot of keyword popularity for border-related topics amongst all text messages with 5+ words. ``Frontera'' is border, and ``troch'' refers to the illegal crossings. The red line indicates March 13, 2020, when Colombia decided to close its land border with Venezuela.}
  \label{figure:coronavirus:5word_proportion_border}
\end{figure}

Figure \ref{figure:coronavirus:5word_proportion_border} plots the popularity of ``frontera'' (border) and ``trocha'' keywords amongst all 5+ word text messages. Both peak in popularity just after March 13, when Colombia announced its border shutdown (which took effect at 5:00 AM local time the next day).

On March 14, nearly 5\% of all text messages discuss trochas, and over 10\% of all text messages involve the border. Of all 5+ word text messages sent before March 13, only around 0.2\% are related to trochas, and only 1.2\% are related to the border. We can look more generally at March 13-15: 2.4\% of all text messages sent involve trochas, compared to 0.3\% of text messages outside this period ($p < 0.001$); 5.8\% of messages between March 13-15 discuss fronteras, compared to 1.3\% of messages outside this period ($p < 0.001$).

\begin{figure}[h]
  \centering
  \includegraphics[width=\textwidth]{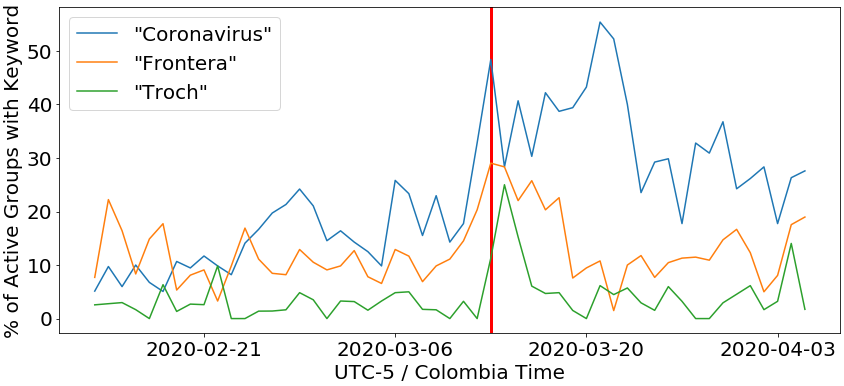}
  \caption{Plot of popularity of border-related topics (\% of active groups where keyword is mentioned that day). ``Frontera'' is border, and ``troch'' refers to the illegal crossings. The red line indicates March 13, 2020, when Colombia decided to close its land border with Venezuela.}
  \label{figure:coronavirus:group_proportion}
\end{figure}

This effect is even larger when analyzing the groups these topics are discussed in. For each day we collected day, we consider groups with messages from that day. Figure \ref{figure:coronavirus:group_proportion} displays what proportion of such groups include mentions of our keywords: on March 14, 25\% of active groups are discussing trochas!

Given our previous exploration of group characteristics, we hypothesize that discussion of trochas is likely linked to group properties like entropy, which estimates how transnational the group is, and proportion of VZ members. We set a dummy variable for groups that discuss trochas starting from March 13, the day the border closure was announced ($n = 23$ of 174 groups); this is positively correlated with proportion of VZ members, group degree, inequality, and discussion of trochas before March 12 ($n = 16$), and negatively correlated with proportion of CO members and concentration.

A Probit model regression on these factors yields the estimates in table \ref{table:coronavirus:probit_group_trocha}. Unsurprisingly, discussion of trochas before the closure makes groups significantly more likely to discuss trochas after the closure; the Probit model estimates a marginal effect (average of marginal effects at each observation) of 0.17. Many groups where trochas are discussed following the closure are news groups or border-related groups that likely discussed the trochas before the closure. But there is an even stronger marginal effect for proportion of VZ members of 0.36 (again, overall marginal effect averaged from each observation), meaning that groups with more VZ members are more likely to discuss trochas.

\begin{table}[h]
\centering
\caption{Probit regression of trocha discussion dummy in a group (since announcement of border closure), on various group characteristics.}
\label{table:coronavirus:probit_group_trocha}
\begin{tabular}{p{0.25\textwidth} p{0.35\textwidth} p{0.10\textwidth} p{0.15\textwidth}}
\toprule
 & \textbf{Coefficient (Std. Err.)} & \textbf{$z$} & \textbf{P-Value} \\
\midrule

Intercept & $-4.2230 (1.119)$ & $-3.775$ & $0.000^{\ast}$ \\[0.2em]

Prev. Discussion of Trochas & $1.5696 (0.480)$ & $3.271$ & $0.001^{\ast}$ \\[0.2em]

Size & $0.0037 (0.003)$ & $1.472$ & $0.141$ \\[0.2em]

Proportion VZ & $3.3837 (0.983)$ & $3.444$ & $0.001^{\ast}$ \\[0.2em]

Proportion CO & $0.9074 (0.728)$ & $1.246$ & $0.213$ \\[0.2em]

Degree & $0.0013 (0.006)$ & $0.200$ & $0.842$ \\[0.2em]

H-H Concentration & $-1.0468 (1.206)$ & $-0.868$ & $0.385$ \\[0.2em]

Gini/Inequality & $2.4242 (1.331)$ & $1.821$ & $0.069$ \\[0.2em]
\midrule
\multicolumn{4}{c}{$n = 174$\,\,(166 d.f.) \quad\quad Pseudo $R^2 = 0.491$}\\
\bottomrule
\end{tabular}
\end{table}

\begin{figure}[h]
  \centering
  \includegraphics[width=\textwidth]{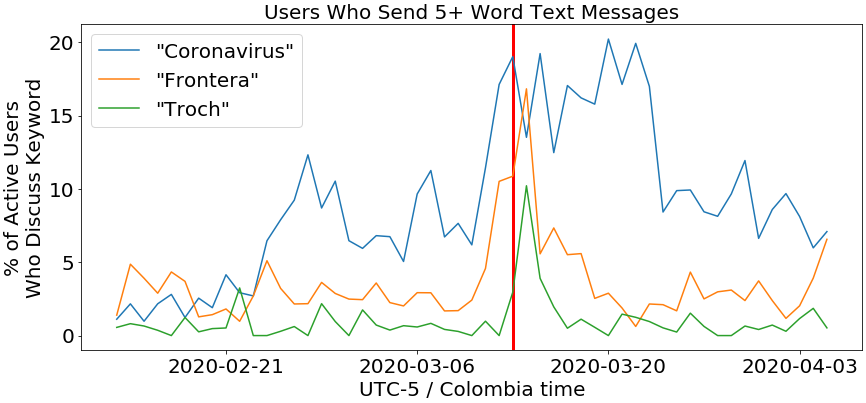}
  \caption{Plot of popularity of border-related topics (\% of active users who mention keyword that day). ``Frontera'' is border, and ``troch'' refers to the illegal crossings. The red line indicates March 13, 2020, when Colombia decided to close its land border with Venezuela.}
  \label{figure:coronavirus:user_proportion_text}
\end{figure}

Analyzing this trend in terms of users reveals the same tremendous rise in trocha interest following the border closure. In figure \ref{figure:coronavirus:user_proportion_text}, we include users who send 5+ word messages on each day, and plot what proportion of such users mention coronavirus, frontera, and trocha. No matter how we understand this trend---whether from the perspective of messages, groups, or users---the conclusion is clear: the coronavirus-related border shutdown unmistakably sparked interest in trochas, and redirected migrants who would've crossed legally to instead attempt irregular crossings.

The effects we've found in WhatsApp users/groups are almost certainly underestimates. Our field work revealed that migrants with more financial resources are more likely to use WhatsApp, but also to cross legally, given the high cost of obtaining Venezuelan documents needed to enter at official crossings. So if even this wealthier, WhatsApp-using subset of migrants has shifted towards using \textit{trochas}, we should expect that much higher proportions of poorer migrants are attempting irregular crossings.

While this effect could've been predicted by almost anyone on the porous border, this is---to our knowledge---the first large-sample evidence of significantly increased interest in \textit{trochas}. The question of how many more irregular crossings actually happened is much more difficult to answer (if not impossible, given who runs \textit{trocha} operations), but any increase has welfare implications for migrants, and political implications for Colombia. Being robbed of smartphones and other valuables is inevitable along \textit{trochas}, and migrants frequently encounter violence and sexual assault; more vulnerable migrants means a greater burden on places like Maicao. We note that coronavirus-related shutdowns have also impacted aid organizations \cite{devex-20200317}, further exacerbating the crisis.

\section{Quarantine}

Coronavirus lockdowns have uprooted life globally; here, we explore how usage patterns have changed across our groups. In Colombia, President Iván Duque announced a 19-day nationwide quarantine on March 20th, which would begin at midnight on March 24th \cite{reuters-quarantine-20200320}. Previously, local officials in cities including Bogotá and Cartagena had announced curfew and isolation measures; the ELN (National Liberation Army), an armed Marxist group long involved in the Colombian civil conflict, even called a ceasefire amidst the coronavirus pandemic \cite{bbc-eln-20200330}.

We separate messages from two relevant periods: a pre-pandemic period that includes all messages on or before March 10 (when most countries were functioning normally; the WHO declared the coronavirus a global pandemic on March 11), and a post-quarantine period, which includes all messages from March 24 and after, when the national lockdown began.

\pagebreak\subsubsection{Nocturnal Activity}

\begin{figure}[h]
  \centering
  \includegraphics[width=0.6\textwidth]{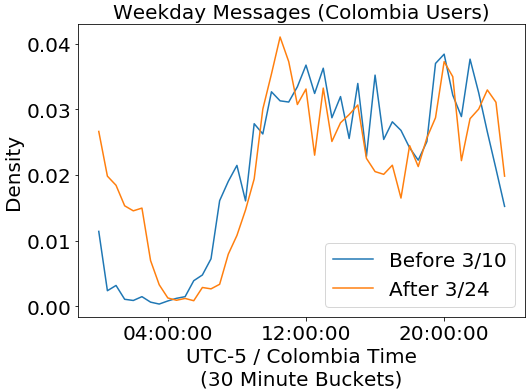}
  \caption{Plots of weekday WhatsApp activity of Colombian users, before and after the coronavirus quarantine.}
  \label{figure:coronavirus:usage_weekday_co}
\end{figure}

Figure \ref{figure:coronavirus:usage_weekday_co} shows a density estimate of WhatsApp usage by Colombian users during weekdays, in both the pre-pandemic and lockdown periods. While some usage patterns remain constant---usage peaks around noon and later around 8 PM---we notice a sharp increase in early AM activity during the lockdown period. Previously, usage would wind down by 12:30-1:00 AM on weekdays, but usage remains high until nearly 3 AM, and there is twice as much activity between 12-1 AM. Message activity also takes longer (around one to two more hours) to rise in the morning, probably because people are taking longer to rise in the morning.

To quantify this discrepancy, we can record a dummy for messages sent between 12:00-5:00 AM; in the pre-pandemic period, this was 1.5\% of all weekday messages, but 7.8\% of weekday messages during the quarantine period ($p < 0.01$).

\begin{figure}[h]
  \centering
  \includegraphics[width=0.6\textwidth]{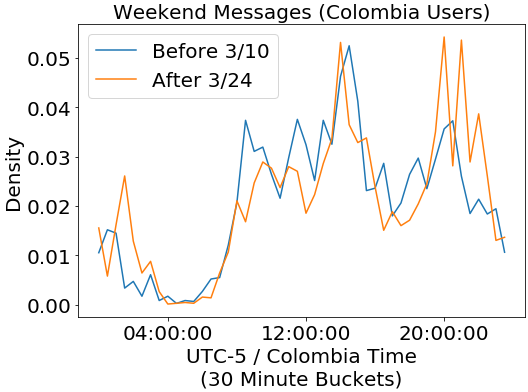}
  \caption{Plots of weekend WhatsApp activity of Colombian users, before and after the coronavirus quarantine.}
  \label{figure:coronavirus:usage_weekend_co}
\end{figure}

Figure \ref{figure:coronavirus:usage_weekend_co} plots the same usage patterns, but on Saturday and Sunday. Usage patterns during the lockdown are more similar to activity before the lockdown; in particular, there is no sharp increase in late-night usage. In the pre-pandemic period, 4.8\% of messages were sent before 5 AM, compared to 7.3\% during the lockdown (difference not statistically significant).

\begin{figure}[h]
\centering
\begin{subfigure}{0.45\textwidth}
  \centering
  \includegraphics[width=\textwidth]{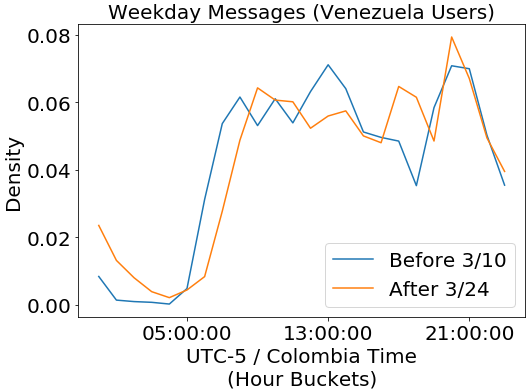}
\end{subfigure}%
\hfill
\begin{subfigure}{0.45\textwidth}
  \centering
  \includegraphics[width=\textwidth]{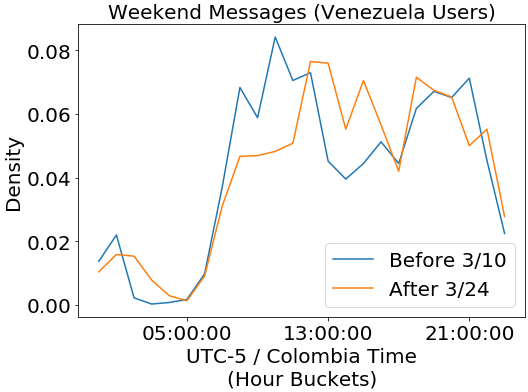}
\end{subfigure}
\caption{Plots of weekday and weekend WhatsApp activity of Venezuelan users, before and after the coronavirus quarantine.}
\label{figure:coronavirus:usage_vz}
\end{figure}

Finally, figure \ref{figure:coronavirus:usage_vz} plots the weekday and weekend activity of Venezuelan users (in hour buckets because there are fewer Venezuelan users and less data in certain buckets). There appears the same trend of more noctural activity on weekdays (and morning activity taking longer to ramp up) during the quarantine period, though with less disparity than with Colombian users; as with Colombian users, weekend activity is relatively unchanged.

\subsubsection{Message Length}

On average, text, audio, and video messages during the lockdown period are longer than messages from before the pandemic. Table \ref{table:coronavirus:message_length} shows average length of messages from both periods, for all messages as well as non-forwarded messages (to better differentiate, say, between professional content producers creating longer content and individual users spending more time on their messages). Across both categories, all messages and non-forwarded (``original'') messages, all message types experience statistically significant increases in length during the quarantine period (with the single exception of non-forwarded audio messages, where the increase is not statistically significant).

\begin{table}[h]
\centering
\caption{Average length of messages from the pre-pandemic (on or before March 10) and quarantine (March 24 and after) periods.}
\label{table:coronavirus:message_length}
\begin{tabular}{p{0.15\textwidth} p{0.25\textwidth} p{0.25\textwidth} p{0.10\textwidth} p{0.15\textwidth}}
\toprule
 & \textbf{Pre-Pandemic} & \textbf{Quarantine} & \textbf{$t$} & \textbf{P-Value} \\
 & (Average Length) & (Average Length) \\
\midrule
\multicolumn{5}{c}{All Messages}\\
\midrule
Text & $20.80$ words & $22.61$ words & $3.55$ & $0.000^{\ast}$ \\[0.2em]

Text & $136.41$ chars. & $146.12$ chars. & 2$.97$ & $0.003^{\ast}$ \\[0.2em]

Audio & $100.00$ secs. & $123.34$ secs. & $5.66$ & $0.000^{\ast}$ \\[0.2em]

Video & $79.06$ secs. & $105.54$ secs. & $10.98$ & $0.000^{\ast}$ \\[0.2em]
\midrule
\multicolumn{5}{c}{Non-Forwarded Messages}\\
\midrule
Text & $15.57$ words & $16.72$ words & $2.59$ & $0.010^{\ast}$ \\[0.2em]

Text & $99.77$ chars. & $107.42$ chars. & $2.67$ & $0.008^{\ast}$ \\[0.2em]

Audio & $72.02$ secs. & $77.89$ secs. & $1.61$ & $0.106$ \\[0.2em]

Video & $72.23$ secs. & $93.13$ secs. & $7.42$ & $0.000^{\ast}$ \\[0.2em]
\bottomrule
\end{tabular}
\end{table}

We might wonder if this is simply a trend in WhatsApp use completely unrelated to the pandemic: over time, perhaps users come to share longer messages. Even though common sense tells us isn't the case, we still perform a falsification test, regressing the lengths of messages on when they're sent (specifically, seconds since 00:00 UTC-5 on February 13, 2020, when we began collecting data).

In these OLS regressions with message length, it turns out that the coefficients on when messages are sent are almost all (very weakly) negative, suggesting that if there's any trend in WhatsApp use, it's that messages become shorter over time. Out of eight falsification tests (regressions of text word length/text character length/audio length/video length on when they were sent, for all messages and for non-forwarded messages), only the regression for non-forwarded audio messages has a (miniscule) positive coefficient, likely the result of random chance.

\subsubsection{General Activity}

We can also examine user activity more generally, by considering how many messages each users sends on each day they're active. Specifically, for each user, we construct a user-date pair for day $i \in [1, 53]$ if they're active on day $i$, and for that user-date pair record how many messages they send on day $i$.

In the pre-pandemic period, on each day they were active, users sent an average of 5.51 messages, while in the quarantine period, users sent on average of 6.09 messages on each day they were active ($t = 3.02$, $p < 0.01$). We perform another falsification test, regressing number of messages for each user-date pair on days since February 13, 2020, which yields a small and non-significant positive coefficient (0.0047, with standard error 0.005).

Messages in the quarantine period also receive more replies, on average, than messages in the pre-pandemic period, at 0.624 compared to 0.572 ($t = 4.88$, $p < 0.001$). To better account for possible discrepancies in content type, we can also examine the average virality of messages in reply cascades during both periods; again, quarantine-period messages in reply cascades are of average virality 1.624, while pre-pandemic messages in reply cascades are of average virality 1.531 ($t = 7.50$, $p < 0.001$). As a falsification test, we regress virality on seconds since 00:00 UTC-5 Feburary 13, 2020 for all messages in reply casacdes, and obtain a very weakly negative coefficient, suggesting there isn't some general trend of increased virality unrelated to the pandemic.

\chapter{Conclusion\label{ch:conclusion}}

There are no straightforward conclusions from our work; there shouldn't be. Instead, what we've found broadly falls along three themes: characterizing public WhatsApp groups as a data source, understanding migrant dynamics within these groups, and theorizing how intervention in WhatsApp groups can best improve the lives of migrants.

\section{Public WhatsApp Groups as a Data Source\label{ch:conclusion:datasource}}

In our field work, we found that smartphones were relatively popular amongst Venezuelan migrants, though estimates for their prevalence varied wildly. Among those with smartphones, however, literally everyone used WhatsApp, primarily for communicating with family and friends in one-on-one chats and small (private) groups---indeed, WhatsApp is typically the primary reason for migrants to own a smartphone, with one migrant even calling the app ``primordial.''

Everyone---elderly people and migrants without smartphones included---knew about WhatsApp and Facebook, and nearly everyone knew about the existence of public groups on these networks. Around 50\% of migrants with smartphones reported being active members of such groups, either currently or previously, and said they turned to these groups for news, personal transactions, and employment opportunities. But even amongst migrants who frequented public groups, few placed significant trust in such groups---everyone had a story of a friend or acquaintance who fell into trouble, usually from fraudulent employment offers.

What this means is that activity in public WhatsApp groups may not be representative of WhatsApp use by migrants: public groups involve more strangers and lower trust than the private groups/chats that matter most to migrants. WhatsApp users, of course, are also not representative of migrants in general; migrants on WhatsApp are typically wealthier and more educated than those without smartphones (the cost of phones and data plans is typically what limits smartphone use).

In spite of all this, our field work suggests that public WhatsApp groups \textit{do} retain some significance in the experience of Venezuelan migrants to Colombia. Migrants may not trust these groups very much, or frequent these groups nearly as much as groups with close friends, but they do turn to public groups, at least once in a while.

More than this, our data suggests that relationships and activity in these groups provide reasonable approximations of overall WhatsApp use by migrants. Public WhatsApp groups with strangers are incredibly different from private chats with friends, but our data tells us that the two are much more closely connected than, say, WhatsApp groups and Facebook groups or WhatsApp groups and Twitter.

In Chapter \ref{ch:members}, we saw that the dynamics of membership in these groups follow patterns we should expect, both from general social networks and our specific context. As with most social/social media networks \cite{networks-book-2010}, we find power-law distributions in group participation, and discover giant connected components in networks of both groups and users.

Most groups had more users from Colombia, an important sanity check since we expect these groups to center on Venezuelan migrants in Colombia; groups with more Venezuelan users were typically larger. Users from Ecuador and Peru were equally well-connected to Venezuelan and Colombian users, while users from Chile were better connected to Venezuelan users---a natural result of Ecuador and Peru sharing land borders with Colombia, and Chile being much farther away.

We also calculated entropy as a measure of geographical heterogeneity within groups. Groups with few Colombian users were relatively diverse, while groups with few Venezuelan users were more homogeneous, making clear that these groups do center on Colombia. Larger groups and more heterogeneous groups are connected to more groups (where we defined connection as sharing one or more users), and the relationship holds when controlling for each factor.

In Chapter \ref{ch:messages}, we saw that text messages follow the power-law distribution we expect of more-personal communications platforms like SMS and private WhatsApp chats (as opposed to more-public communication like Twitter tweets, where length peaks at around 10 words). We modeled text topics, and found that they center on topics we would expect migrants to discuss, like general greetings, Venezuelan politics, and the coronavirus. We also noticed power-law distributions in average group activity, measured in number of messages per day, which fits the common assumption of  exponential distribution of user participation (popularly known as the 80/20 rule) \cite{guo-content-2009}.

We end the restating of results here, but the list goes on in further chapters. Many of these patterns are obvious and expected, fitting what we know of online social media networks and the context of Venezuelan migrants to Colombia. There is no novel discovery here; instead, these findings should reassure us that public WhatsApp groups aren't as skewed, distorted, or misrepresentative as we might initially fear.

In other words, public WhatsApp groups are used by only a subset of migrants, and even that subset uses these groups differently than groups with friends and family, but public WhatsApp groups \textit{do} allow us to meaningfully research the complicated dynamics of the Venezuelan migrant crisis.

\section{Migrant Dynamics within WhatsApp Groups\label{ch:conclusion:dynamics}}

We found a broad range of results related to how migrants connect to each other in these groups. Some of those results we discussed in Section \ref{ch:conclusion:datasource}, but below we discuss several more.

We found that larger groups and more geographically heterogeneous groups were less concentrated (controlling for these and other factors in an OLS regression) in Chapter \ref{ch:messages}, while geographically heterogeneous groups were also more unequal. We put the latter in the context of cross-border groups where many transient users enter to conduct one-time business or ask one-time questions, while a contingent of stable users maintain the group; such a structure which would produce geographically heterogeneous but highly unequal groups.

Later in the same chapter, we found that while most multimedia content is only shared once, content that is first shared in less concentrated, highly unequal groups is more likely to be re-shared. We found that the average number of replies, and even the average virality \textit{of} replies (which allows us to generalize across groups with many/few replies or with different content types), are both positively linked to group geographic heterogeneity, and both negatively linked to group concentration. We discuss these patterns more later on, in Section \ref{ch:conclusion:intervention}.

In Chapter \ref{ch:misinformation}, we found that fake news is much longer than other text messages, and that fake news and scams receive replies at lower rates than other messages, perhaps illustrating how users perceive and respond to misinformation differently than other messages. We showed that Venezuelan users are much more likely to share fake news, at almost double the rate of Colombians, but that Colombian users share economic scams at much higher rates. Concentrated groups are more likely to breed fake news and internet scams, but inequality in groups was linked to lower prevalence of misinformation amongst messages. The former conclusion aligns nicely with common understandings of ``echo chambers''; the latter conclusion is a bit counterintuitive, but likely stems from message-poor users either chiming in with alternate perspectives, or such users subconsciously causing other members to filter what they share.

In Chapter \ref{ch:coronavirus}, we explored dynamics related to the coronavirus pandemic, and found that interest in \textit{trochas}, illegal crossings controlled by armed criminal groups, proliferated immediately after Colombia announced it would close its border with Venezuela. Groups with a higher proportion of members from Venezuela were more likely to discuss these illegal crossings; on March 14, the day the border closure took effect, nearly 10\% of text messages involved trochas, and trochas were being discussed in 25\% of active groups. Though nearly anyone on the border would have predicted increased interest in \textit{trochas} following the border closure, this is the first large-sample evidence of such interest, and we argued that this is only an underestimate.

We also explored how the coronavirus shifted usage patterns in WhatsApp groups. Particularly, all types of messages were longer during the quarantine period than in pre-pandemic times, and there was significantly more late-night activity on weekdays (but not weekends). These may seem like obvious findings, but they allow us to put numbers on the efficacy of shutdown measures in a country where traditional data sources are less effective. They allow us to measure, at large-scale and with little cost, if migrants---most of whom work in informal roles, and many of whom live in informal settlements---are staying home or reducing their hours on the street. If we had more detailed location information, we might be able to test, for example, if the messaging activity of migrants in Bogota, which was one of the first places to begin quarantine measures, changed earlier than in other cities.

\section{Intervention in WhatsApp Groups\label{ch:conclusion:intervention}}

In interviews with both migrants and aid organizations, we heard very few stories of official actors---aid organizations and governments---paying attention to public WhatsApp groups. In the previous section, we discussed important analytical results obtained from these groups, and those are certainly reasons to at least monitor groups. More than this, however, there are also reasons for official actors to consider actively \textit{intervening} in public WhatsApp groups. These reasons fall along two broad avenues: reducing harms and sharing official information.

The first centers on misinformation in these groups which, beyond directly harming victims who believe false information or fall prey to internet scammers, lowers the trust and social responsibility shared in these groups. Migrants may not trust true information about crossing the border in a group with many internet scams, or migrants may hesitate to ask about medical clinics in a group that shares fake coronavirus cures; this is exactly broken windows theory.

In Chapter \ref{ch:misinformation}, we shared a rather-successful methodology for identifying fake news and economic scams using only public fact-checking sources and very limited manual verification. We then characterized users and groups among which fake news is most prevalent, and showed how these results differed for scams. Highly concentrated groups, for example, are more likely to be breeding grounds for both fake news and scams, making it worthwhile to target interventions at such groups. And, as we already stated, Venezuelan users are much more likely to share fake news, while scams come disproportionately from Colombian users.

We then showed how both fake news and scams often involved slightly-altered variants, and demonstrated how automated approaches can be taken to flag scam messages: we trained several machine learning classification methods on tokenized messages. This involved a nuanced discussion of why the underlying language structure of scams makes certain classifiers naturally better than others: logistic regression fails to take into account non-linear relationships between tokens, for example, while nearest neighbor classifiers allow detection of subtly altered scams.

The second avenue of possible intervention---sharing useful information in public WhatsApp groups---relies on understanding what characteristics of groups spur dissemination of information. When disseminating official information, our ultimate goal is for users in public groups to then forward this information to private chats, reaching the 50\% of migrants on WhatsApp who don't use public groups and would otherwise be inaccessible.

In Chapter \ref{ch:messages}, we showed that images that first appear in diverse, unequal, and less concentrated groups are more likely to be re-shared; the same result held for images and videos that were shared for longer periods. When analyzing messages in reply cascades (i.e., messages with replies or that are replies) in Chapter \ref{ch:replycascades}, we showed that messages in more diverse and less concentrated groups had greater virality, their dissemination being more decentralized and organic. These findings on information spread should shape how official actors disseminate information, spurring them to focus on geographically diverse groups with many ``message-poor'' members. This second conclusion may be somewhat counter-intuitive, but we can imagine silent users as the best spreaders of information to other channels.

We also showed in Chapter \ref{ch:messages} that while text messages are generally short, over half of audio and video messages are longer than 30 seconds, possibly offering an alternative approach to directly sending textual information.

\section{Limitations and Future Work}

In Section \ref{ch:datamethod:data-limitations}, we discussed how our methodology for collecting data only recorded the cryptographic hashes of images and videos, driven by a desire to both reduce our project's technical burden, and to follow the advice of researchers who've advised against saving multimedia content that may be obscene and/or illegal.

This necessarily means that our analysis of multimedia content is less interpretable and effective than our analysis of text. In Chapter \ref{ch:misinformation}, we detected countless variants of fake news and scams, and it's certain that images, audio recordings, and videos are also slightly altered as they're shared in our groups. An approach like perceptual hashing can help us detect image/audio/video variants in the same way that cosine similarity detects variants of texts, and image recognition and OCR techniques may grant us interpretability of multimedia content. Yet the unassailable gold standard is simply manual processing and labeling of multimedia content, which is a realistic possibility given services like Amazon's Mechanical Turk.

Even in our analyses of text, our techniques were rudimentary, only using tokens obtained from basic text pre-processing steps. But there are important textual relationships in our context. A number of scams, for example, involve fake links that appear to be links to join WhatsApp groups, with the domains \url{whatsbpp.com} or \url{whatsclpp.com} and so on; detecting URLs that are one letter off from \url{whatsapp.com} would nearly perfectly identify these variants. More generally, natural language is much more complicated than the bag of words approach we took, and sophisticated methods exist to analyze sentence and discourse structure and meaning. As other researchers of textual misinformation have noted, many in the NLP field ``have proposed learning methods to automatically detect fake messages ranging from lexical to deep learning approaches exploring linguistic and network features'' \cite{resende-textual-2019}.

We did relatively little work with temporal trends in our data, only examining patterns that arose during the coronavirus pandemic. Our use of outside sources was also limited, with only the two public fact-checking databases we used to identify fake news. Both of these factors necessarily mean that the power of our data is limited, restricted to insights mostly derived from the data itself. Yet things like crime, xenophobia, and even actual migration counts are important aspects of the migrant crisis, and one of the highest possibilities of WhatsApp data would be stronger connections to these other themes. Such a result would require us to bring in outside sources, like newspaper and government databases, and expand the time period for which we collected to data, to better parse out meaningful trends amongst much noise.

Finally, there are improvements abound to our methodology for collecting data. Because of time and resource constraints, we only looked in a small set of Facebook groups for links to groups, but links surely exist in many more groups, as well as elsewhere on the internet. After joining groups, we could have taken steps to mitigate the risk of us being kicked out of groups, as well as reduce any possible effects of us joining groups. During this project, we set our names to common Spanish female names, and used profile pictures depicting Latin American women (per \cite{slonim-gender-2010}, ``most studies [of strangers] find females more trustworthy than men''); a more rigorous approach might involve obtaining Colombian telephone numbers, and perhaps sending an occasional message in groups that require members to introduce themselves and/or stay active.

There are many more shortcomings of our work---too many to list---but as the last thought in this thesis, we re-emphasize that all of our work is preliminary. We've encountered a number of interesting findings in various directions, but there is so much more that can be done in \textit{any} of these directions.

\appendix 
\chapter{Technical Implementation\label{ch:appendicies:implementation}}

\section{Collecting Data from Groups}

Our process of traversing groups involves loading WhatsApp Web, clicking each group in the sidebar, and then in each group recording the group’s members and logging the group’s messages.

Even the first step, clicking through each group, turned out to much more difficult than it sounds. In the underlying HTML, WhatsApp Web doesn’t load a user’s entire sidebar at a time, only the 15 or so chats that are currently visible on screen. Initially, we attempted to navigate through all visible groups and then scroll down, using the bottom-most group to determine scroll displacement, though even that was complicated by the fact that WhatsApp randomly orders the 15 visible groups in the HTML (as opposed to ordering them top-to-bottom).\footnote{The solution is to check each HTML element (which represents a group) for a webpage coordinate, and find the bottom-most group by the greatest Y displacement.} But the sidebar, ordered by chats with the most recent activity, constantly changes when new messages are received. The first 15 groups may be drastically different 10 minutes later.

We concluded it was necessary, then, to click through groups based on their unique characteristics, and not simply by their position in the sidebar. Using the unique group identifier found in group icon links, as described in the section on joining groups, requires waiting for profiles picture to load every time we poll to see if a group has already been checked; with $n$ groups, this is significantly more than $n$ times, and close to $\frac{n^2}{2}$ (say the group order never changes, then to determine the first unchecked group each time, we must first check the unique identifiers of all the groups we’ve already checked\footnote{We avoid taking any shortcuts like caching position of last checked group, since sidebar order can always change drastically. A programmer with more time on their hands would probably create a refined solution based on initially taking shortcuts and at the end checking for possibly missed groups.}). Waiting for the profile picture identifier to load turns out to be extremely slow, so we instead relied on only checking title HTMLs, which load much more quickly. In our case, there were no duplicates, but this is certainly not guaranteed; checking for profile picture identifiers and falling back to title HTMLs would be less likely to have issues.

Once we click on a group, we record the members in that group. The subtitle, which is directly visible (and hence recordable) from the main group page, typically contains a comma-separated list of all members’ phone numbers; sometimes, however, it has other messages, like “+57 XXX XXX is typing…” or “Click here for group info.” Our script tries to capture the member list directly from the subtitle for a certain period (long enough so that anyone typing would stop typing), and if that fails, reverts to clicking on a scrollable member list in the sidebar. The latter process takes significantly longer, since the process must iterate between scrolling through the members list and waiting for members’ details to load, 15 or so members at a time.

Finally, we read messages. The traverseGroups script first jumps to the latest message (by default, WhatsApp Web scrolls to the earliest unread message). WhatsApp Web loads messages in reverse chronological order, so only the most recent messages are present in the HTML. Earlier messages are loaded as a user (or bot) scrolls up in the chat. We scroll up until we reach messages from a certain time, passed as argument into the function (i.e., we scroll until the time when we last collected data).

\section{Preprocessing Message Data}

\subsection{Duplicate Messages}

traverseGroups, the script we use to collect message data from groups every few hours, outputs a file for each group/read-time pair with the messages from that group read at that time. Duplicate messages arise in these files for two reasons:
\begin{enumerate}
  \item Our methodology involves two independent WhatsApp accounts/smartphones joining and collecting data from each group.
  \item The ``scrolling up'' process described above is imperfect. Namely, if traverseGroups checks group X at time $t$ hours and later at $t+3$ hours, but group $X$ has no new messages in $[t, t+3]$, then traverseGroups will read and record duplicate messages. At $t+3$, traverseGroups will not scroll up, but it will still read the messages it read previously.\footnote{This is due to WhatsApp Web's design, which makes ``checking for messages since at least 7:00 AM'' much simpler than ``checking for messages \textit{only} since 7:00 AM''.}
\end{enumerate}

Duplicate messages can be identified as messages with the exact same contents, sent by the same user at the same datetime in the same group. It's easy enough to identify such duplicates with the standard Python data analysis library \textit{pandas}, but we must take care to preserve true duplicates. Namely, a user may actually send two identical messages at the same time in the same group (whether intentionally or not). We can identify such true duplicates as duplicates that are read by the same traverseGroups process (server/read-time pair). For example, if the server collecting data from smartphone $A$ at datetime $d$ reads duplicate messages in a group, then those are true duplicates. We identify them since they they share the server/read-time pair $(A, d)$; the sources of duplicates described above either involve different servers or different read-times.

A final nuance is that we may read each set of true duplicates multiple times. For example, if two of our WhatsApp accounts are in the same group where a true duplicate pair is sent, both servers will collect that true duplicate pair. Or one of the servers may, at a later time that traverseGroups is run, scroll up enough to again read the message. We account for this by, for each true duplicate pair, only preserving its occurrence the first time it's read by any server/time pair.

In our data, true duplicate messages account for just over 1\% of all messages, not an insignificant amount.

\subsection{Groups Whose UIDs Change\label{ch:appendicies:implementation:groupschange}}

As we described in Section \ref{ch:datamethod:identifygroups}, on uniquely identifying groups, WhatsApp Web's design and intended use case make it difficult to uniquely identify groups. We settled on a heuristic, which we called \textit{uid}, based on (for groups with a profile picture) the group's unique identifier (if it has a profile picture), discoverable in the link to its profile picture, or (for groups without a profile picture) the cryptographic hash of its title.

As we said, this method isn't foolproof, and in general there isn't a foolproof way to uniquely identify groups. For example, in the span of an hour, users in a group may change its title, remove its profile picture, send hundreds of new messages, and completely change the group's membership---it'd be impossible to link this modified group to the original group. Less extreme cases where our \textit{uid} heuristic fails include a group without profile picture changing its title, or a group removing its profile picture and changing its title.

To discover groups that change their \textit{uid}, we rely on text similarity techniques, as we did to identify misinformation in Chapter \ref{ch:misinformation:labeling}. We tokenize all messages (removing stopwords and stemming all words), and for each group create a feature vector of tokens sent in messages in that group. We then compute cosine similarity across groups, and for all group pairs with cosine similarity $> 0.5$, do the following:
\begin{enumerate}
  \item Compute the number of identical messages (same content, same user, same datetime) across the two groups. After a group changes \textit{uid}, traverseGroups may again read the same messages, because of the scrolling nuance described previously.
  \item If this number is over 60\% of the minimum total number of messages of each group, we merge the groups.
\end{enumerate}

We ended up finding 23 groups where the $\textit{uid}$ had changed, likely because they changed their titles or added/removed profile pictures. Most of the groups we merged had a 100\% match in the second step, meaning that in the duplicate group with fewer messages, all of the messages were identical to the original group.

After merging groups, we again needed to remove duplicate messages (while again preserving true duplicates).

\section{Code}

All code for this thesis is provided at \url{https://github.com/admchg/thesis-production}.

\chapter{Field and Interview Notes\label{ch:appendicies:field}}

For brevity and privacy, this section has been left out of the online public version. Please contact us to request a copy with field and interview notes included.

\singlespacing
\bibliographystyle{unsrt}

\cleardoublepage
\ifdefined\phantomsection
  \phantomsection  
\else
\fi
\addcontentsline{toc}{chapter}{Bibliography}

\bibliography{thesis}

\end{document}